\documentclass{jfm}
\usepackage[utf8]{inputenc}

\title[Asymptotics of wall-turbulence fluctuations]
{Reynolds number asymptotics of wall-turbulence fluctuations}

\author[Xi Chen and Katepalli R. Sreenivasan] {Xi Chen$^{1}$ and Katepalli R. Sreenivasan$^2$}

\affiliation{$^1$ Key Laboratory of Fluid Mechanics of Ministry of Education, Beihang University (Beijing University of Aeronautics and Astronautics), Beijing, China
\\$^2$ Tandon School of Engineering, Courant Institute of Mathematical Sciences, Department of Physics, New York University, New York, USA}

\pubyear{20xx} \volume{000} \pagerange{000--000}
\date{?; revised ?; accepted ?. - To be entered by editorial office}

\usepackage{graphicx}
\usepackage{xcolor}
\usepackage{subfig}

\usepackage{authblk}
\usepackage[]{natbib}
\usepackage{numprint}
\usepackage{graphicx}
\usepackage{amssymb}
\usepackage{mathtools}
\usepackage{indentfirst}
\usepackage{multirow}

\begin{document}

\maketitle

\begin{abstract}

In furtherance of our earlier work (Chen \& Sreenivasan, {\it J. Fluid Mech.} {\bf 908}, 2021, p. R3; {\bf 933}, 2022, p. A20---together referred to as CS hereafter), we present a self-consistent Reynolds number asymptotics for wall-normal profiles of variances of streamwise and spanwise velocity fluctuations as well as root-mean-square pressure, across the entire flow region of channel and pipe flows and flat-plate boundary layers. It is first shown that, when normalized by peak values, the Reynolds number dependence and wall-normal variation of all three profiles can be decoupled, in excellent agreement with available data, sharing the common inner expansion of the type $\phi^+(y^+)=f_0(y^+)+f_1(y^+)/Re^{1/4}_\tau$, where $\phi^+$ is one of the quantities just mentioned, and the functions $f_0$ and $f_1$ depend only on $y^+$. Here, the superscript $+$ indicates normalization by wall variables. Secondly, by matching the above inner expansion and the inviscid outer flow similarity form, a bounded variation $\phi^+(y^\ast)=\alpha_\phi-\beta_{\phi}y^{\ast{{1}/{4}}}$ is derived beyond the peak where, for each $\phi^+$, the constants $\alpha_\phi$ and $\beta_{\phi}$ are independent of $Re_\tau$ and $y^\ast$ ($=y^+/Re_\tau$, $Re_\tau$ being the Reynolds number based on the friction velocity)---also in excellent agreement with simulations and experimental data. One of the predictions of the analysis is that, for asymptotically high Reynolds numbers, a finite plateau $\phi^+\approx\alpha_\phi$ appears in the outer region. This result sheds light on the intriguing issue of the outer shoulder of the variance of the streamwise velocity fluctuation, which should be bounded by the asymptotic plateau of about 10. 
\end{abstract}

\section{Introduction}

The Reynolds number dependencies of the variances of streamwise and spanwise velocity fluctuations as well as pressure fluctuations are thought to present exceptional challenges for the classical notion of wall scaling \citep{marusic2010wall,Smits2011ARFM}. A salient example is that, when scaled in wall units, the peak values of these quantities near the wall grow with increasing Reynolds number (though the peak locations are remarkably invariant; see, e.g., \cite{Sreenivasan1989}). In CS, the growth of these peaks was cast as a finite Reynolds number effect, and it was shown that a bounded growth model (discussed below) fits the data better. In this paper, we turn attention to wall-normal profiles of the variances of these fluctuations. This work is an alternative to the attached-eddy hypothesis by \cite{Townsend1956}, which ascribes a logarithmic decay for fluctuations in the outer flow as
\begin{equation}\label{eq:townsend}
\phi^+(y^\ast) = B_\phi-A_{\phi} \ln y^\ast.
\end{equation}
Here, $\phi^+$ represents the variance of $\langle uu \rangle^+$ or $\langle ww \rangle^+$; the superscript $+$ indicates normalization by ${u_\tau}$ and $\nu$, and $u,v,w$ for fluctuation velocities in the streamwise ($x$), wall-normal ($y$) and spanwise or azimuthal ($z$) directions; $y^\ast=y/\delta$ where $\delta$ is the flow thickness; the slope $A_\phi$ and intercept $B_\phi$ are constants independent of $y^\ast$ and the friction Reynoldss number $Re_\tau = u_\tau \delta/\nu$, where $u_\tau\equiv (\tau_w/\rho)^{1/2} $ is the friction velocity, but may depend on $\phi$.

The rationale behind (\ref{eq:townsend}), as discussed by \cite{Marusic2019ARFM}, is that the number density of the attached eddies that contribute to turbulent fluctuations varies inversely with $y^\ast$, and an integration with respect to $y^\ast$ leads to the total fluctuation intensity given by (\ref{eq:townsend}). Some consequences of this idea have been explored in laboratory measurements (EXP) \citep{Metzger2001,Hultmark2012PRL,vincenti2013,LongPipe2017,Samie2018JFM,Ono2022} as well as direct numerical simulations (DNS) \citep{WuxiaohuaTBL2009,SO2010,Jimenez2010,Moser2015,Pirozzoli2021,hoyas2022,Yao2023JFM}. The resulting findings have been discussed in terms of mixed scaling \citep{Degraff2000}, multi-regime of the power-law spectrum \citep{Vassilicos2015,Vassilicos2017}, inner-outer interactions \citep{marusic2017PRF} and random addictive process \citep{Xiang_12}. The notion of attached eddies has been extended to study high-order moments of single point velocity fluctuations \citep{MM2013} as well as to velocity structure functions \citep{desilva2015}.

Pressure fluctuations have also received attention in the past sixty years \citep{Bradshaw1967,Klewicki2008,Panton2017}, in part because of their importance for aircraft cabin noise. By extending Townsend's attached-eddy hypothesis, \cite{Bradshaw1967} obtained a $k^{-1}$ spectrum by an inner-outer matching in wavenumber space and hence an $\ln Re_\tau$ growth of wall pressure fluctuation. The $k^{-1}$ spectrum so deduced is marginally detected in laboratory boundary layers at $Re_\tau\approx6000$ \citep{Tasuji2007}, but not in the DNS data so far. 
This unsatisfactory situation prompted \cite{Panton2017} to develop alternative matching analysis in the spatial domain, also yielding the log-profile of (\ref{eq:townsend}). This is reminiscent of \cite{Hultmark2012JFM}  who derived the $\ln y^\ast$ variation in pipes by matching $\langle uu\rangle^+$ between the inner and outer regions. It is worth noting that the $Re_\tau$ effects included in these models are not part of Townsend's original attached-eddy hypothesis.

While the above works suggest a boundless growth of turbulence peaks as $Re_\tau\rightarrow\infty$, CS argued that the observed variations are bounded at very high Reynolds numbers and follow a defect law of the type
\begin{equation}\label{eq:CS2022a}
\phi^+_p (Re_\tau)=\phi^+_\infty-c_{\phi, \infty}Re^{-1/4}_\tau.
\end{equation}
Here, $\phi^+_\infty$ the asymptotically bounded value of $\phi^+_p (Re_\tau)$ and $c_{\phi,\infty}$ are the fixed coefficients. The underlying physics depends on the slight imbalance that exists between wall dissipation and maximum production in the turbulent energy budget at any finite Reynolds number, and on their tendency to eventually balance each other. Subsequently, \cite{Monkewitz2021} showed that an asymptotic expansion of $\langle uu\rangle^+$ profiles with the $Re_\tau^{-1/4}$ gauge function from CS reproduced data better than $\ln Re_\tau$ \citep{Smits2021a}. Recent measurements of \cite{Ono2022} in pipes for $Re_\tau$ ranging from $990$ to $20750$ are also supportive of the bounded behavior. The results of CS have been checked against DNS data in the open channel \citep{YaoChenHussain2022JFM} and compressible channel \citep{vallet2023}, indicating the universality of the bounded behavior for different flow conditions. Indeed, \cite{hoyas2022} reported that the wall pressure fluctuation in their DNS channel data for $Re_\tau$ up to $10^4$ might also be bounded.  

Yet, to differentiate between the two sets of results beyond doubt, one clearly requires much higher Reynolds numbers than currently covered (or likely to be covered for the foreseeable future) in laboratory experiments or DNS \citep{Nagib2022APS}. Measurements in the atmospheric boundary layer \citep{Metzger2001,Metzger2007} might be thought of as helpful but various uncertainties characteristic of field measurements prevent a decisive conclusion there also. At the current stage, new theoretical ideas are highly desired to provide additional insights. As noted by \cite{Klewicki2022}, the bounded growth, once accepted, would necessitate a reassessment of a number of earlier empirical findings. In this spirit, we obtain an alternative to (\ref{eq:townsend}) by using the bounded behavior of (\ref{eq:CS2022a}), providing a more complete description of the asymptotic behavior of wall turbulence (including pressure).

Specifically, we first decouple the $Re_\tau$ dependence from the wall-normal variation for the root-mean-square (rms) profiles by using peak values for normalization. Then we develop a matching procedure between the inner viscous and outer inviscid regions, yielding in the outer flow a defect law of the type  
\begin{equation}\label{eq:CS:outer28}
\phi^+(y^\ast)=\alpha_\phi-\beta_\phi y^{\ast1/4}.
\end{equation}
Here, $\phi^+$ represents not only $\langle uu \rangle^+$ and $\langle ww \rangle^+$ but also the rms of pressure fluctuation $p'^+$ (superscript prime denotes the rms). 

To verify (\ref{eq:CS:outer28}), 
DNS data sets are collected for those with clear $Re_\tau$ trend for $\langle uu \rangle^+$, $\langle ww \rangle^+$ and $p'^+$, all publicly available. In particular, we use the DNS on channels by \cite{Moser2015} for $Re_\tau$ from $550$ to $5200$, on pipes by \cite{Pirozzoli2021} for $Re_\tau$ from $500$ to $6000$, and on TBLs by \cite{schlatter2009,schlatter2010simulations} for $Re_\tau$ from $490$ to $1270$. Higher $Re_\tau$ data in the literature \citep{Sillero2013,hoyas2022} are also included for comparison. For experiments, we select $\langle uu \rangle^+$ data from the Princeton pipe by \cite{Hultmark2012PRL} for $Re_\tau$ from $5411$ to $98187$, from the Princeton TBLs by \cite{vallikivi2015} for $Re_\tau$ from $4635$ to $25062$, and from the Melbourne TBLs by \cite{Samie2018JFM} for $Re_\tau$ from $6000$ to $20000$. Channel experiments are not collected here because their (limited) $Re_\tau$ variation has been covered by the DNS of \cite{Moser2015}. Note that the data uncertainty, especially concerning the probe resolution in experiments and grid resolution in the DNS, are not addressed in this paper (see CS for a brief discussion). We do wish to state, however, that there is much need for better-resolved data.   

The paper is organized as follows. Section 2 presents data collapse for the inner flow region, which leads to the uniform expansion scheme presented there. Section 3 begins with the verification of inviscid similarity in the outer region, followed by the derivation of the defect decay, using comprehensive data comparisons. Section 4 is devoted to a discussion of the geometry effect. A perspective and summary of the results are given in section 5.

\section{$Re_\tau$-scaling for near wall region}
When scaled by wall variables, an asymptotic expansion for $\langle uu\rangle^+$,
$\langle ww\rangle^+$, and $p^{'+}$, represented by $\phi^+$, can be written as
\begin{equation}\label{Eq:MonTay}
\phi^+(y^+,Re_\tau)=f_0(y^+)+f_1(y^+)g(Re_\tau)+f_2(y^+)g^2(Re_\tau)+h.o.t.,
\end{equation}
where $g$ is the gauge function of $Re_\tau$; $f_0$, $f_1$ and $f_2$ (as well as $f$ introduced below in (2.2)) are general functions depending merely on the wall-normal distance $y^+$, and $h.o.t.$ indicates high order terms. For the streamwise mean velocity $\phi^+=U^+$, a first order truncation of (\ref{Eq:MonTay}) is fairly accurate near the wall. 
as already remarked, for turbulent fluctuations, the wall scaling has received challenges, reflected in the notable $Re_\tau$-dependence for the near wall $\langle uu \rangle^+$ and $\langle ww \rangle^+$ as well as pressure fluctuation $ p'^+$. Note that $\langle uv \rangle^+$ and $\langle vv \rangle^+$ are thought as `active' motions by \cite{Townsend1956}, with bounded maximum values, with the latest evidence being given, for example, in \cite{Smits2021a} and \cite{YaoChenHussain2022JFM}. In the rest of this section, we first show data collapse of $\langle uu \rangle^+$, $\langle ww \rangle^+$ and $p'^+$ after normalization by their corresponding peak values, and then summarize a common expansion for these quantities, which is actually a second-order truncation of (\ref{Eq:MonTay}).

\subsection{Data collapse for the inner flow region}

\begin{figure}
\centering
\subfloat[]{\includegraphics[trim = 0.5cm 10cm 15cm 1cm, clip, width = 6.5 cm]{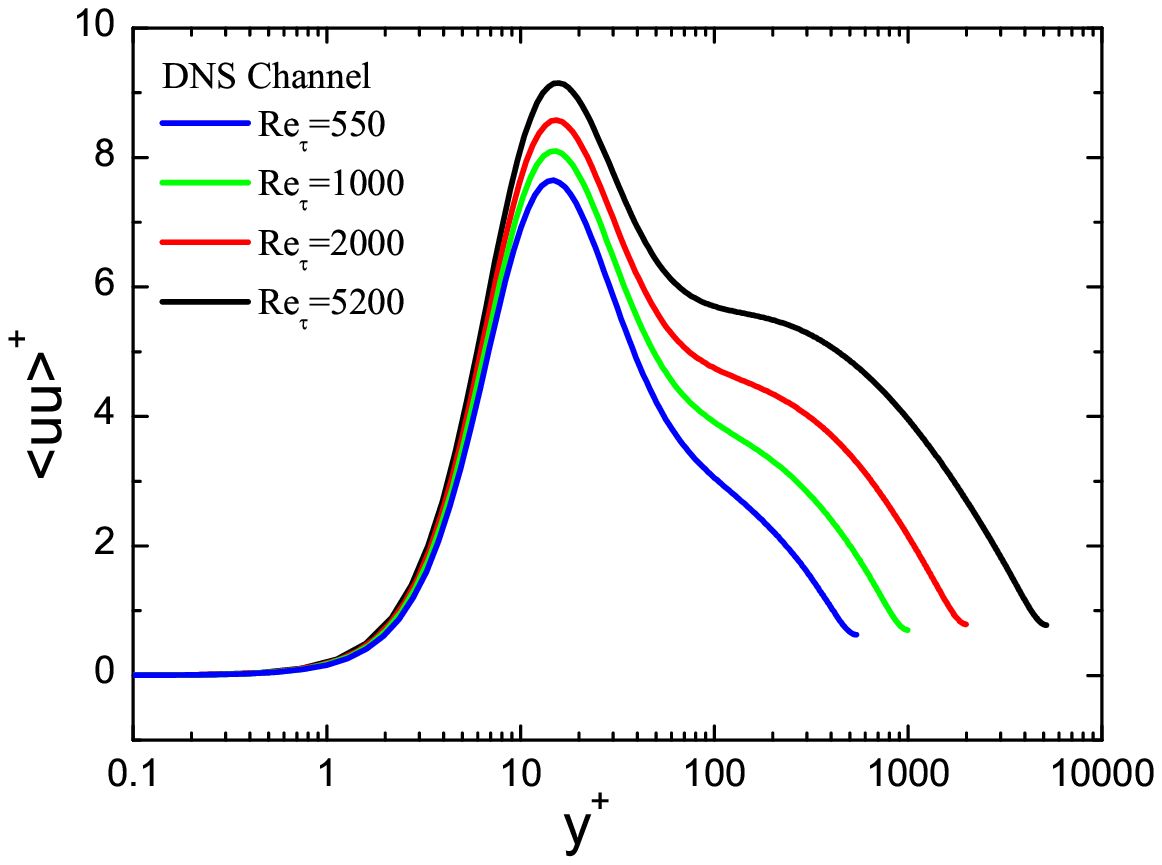}}
\subfloat[]{\includegraphics[trim = 0.5cm 10cm 15cm 1cm, clip, width = 6.5 cm]{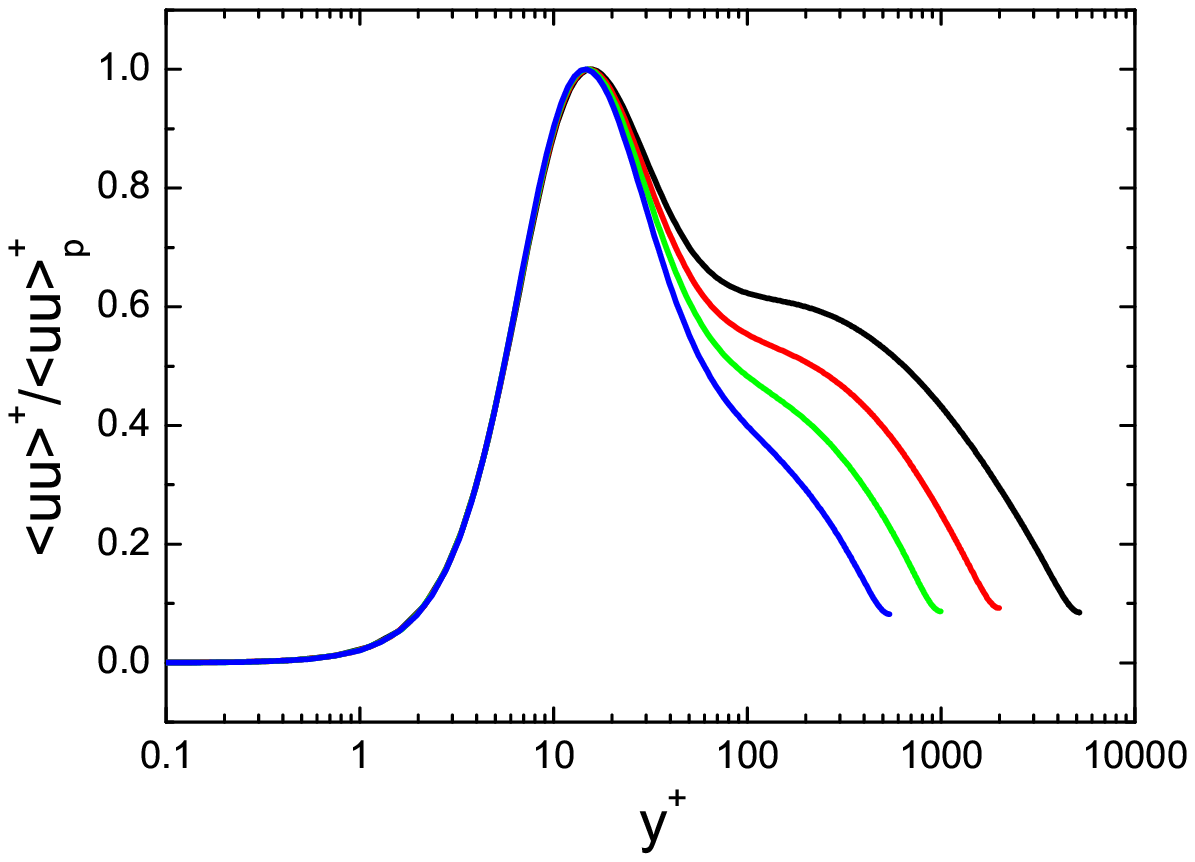}}\\
\subfloat[]{\includegraphics[trim = 0.5cm 10cm 15cm 1cm, clip, width = 6.5 cm]{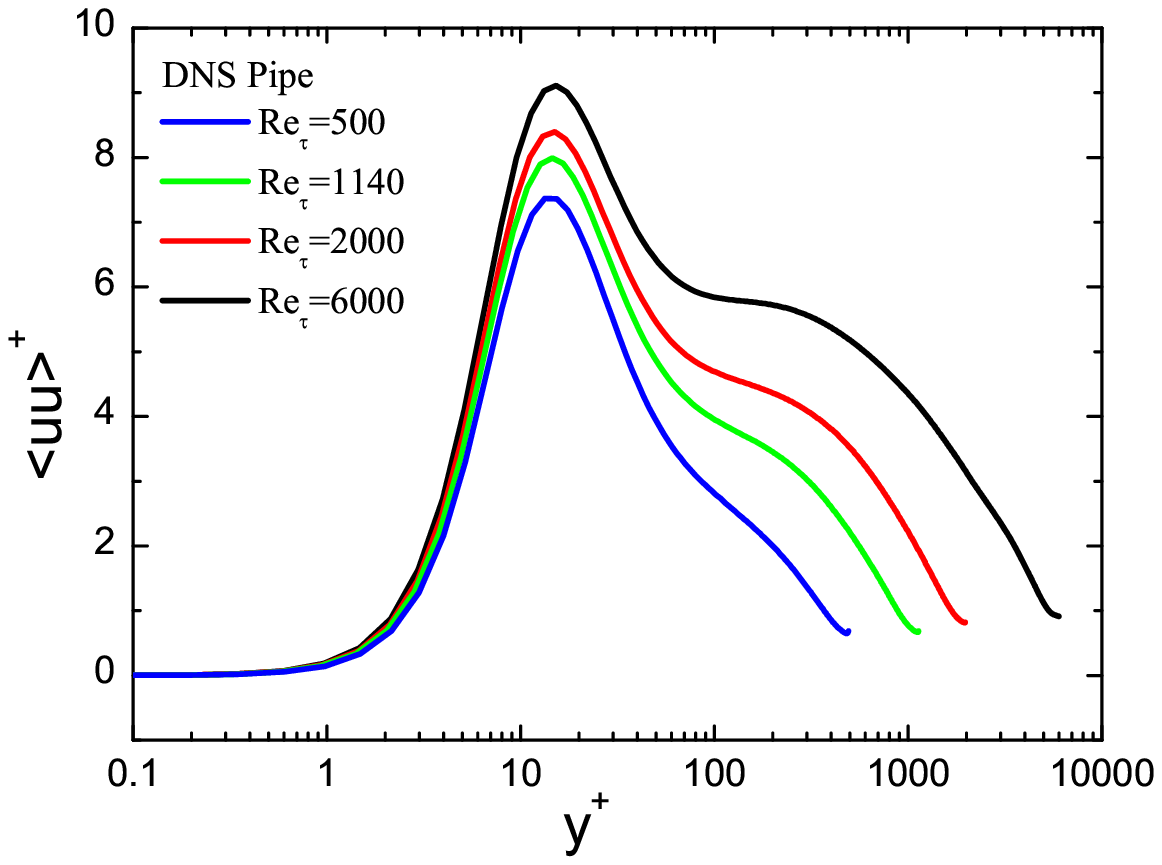}}
\subfloat[]{\includegraphics[trim = 0.5cm 10cm 15cm 1cm, clip, width = 6.5 cm]{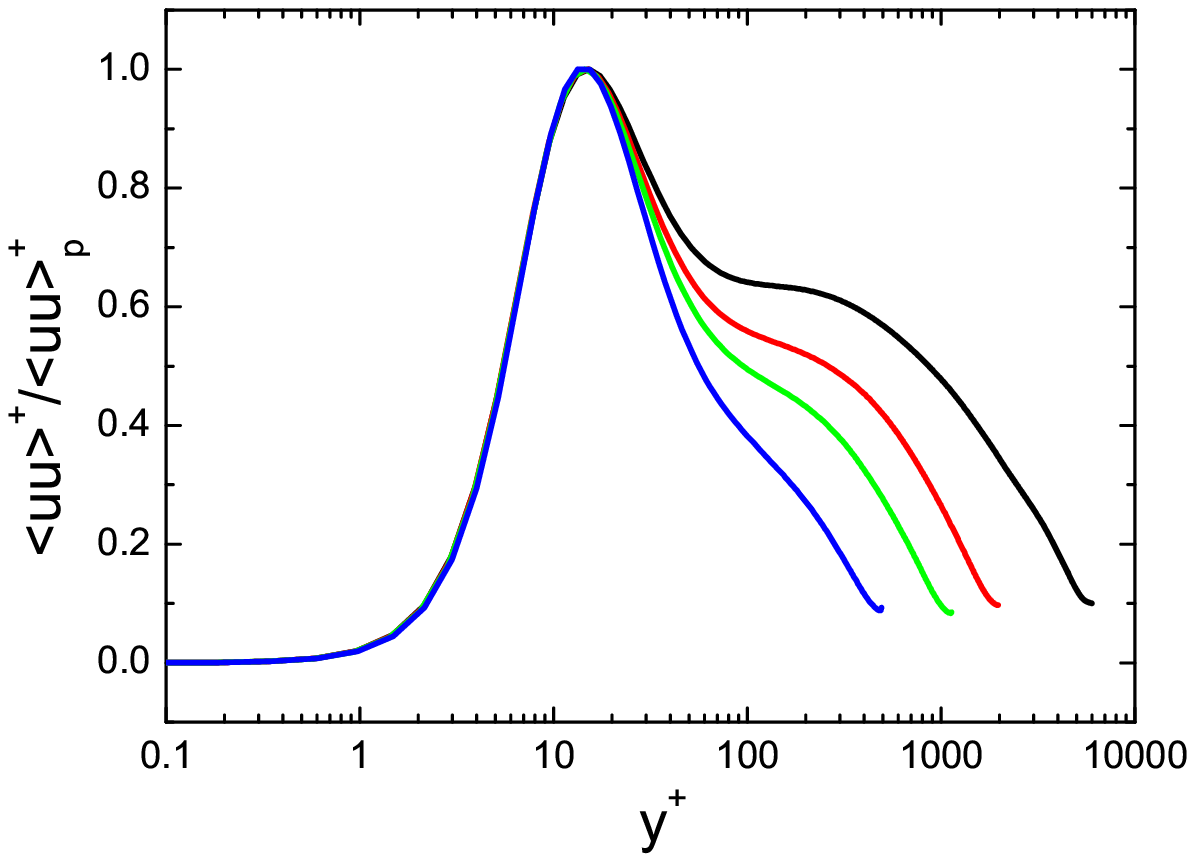}}\\
\subfloat[]{\includegraphics[trim = 0.5cm 10cm 15cm 1cm, clip, width = 6.5 cm]{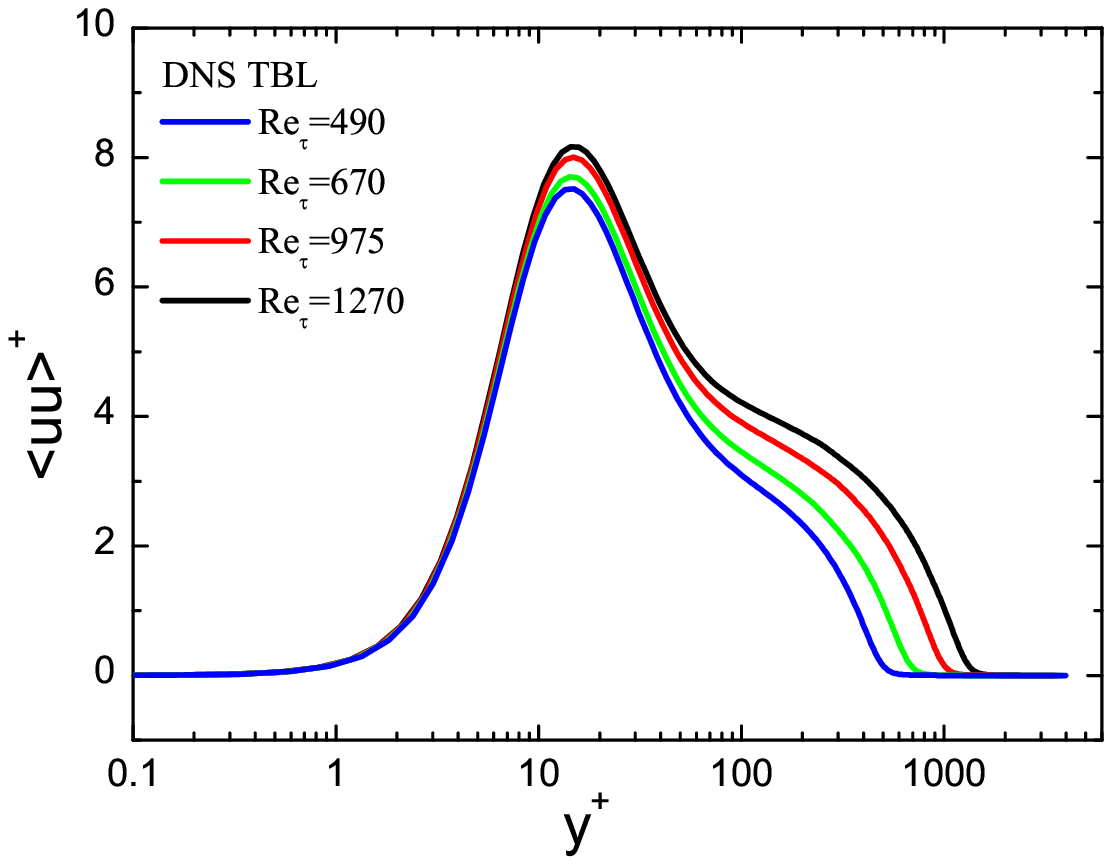}}
\subfloat[]{\includegraphics[trim = 0.5cm 10cm 15cm 1cm, clip, width = 6.5 cm]{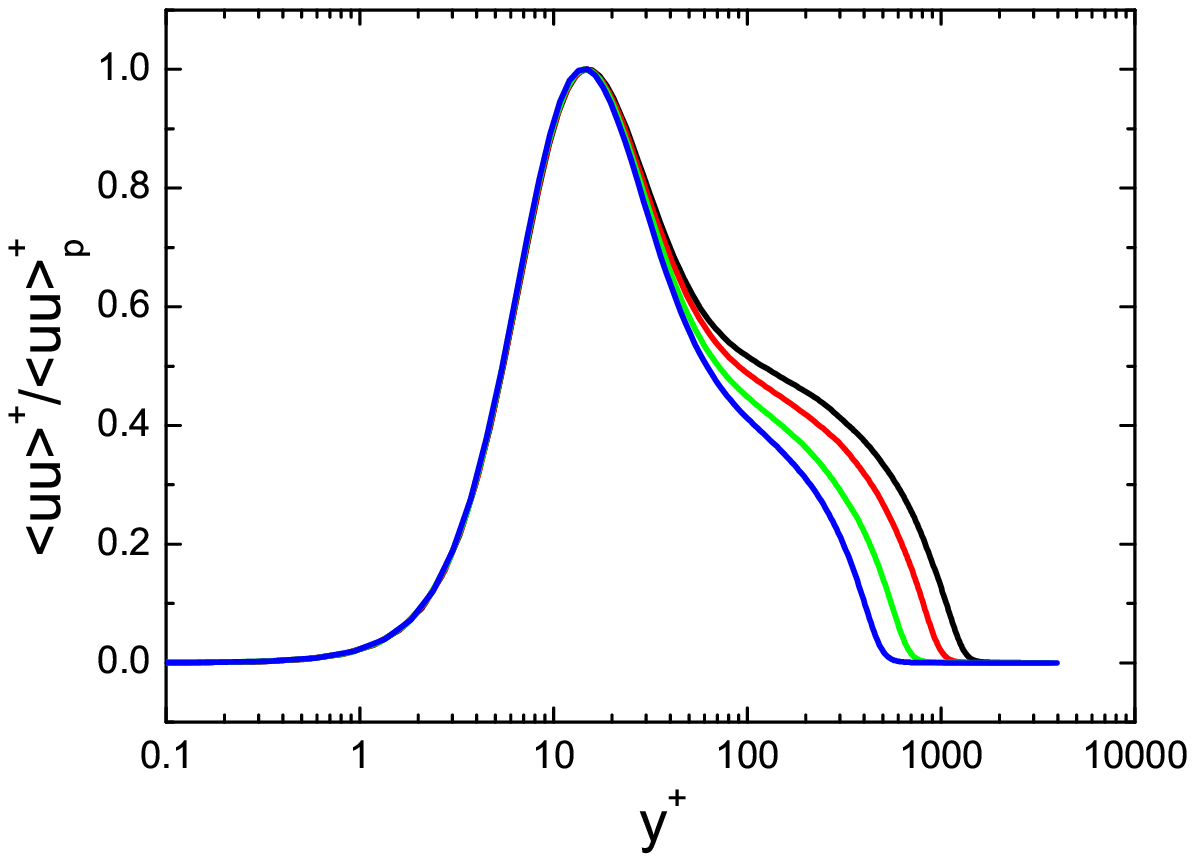}}
\caption{Wall-normal dependence of streamwise velocity fluctuation scaled in viscous units (abscissa in logarithmic scale) for a series of $Re_\tau$'s in channels (top panels), pipes (middle panels) and TBL flows (bottom panels). Left column: $\langle u u\rangle^+$ versus $y^+$. Right column: $\langle u u\rangle^+$ normalized by its (inner) peak value $\langle u u\rangle^+_p$, showing very good collapse. Colored lines indicate DNS data at different $Re_\tau$'s marked in the figure legends, for channels by  \cite{Moser2015}, for pipes by \cite{Pirozzoli2021}, and for TBLs by \cite{schlatter2009,schlatter2010simulations}.}\label{fig:uu}
\end{figure}

Figure \ref{fig:uu} shows the profiles of $\langle uu \rangle^+$, top panels for the channel, middle for the pipe and bottom for the TBL. While the left column displays marked $Re_\tau$ variations, the right column illustrates excellent data collapse after normalization by peak values. That is,
\begin{equation}\label{Eq:CS21A}
\langle u u\rangle^+(y^+,Re_\tau)=\langle u u\rangle^+_p(Re_\tau) f(y^+).
\end{equation}
Note that according to CS, the peak location is an invariant at $y^+_p\approx15$. This is generally accepted as correct (at least since \cite{Sreenivasan1989}); see \cite{Smits2021a}. On the other hand, \cite{Pirozzoli2021} commented that the invariant peak location in CS is violated by their pipe data, which shows that $y^+_p$ slightly increases from 14.28 at $Re_\tau\approx500$ to 15.14 at $Re_\tau=6000$. Nevertheless, using a finer near-wall resolutions than in \cite{Pirozzoli2021}, a later study by \cite{Yao2023JFM} found no such variation of $y_p^+$ with $Re_\tau$ ($y^+_p=15.07, 15.03, 15.50$ for $Re_\tau=180,2000,5000$, respectively). This range of variation is typically seen by others as well \citep{Moser1999,Jimenez2010,Chin2013JFM}, but is regarded as small here, owing possibly to secondary reasons such as the grid and probe resolution.

\begin{figure}
\centering
\subfloat[]{\includegraphics[trim = 0.5cm 10cm 15cm 1cm, clip, width = 6.5 cm]{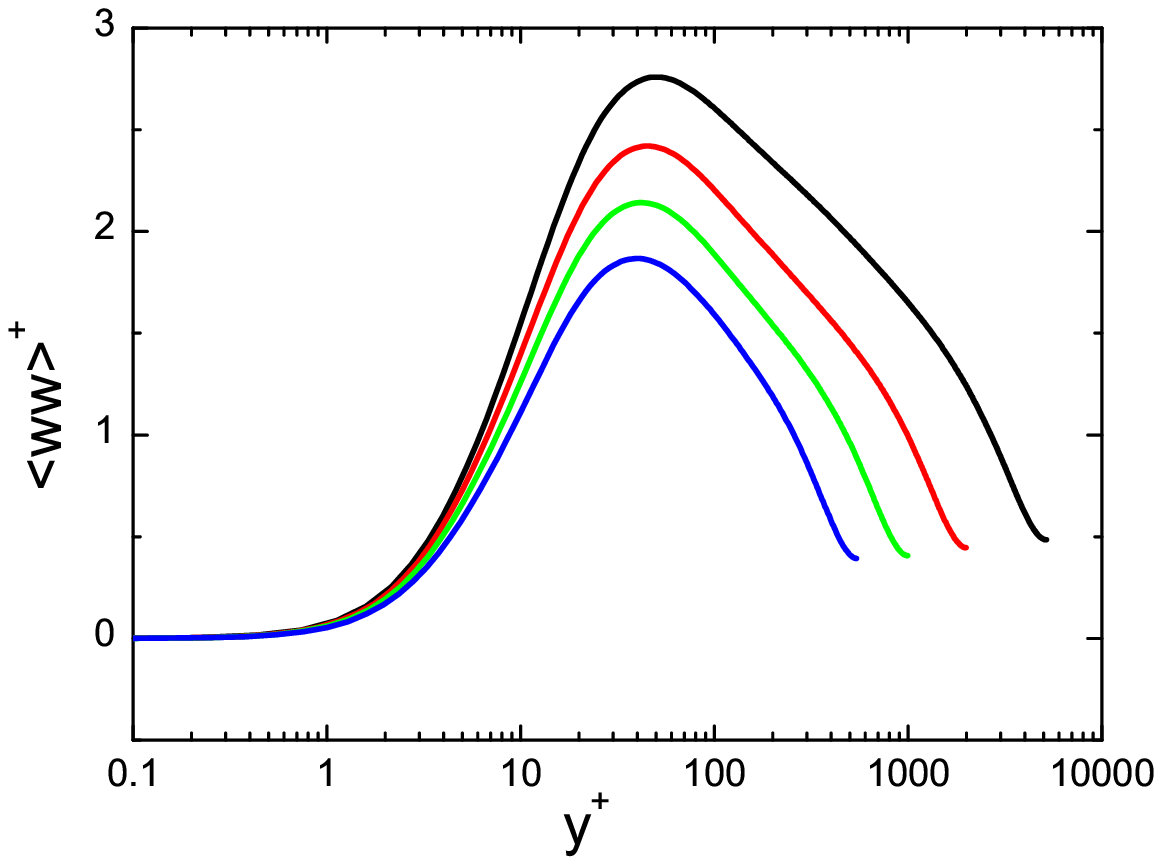}}
\subfloat[]{\includegraphics[trim = 0.5cm 10cm 15cm 1cm, clip, width = 6.5 cm]{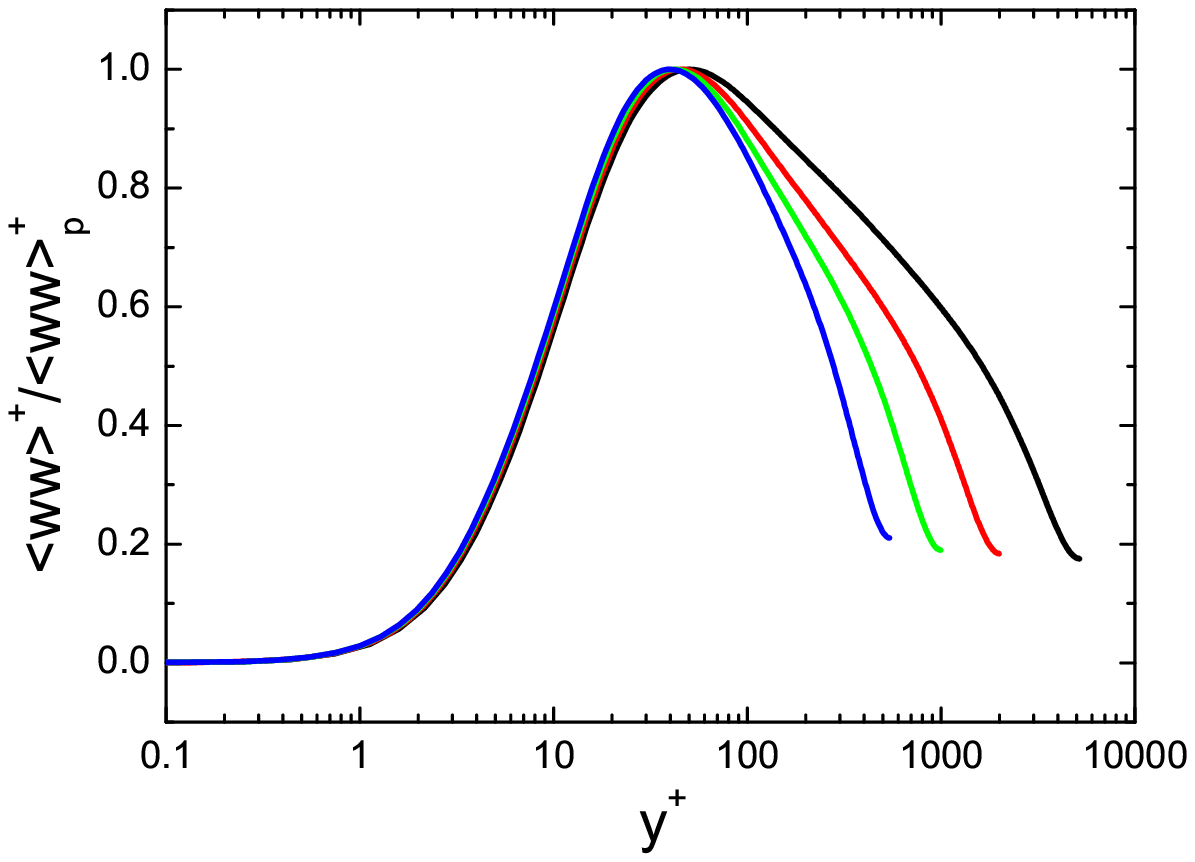}}\\
\subfloat[]{\includegraphics[trim = 0.5cm 10cm 15cm 1cm, clip, width = 6.5 cm]{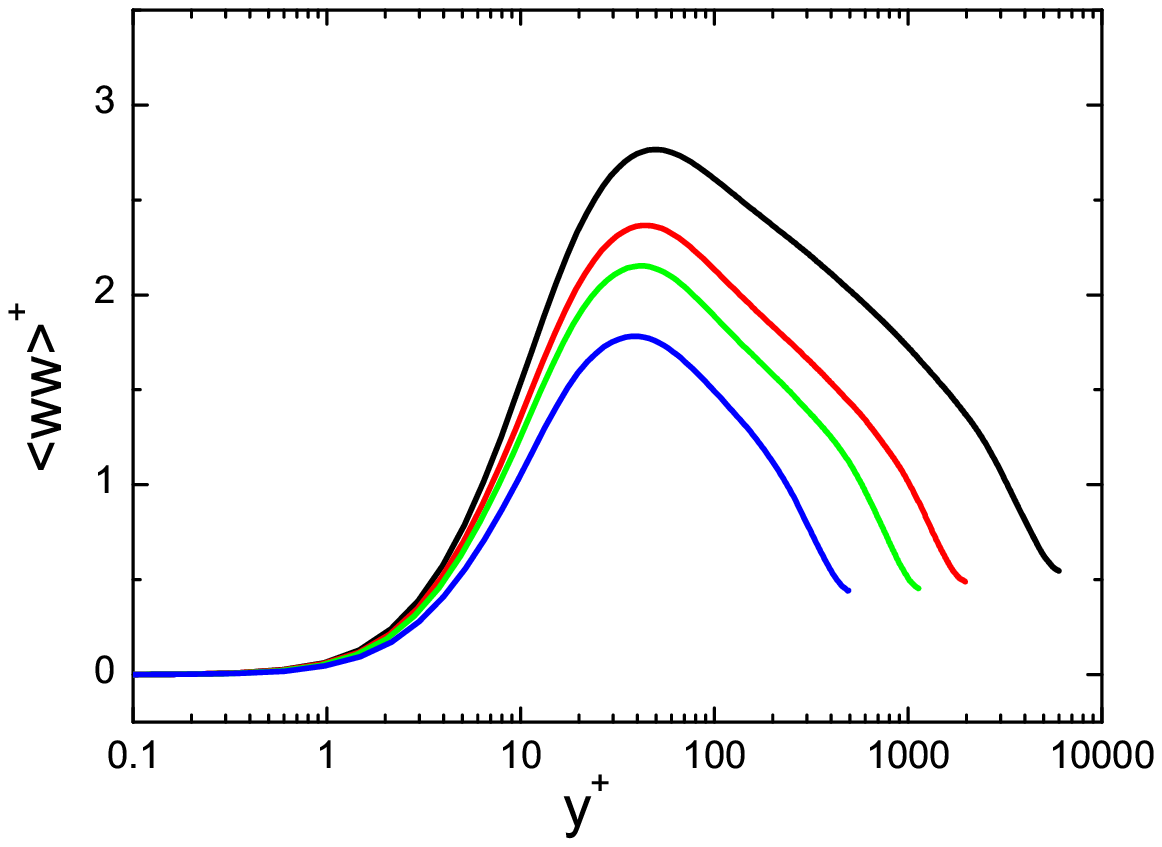}}
\subfloat[]{\includegraphics[trim = 0.5cm 10cm 15cm 1cm, clip, width = 6.5 cm]{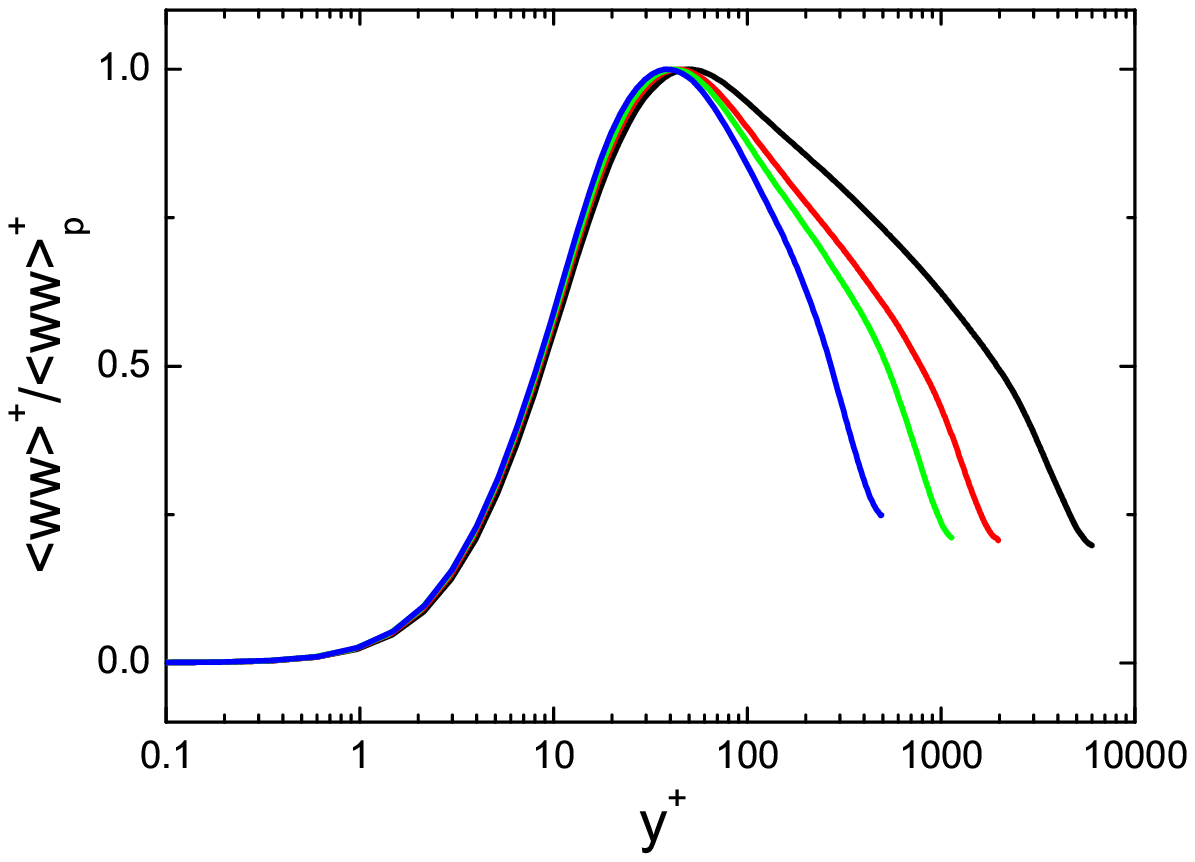}}\\
\subfloat[]{\includegraphics[trim = 0.5cm 10cm 15cm 1cm, clip, width = 6.5 cm]{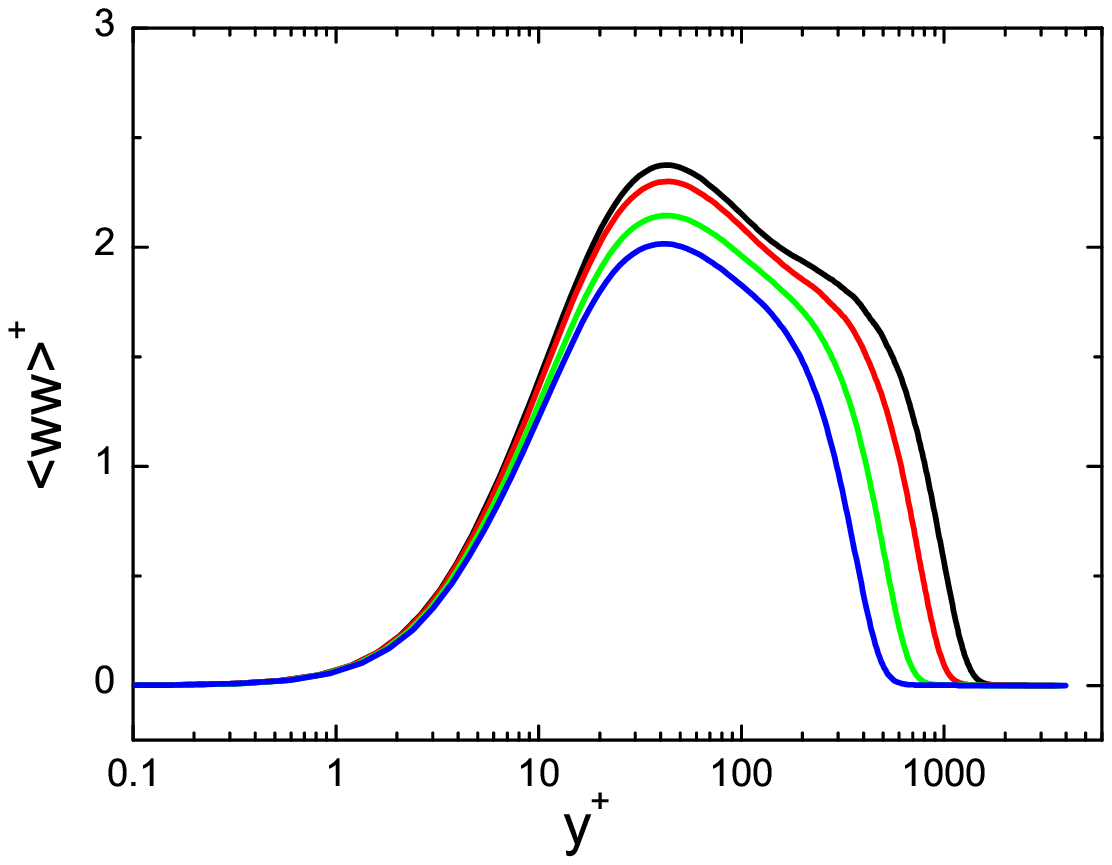}}
\subfloat[]{\includegraphics[trim = 0.5cm 10cm 15cm 1cm, clip, width = 6.5 cm]{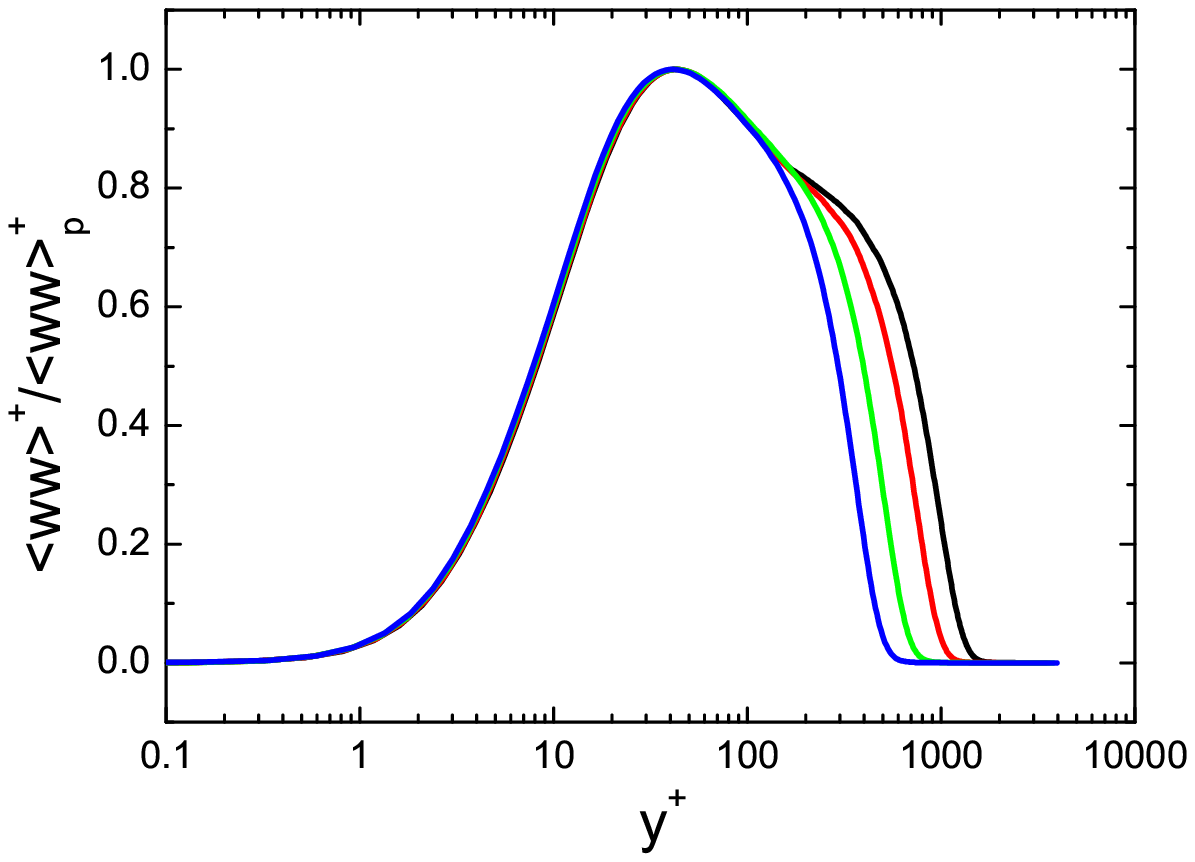}}\\
\caption{The same plots as in figure \ref{fig:uu} but for the spanwise velocity fluctuations. That is, top panels are for channels, middle for pipes and bottom for TBLs; left column for $\langle ww\rangle^+$ while right for $\langle w w\rangle^+$/$\langle ww\rangle^+_p$ versus $y^+$. Lines represent the same DNS data as in figure \ref{fig:uu}.}\label{fig:ww}
\end{figure}

Similar to (\ref{Eq:CS21A}), data collapse for the spanwise velocity fluctuation is achieved via
\begin{equation}\label{Eq:CS21B}
\langle w w\rangle^+(y^+,Re_\tau)=\langle w w\rangle^+_p(Re_\tau) h(y^+),
\end{equation}
where $\langle w w\rangle^+_p$ is the peak value, and $h$ is a $y^+$-dependent function. As shown in figure \ref{fig:ww}, different $Re_\tau$ curves are in close agreement with each other, with the self-preserving range from the wall to the peak (at $y^+\approx45$) or beyond. We note a marginal $Re_\tau$ dependence on this peak location; it is not clear whether they arise from numerical uncertainty or physical modulation by outer flow structures.

\begin{figure}
\centering
\subfloat[]{\includegraphics[trim = 0.5cm 10cm 15cm 1cm, clip, width = 6.5 cm]{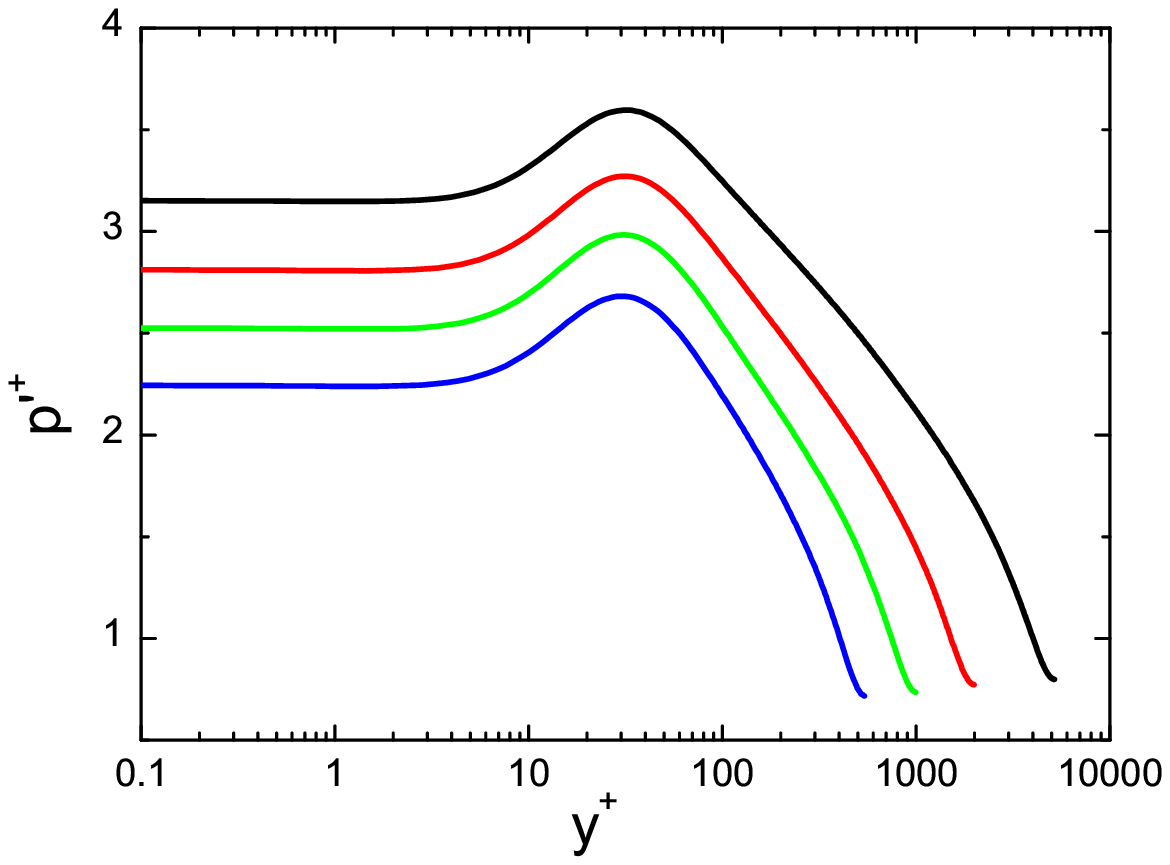}}
\subfloat[]{\includegraphics[trim = 0.5cm 10cm 15cm 1cm, clip, width = 6.5 cm]{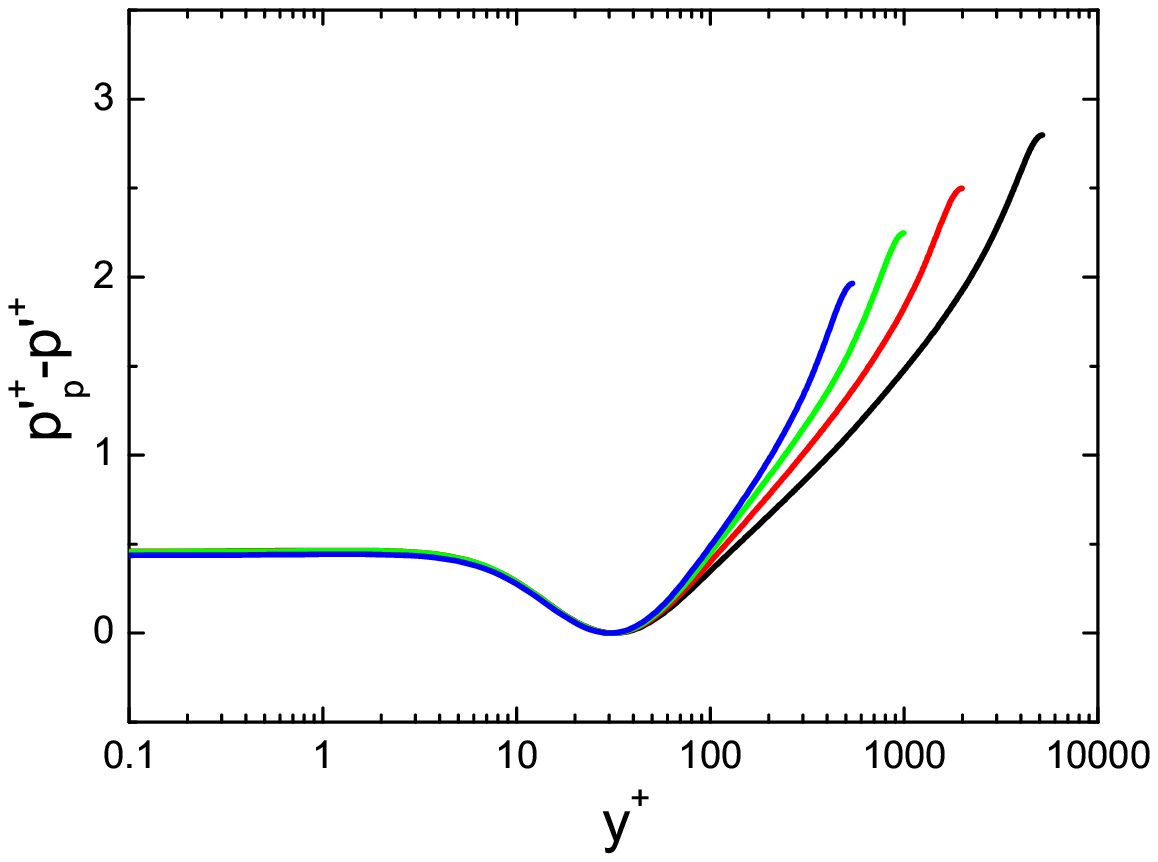}}\\
\subfloat[]{\includegraphics[trim = 0.5cm 10cm 15cm 1cm, clip, width = 6.5 cm]{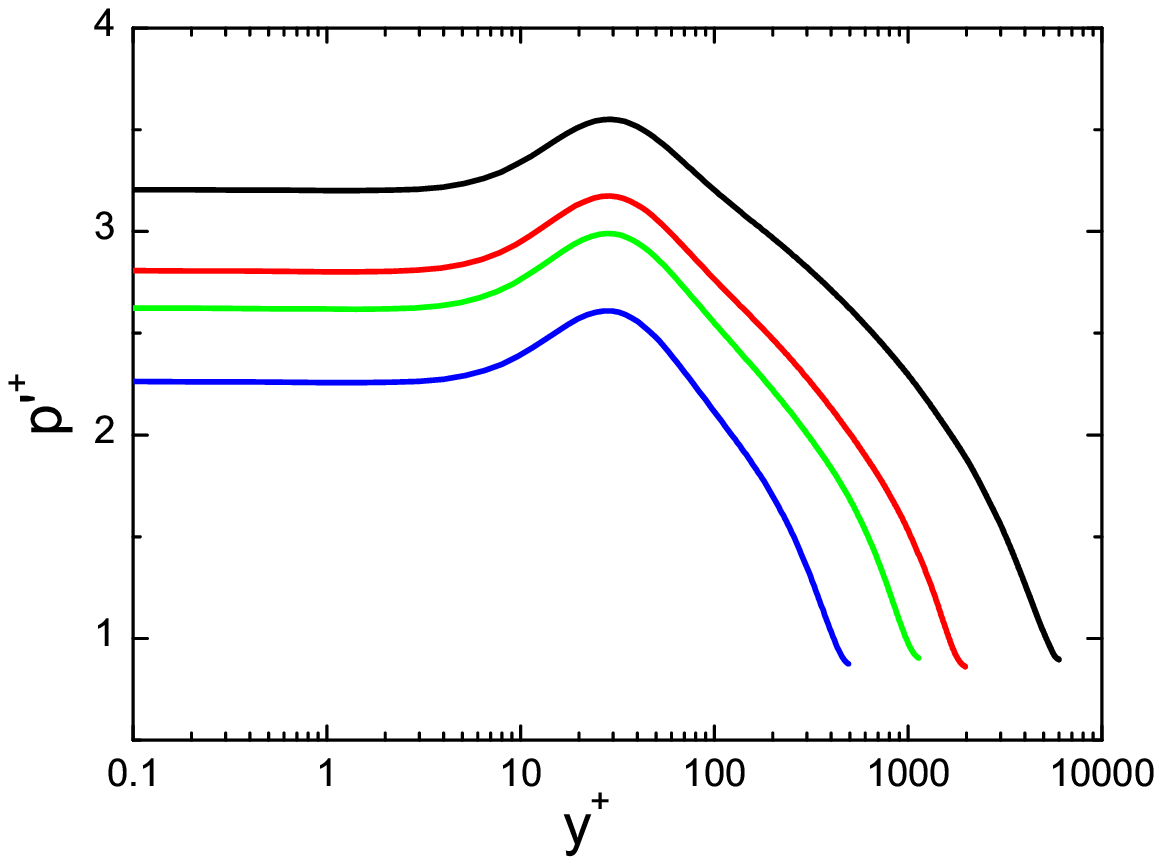}}
\subfloat[]{\includegraphics[trim = 0.5cm 10cm 15cm 1cm, clip, width = 6.5 cm]{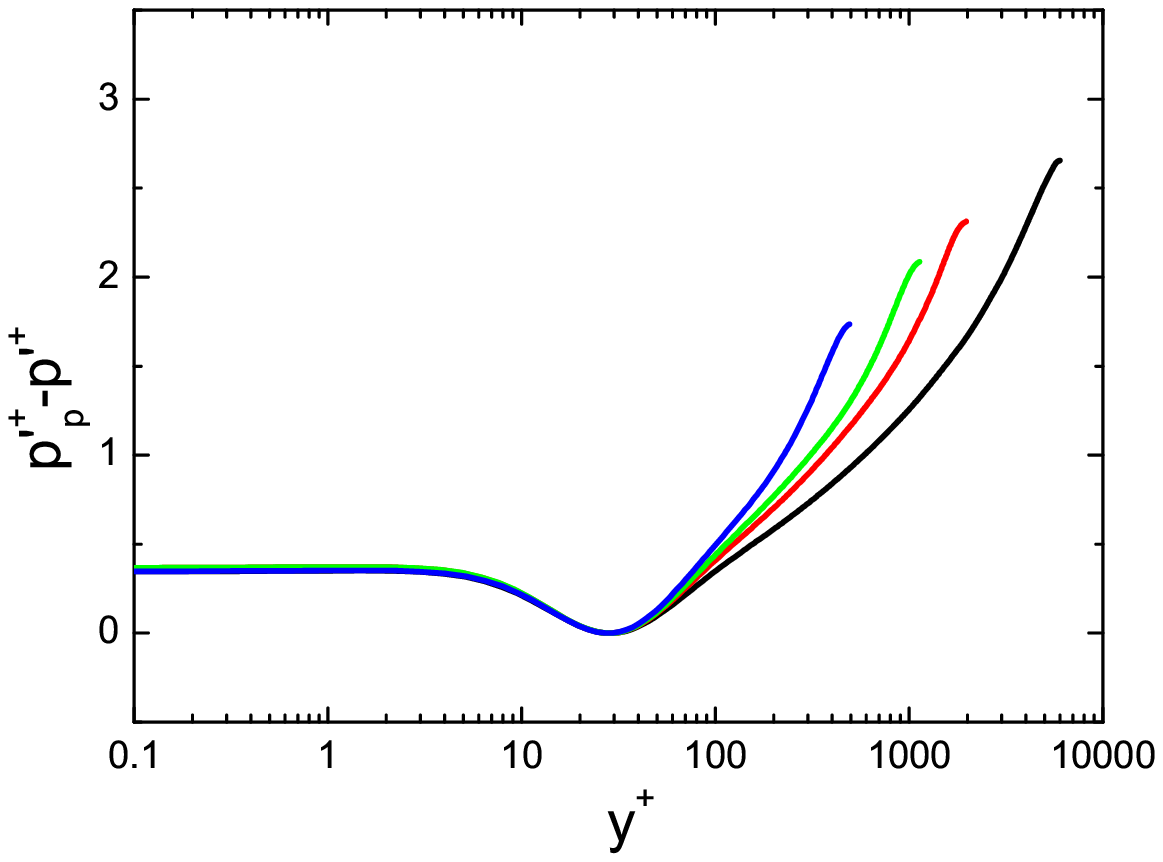}}\\
\subfloat[]{\includegraphics[trim = 0.5cm 10cm 15cm 1cm, clip, width = 6.5 cm]{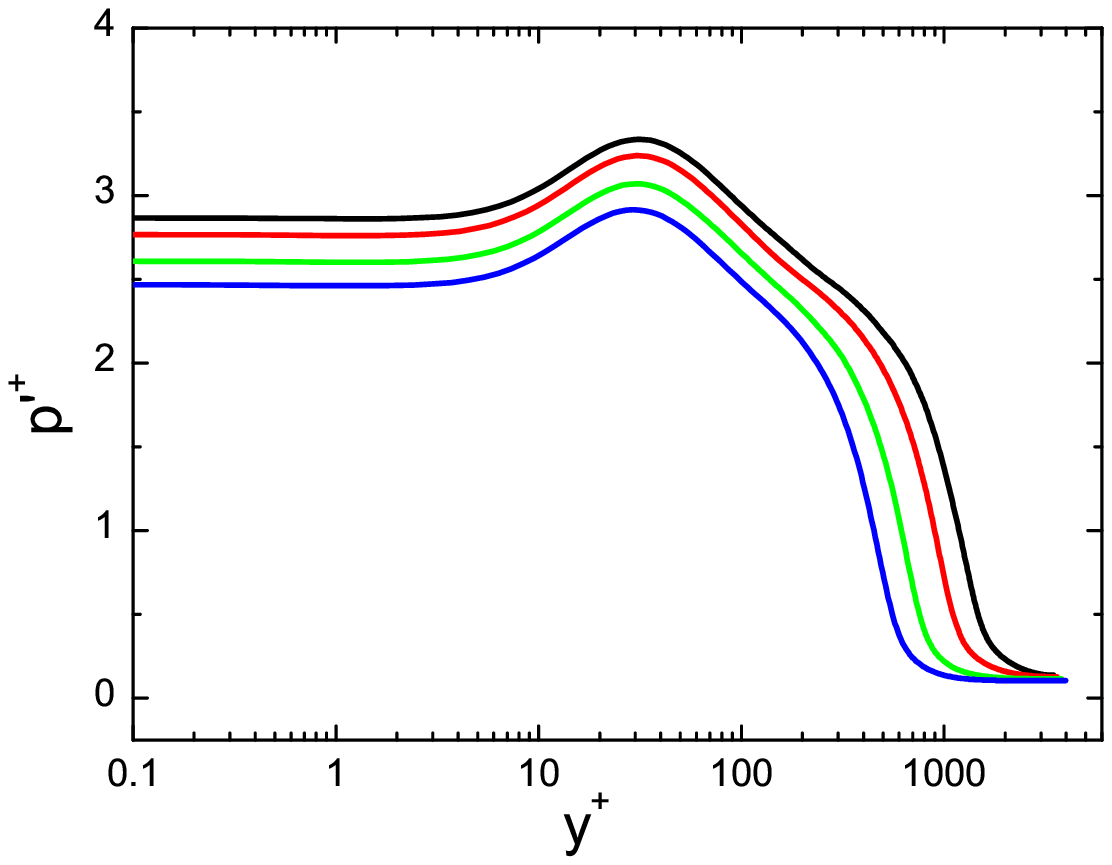}}
\subfloat[]{\includegraphics[trim = 0.5cm 10cm 15cm 1cm, clip, width = 6.5 cm]{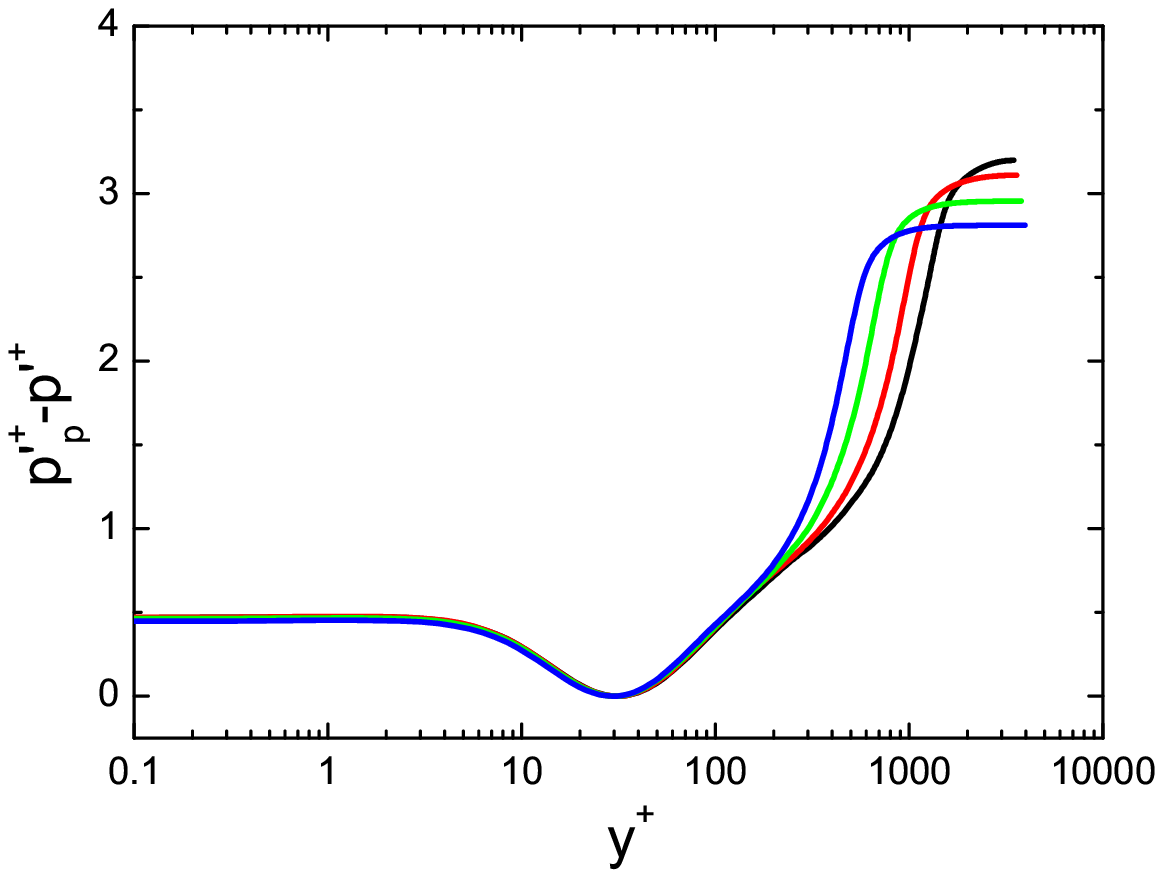}}\\
\caption{Wall-normal dependence for the rms of pressure fluctuation $p'^+=\langle pp\rangle^{+1/2}$ in channels (top panels), pipes (middle panels) and TBL flows (bottom panels). Left column for $p'^+$ while right for $p'^+_p-p'^+$  versus $y^+$. Lines are the same DNS data as in figure \ref{fig:uu}.}\label{fig:p}
\end{figure}

Coming now to pressure fluctuations, the left column of figures \ref{fig:p} shows $Re_\tau$ dependence of $p'^+$. The best collapse is obtained by plotting $p'^+_{p}-p'^+$, as shown in the right column. On this basis, we may write 
\begin{equation}\label{Eq:CS21C}
p'^+(y^+, Re_\tau)=p'^+_{p}(Re_\tau)-j(y^+),
\end{equation}
where $j$ is (in general) a $y^+$-dependent function. The collapse extends from wall to the peak (at $y^+\approx30$), with $p'^+_p-p'^+_w$ a constant around 0.4. This constancy inspires us to postulate (\ref{Eq:CS21C}). From (\ref{Eq:CS21C}) one has $p'^+/p'^+_{p}=1-j(y^+)/p'^+_{p}(Re_\tau)$, in which an increasing $p'^+_{p}$ with $Re_\tau$ would eventually spoil the data collapse if one plotted $p'^+/p'^+_{p}$. That is the reason why (\ref{Eq:CS21A}) or (\ref{Eq:CS21B}) is not applied to $p'^+$. Note that \cite{Panton2017} attempted another data collapse by using $\langle pp\rangle^+_w-\langle pp\rangle^+$, but it is not as satisfactory as (\ref{Eq:CS21C}) in figure \ref{fig:p}, as discussed later in section 3.

\subsection{Summary for the near wall scaling}
The above comparisons demonstrate that the $Re_\tau$ and $y^+$ dependencies could be decoupled after a proper normalization by peak values. Recalling (1.2)
for the $Re_\tau$-scaling of the peak values
and substituting it for $\langle uu\rangle^+_p$,  $\langle ww\rangle^+_p$ and $p'^+_p$ into (\ref{Eq:CS21A}), (\ref{Eq:CS21B}) and (\ref{Eq:CS21C}), respectively, one has a uniform expansion
\begin{equation}\label{eq:CS:inner}
\phi(y^+)=\phi_0(y^+)+\phi_1(y^+)/Re^{1/4}_\tau,
\end{equation}
where $\phi_0(y^+)=\phi^+_\infty f(y^+)$ and $\phi_1(y^+)=-c_{\phi, \infty} f(y^+)$ for $u$; $\phi_0(y^+)=\phi^+_\infty h(y^+)$ and $\phi_1(y^+)=-c_{\phi, \infty} h(y^+)$ for $w$; $\phi_0(y^+)=\phi_\infty-j(y^+)$ and $\phi_1(y^+)=-c_{\phi, \infty}$ for pressure fluctuations. 

Note that (\ref{eq:CS:inner}) is a specific case of
\begin{equation}\label{Eq:Mon1}
\phi^+(y^+,Re_\tau)=f_0(y^+)+f_1(y^+)g(Re_\tau),
\end{equation}
which is a second-order truncation of (\ref{Eq:MonTay}) that was initiated first by \cite{Spalart2021} and \cite{Monkewitz2021}. If $f_0=0$, the above (\ref{Eq:Mon1}) reduces to
\begin{equation}\label{Eq:Smits}
\phi^+(y^+,Re_\tau)=f_1(y^+)g(Re_\tau),
\end{equation}
which is the scaling proposed by \cite{Smits2021a} for $\langle uu \rangle^+$ with $g(Re_\tau)=\epsilon^+_{x-w}$ (streamwise wall dissipation). However, \cite{Smits2021a} found that their proposal did not work as well for $\langle w w\rangle^+$, which they speculated was due to different superposition and modulation enforced by outer flow structures. Here, we show that replacing wall dissipation by peak value, i.e. $g=\phi^+_p$, (\ref{Eq:Smits}) applies for both $\langle uu \rangle^+$ and $\langle ww \rangle^+$. Even so, (\ref{Eq:Smits}) is not proper for pressure due to a constancy $p'^+_p-p'^+_w$ as explained earlier. Thus, a nonzero $f_0$ is needed in (\ref{Eq:Mon1}) when taking pressure into consideration, which is missed in \cite{Smits2021a}.

Finally, we recall that from (\ref{Eq:Mon1}), \cite{Monkewitz2021} developed a composite model for $\langle uu\rangle^+$, which shows that $g=Re^{-1/4}_\tau$ yields a better data description than the alternative $g=\ln Re_\tau$ by \cite{Smits2021a}. The gauge function $g=Re^{-1/4}_\tau$ in (\ref{Eq:Mon1}) restores wall scaling for asymptotically high $Re_\tau$, and we will use it below to derive an outer decay profile, which has not not achieved before.

\section{$Re_\tau$-scaling for outer region and the defect law for fluctuations}

Similar to (\ref{Eq:MonTay}) for the inner region, an asymptotic expansion for the outer flow reads
\begin{equation}\label{Eq:outer}
\phi^+(y^\ast,Re_\tau)=F_0(y^\ast)+F_1(y^\ast)G(Re_\tau)+F_2(y^\ast)G^2(Re_\tau)+h.o.t.,
\end{equation}
where $y^\ast=y/\delta=y^+/Re_\tau$ is the outer unit; $F_0$, $F_1$ and $F_2$ are general functions depending on $y^\ast$, and $G(Re_\tau)$ is the gauge function. In analogy to \emph{law of the wall}, the inviscid outer flow similarity corresponds to the first order truncation in (\ref{Eq:outer}), i.e.
\begin{equation}\label{Eq:outer2}
\phi^+(y^\ast,Re_\tau)=F_0(y^\ast).
\end{equation}
When we take $\phi^+=U^+_e-U^+$, the resulting equation for the mean velocity is known as the \emph{velocity defect law}. Here, the subscript $e$ indicates the value at $y=\delta$ or $y^\ast=1$, i.e., the centerline for channel and pipe flows, and the boundary layer edge for TBL. This law has been tested extensively in the literature---see \citet{nagib2007} and \citet{she2017JFM} for recent efforts. We shall now develop the equation for the fluctuating quantities.

\begin{figure}
\centering
\subfloat[]{\includegraphics[trim = 0.75cm 10cm 15.5cm 1cm, clip, width = 4.5 cm]{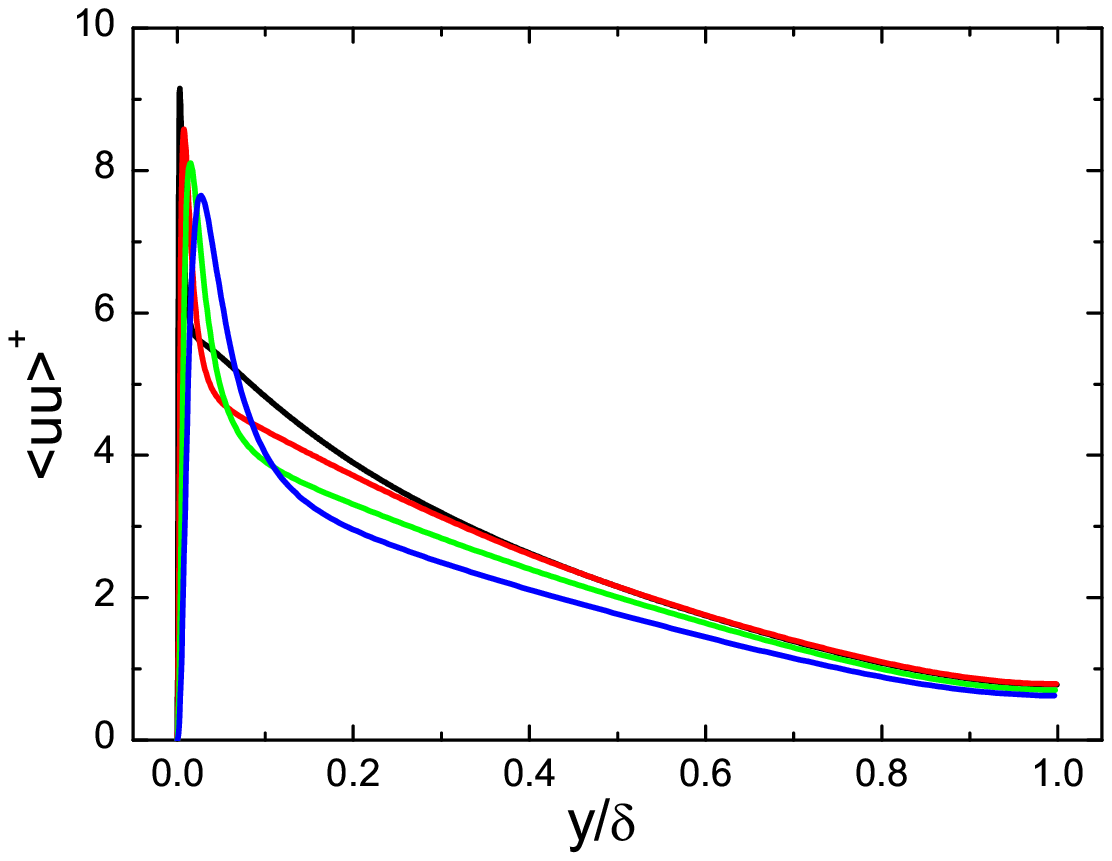}}
\subfloat[]{\includegraphics[trim = 0.75cm 10cm 15.5cm 1cm, clip, width = 4.5 cm]{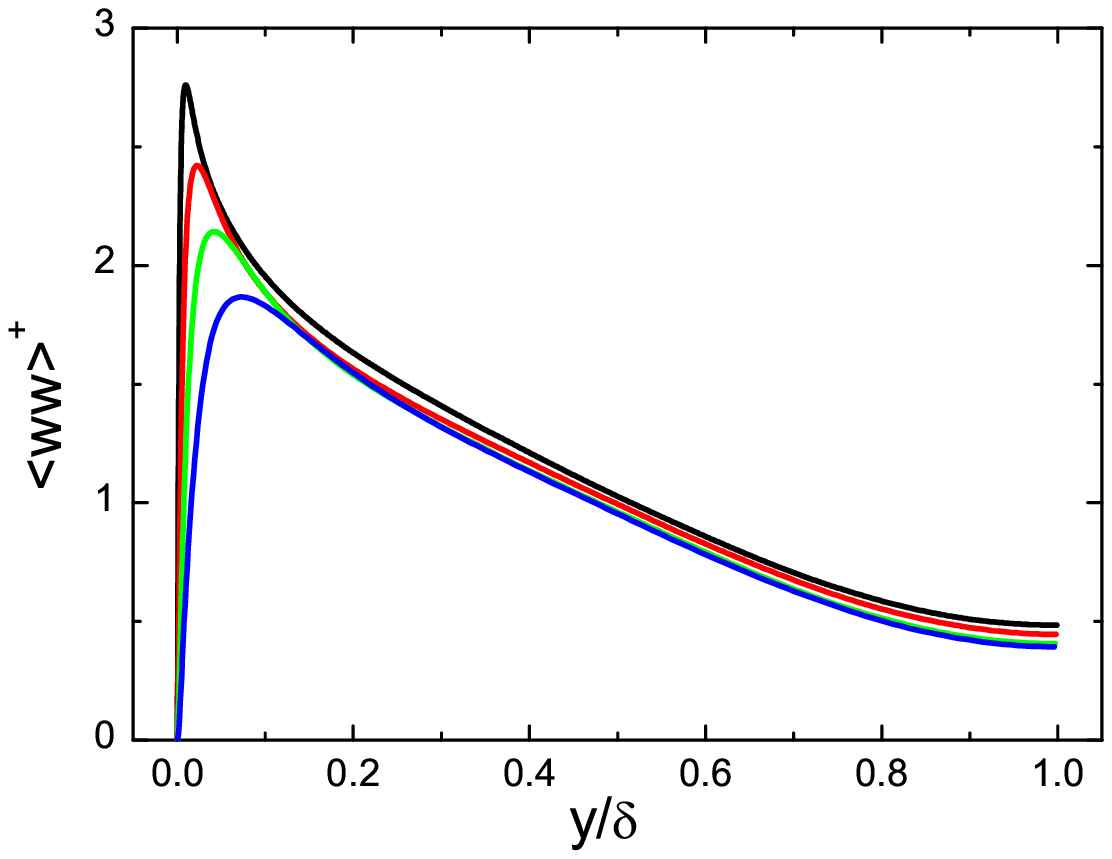}}
\subfloat[]{\includegraphics[trim = 0.75cm 10cm 15.5cm 1cm, clip, width = 4.5 cm]{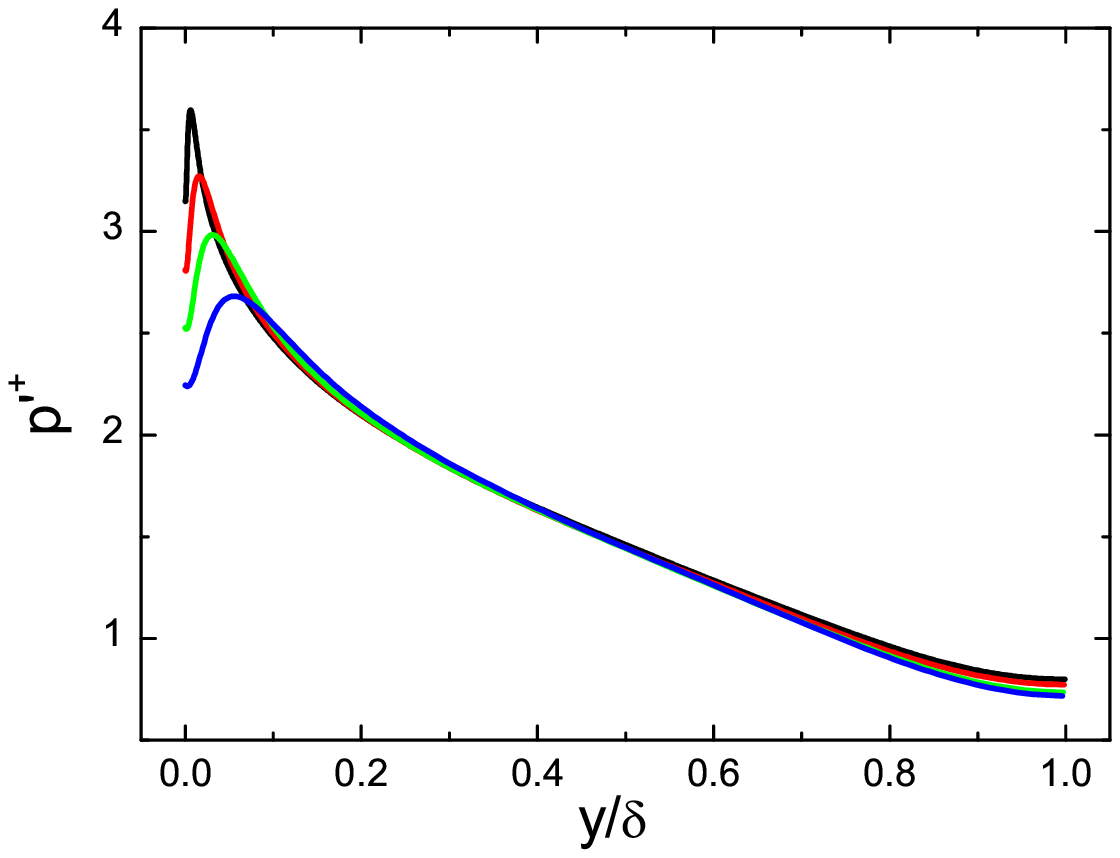}}\\
\subfloat[]{\includegraphics[trim = 0.75cm 10cm 15.5cm 1cm, clip, width = 4.5 cm]{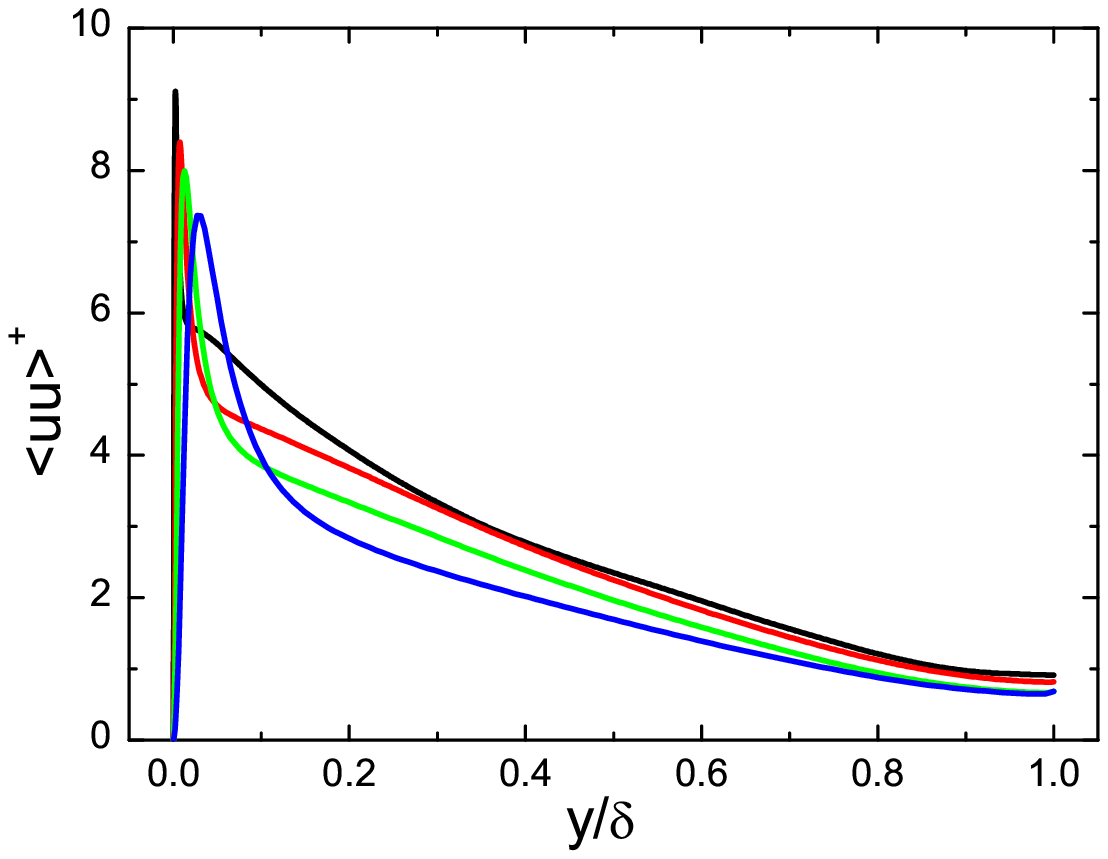}}
\subfloat[]{\includegraphics[trim = 0.75cm 10cm 15.5cm 1cm, clip, width = 4.5 cm]{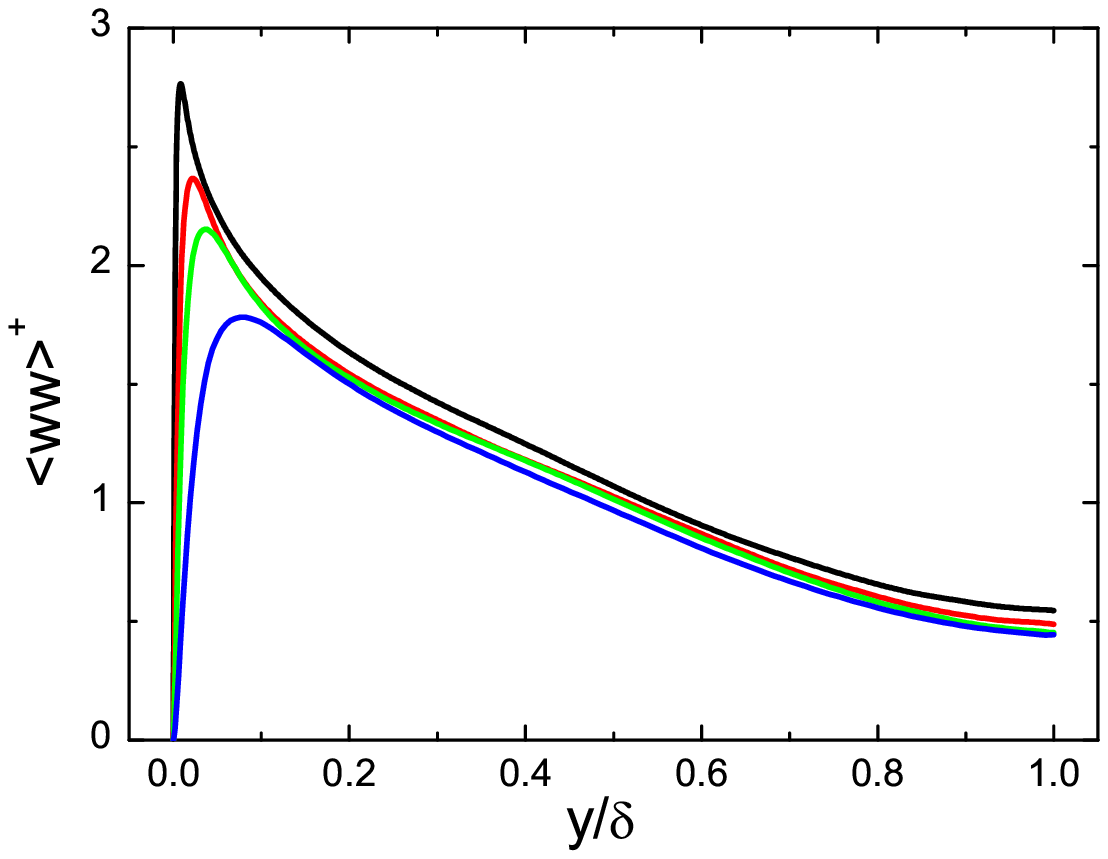}}
\subfloat[]{\includegraphics[trim = 0.75cm 10cm 15.5cm 1cm, clip, width = 4.5 cm]{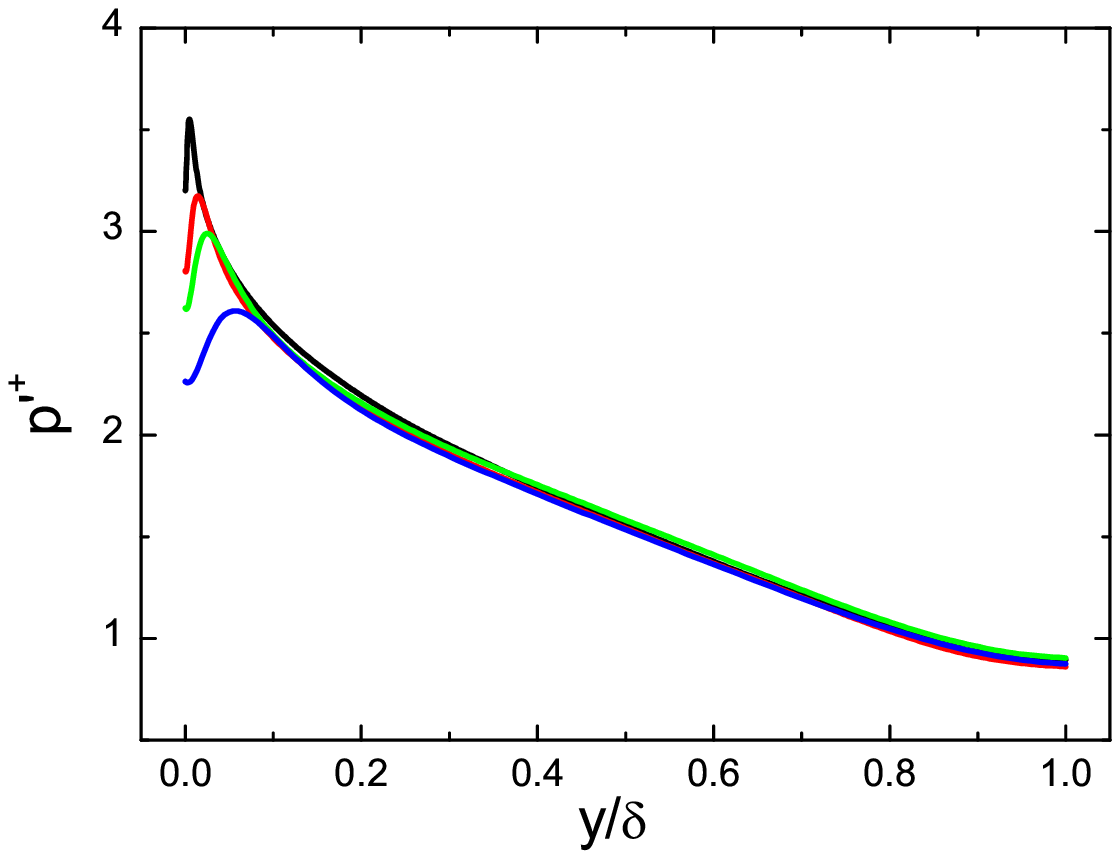}}\\
\subfloat[]{\includegraphics[trim = 0.75cm 10cm 15.5cm 1cm, clip, width = 4.5 cm]{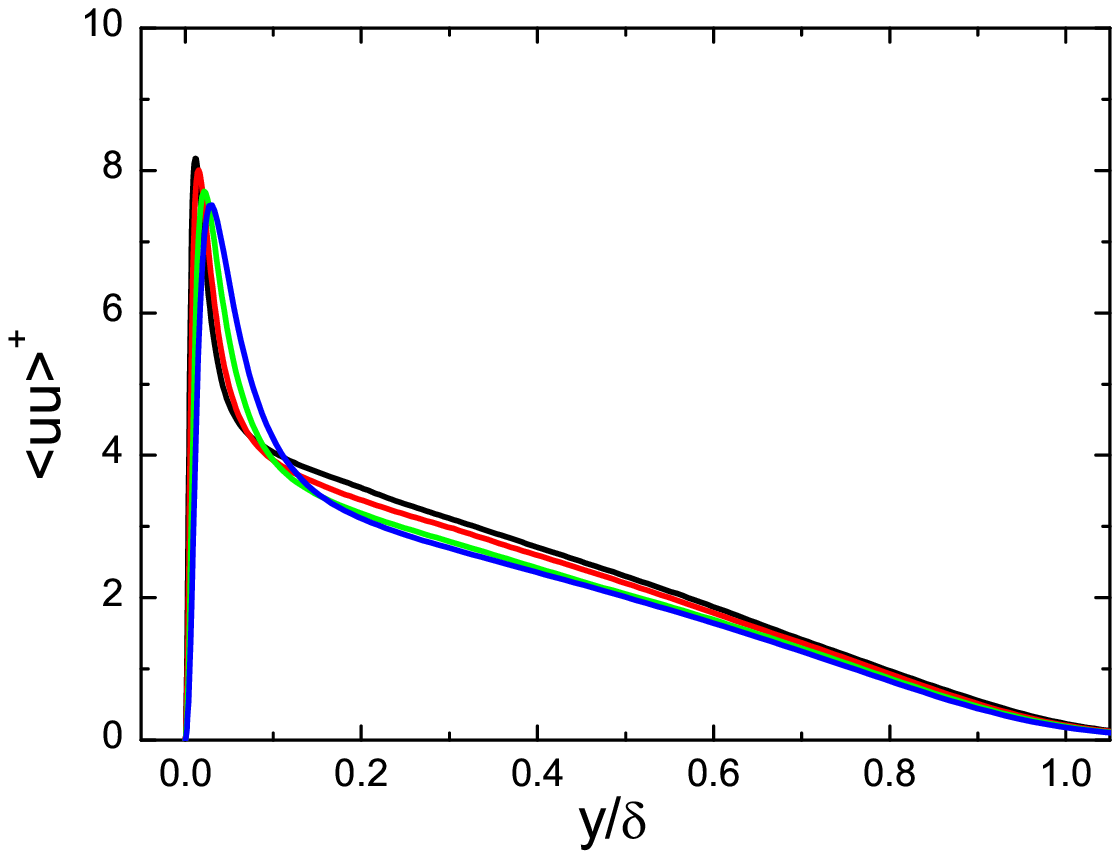}}
\subfloat[]{\includegraphics[trim = 0.75cm 10cm 15.5cm 1cm, clip, width = 4.5 cm]{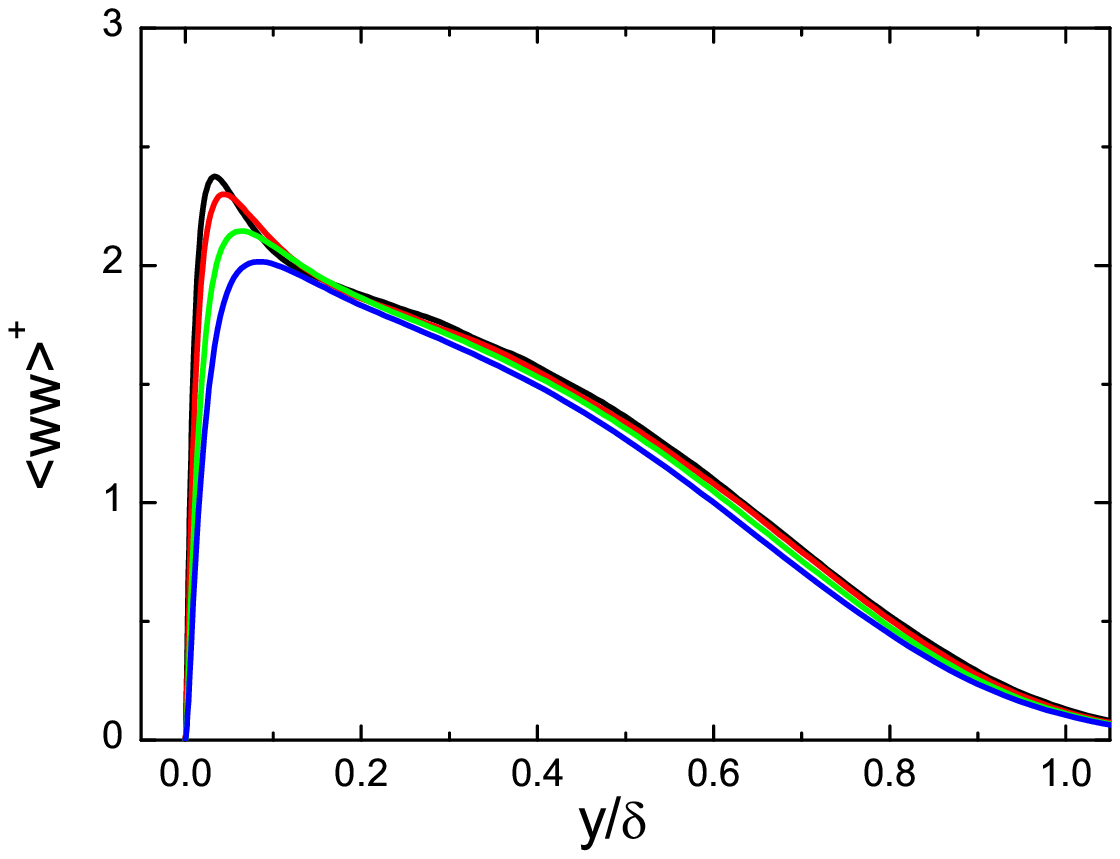}}
\subfloat[]{\includegraphics[trim = 0.75cm 10cm 15.5cm 1cm, clip, width = 4.5 cm]{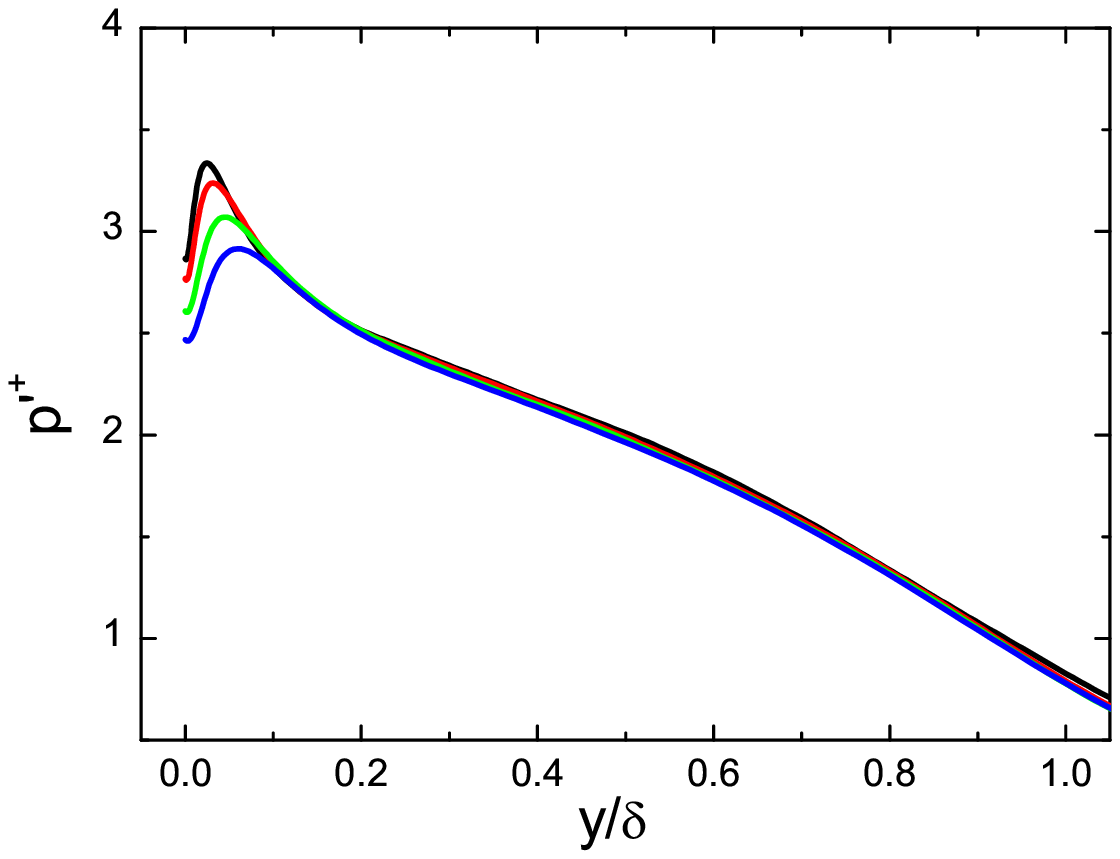}}
\caption{Wall-normal dependence of turbulence fluctuations in outer length unit $y^\ast=y/\delta$ (the abscissa in linear scale). Top panels for channels, middle for pipes and bottom for TBLs. Left column for $\langle uu\rangle^{+}$; middle for $\langle ww\rangle^{+}$ and right for $p'^{+}$. Lines are the same DNS data as in figure \ref{fig:uu}.}\label{fig:outer}
\end{figure}

An assessment of outer similarity for the three profiles is shown in figure \ref{fig:outer}, with the top panels for the channel, middle for the pipe and bottom for the TBL. It is remarkable that the pressure fluctuation displays an excellent data collapse from $y^\ast=0.1$ to $y^\ast=1$. In the same flow region, the spanwise velocity variance data also collapse well, with a deviation of the order of 0.1 (scaled on $u^2_\tau$). For streamwise velocity, the outer similarity holds better towards the outer edge though discernible $Re_\tau$ dependence exists for small $Re_\tau$ profiles. For $Re_\tau > 1000$, the streamwise variance profiles also collapse together closely (figures 5-6), thus supporting the inviscid similarity of (\ref{Eq:outer2}) at high $Re_\tau$.

\subsection{Matching for the defect law for fluctuations}

Based on the above inner viscous expansion and outer inviscid similarity, we now develop a matching procedure to derive the analytical form in the intermediate zone. To start with, one may invoke the derivation of \cite{Millikan} for the log-law, which has been extended (for example) for $\langle uu\rangle^+$ by \cite{Hultmark2012JFM} and for $\langle p p\rangle^+$ by \cite{Panton2017}. Nevertheless, different orders of matching would lead to different scaling proposals, and here we present a short account of matching to obtain the defect law for fluctuations. Later we will address the question of how the current defect law is consistent with Millikan's matching analysis.

Specifically, as (\ref{Eq:Mon1}) applies near the wall (with $g=Re_\tau^{-1/4}$) and (\ref{Eq:outer2}) holds in the outer, we match (\ref{Eq:outer2}) directly with (\ref{Eq:Mon1}) in the outer region. That is,
\begin{eqnarray}
\label{eq:CS:profile2}
F_0(y^\ast)=\phi^+(y^+)&=&f_{0}(y^+)+f_{1}(y^+)/Re^{1/4}_\tau\nonumber\\
&=&f_{0}(y^+)+{h}_{1}(y^+)y^{\ast{1/4}}
\end{eqnarray}
with ${h}_{1}(y^+)=f_1(y^+)/y^{+1/4}$. Note that towards the wall, $f_{0}=\phi_0$ and $h_{1}=\phi_{1}/y^{+1/4}$ so that (\ref{eq:CS:profile2}) approaches (\ref{eq:CS:inner}); $f_0$ and $h_1$ are to be determined towards the outer region. At this stage, assuming the scale-separation between $y^\ast$ and $y^+$ for asymptotically high $Re_\tau$, one has $\partial F_0(y^\ast)/\partial y^+=0$ in (\ref{eq:CS:profile2}), which leads to
\begin{subequations}\label{eq:match:CS}
\begin{equation}\label{eq:f0}
f_{0}(y^+)=c_0;
\end{equation}
\begin{equation}\label{eq:f1}
{h}_{1}(y^+)=c_1,
\end{equation}
\end{subequations}
where $c_0$ and $c_1$ are constants independent of $y^\ast$, $y^+$ and $Re_\tau$, but may depend on the precise quantity $\phi$. Denoting $c_0=\alpha_\phi$ and $c_1=-\beta_\phi$, we obtain (\ref{eq:CS:outer28}) from (\ref{eq:CS:profile2}) and (\ref{eq:match:CS}).

It is readily verified that  (\ref{eq:CS:outer28}) matches (\ref{Eq:outer2}) in the outer region and  (\ref{Eq:Mon1}) in the inner, hence offering a common description for the overlap region. It should be mentioned that a logarithmic decay of (\ref{eq:townsend}) can also be obtained if the gauge function $Re^{-1/4}_\tau$ in (\ref{eq:CS:profile2}) is replaced by $\ln Re_\tau$. In other words, mere matching analysis cannot preclude (\ref{eq:CS:outer28}) or (\ref{eq:townsend}), unless the gauge function is justified rigorously. 

Before turning to the data in evidence for (\ref{eq:CS:outer28}), we discuss how the defect law conforms with the classical matching analysis. In fact, by following the matching of \cite{Millikan}, as discussed for example in \cite{pope2000turbulent}, one has
\begin{eqnarray}\label{Eq:EP2}
\frac{\partial \phi^+}{\partial \ln {y^\ast}}=A_0+A_1(Re_\tau)+h.o.t.,
\end{eqnarray}
where $A_0$ and $A_1$ are coefficients independent of $y^+$ or $y^\ast$. Note that $A_1$ involving the $Re_\tau$-dependence is a higher-order modification compared to the leading order coefficient $A_0$, which has been introduced for the streamwise mean velocity  by \cite{wosnik2000} and \cite{Monkewitz2023} . Here, particularly for $A_0<0$, (\ref{Eq:EP2}) yields the logarithmic decay, e.g. for $\langle uu\rangle^+$ as in \cite{Hultmark2012JFM}, and for $\langle p p\rangle^+$ as in \cite{Panton2017}.   

On the other hand, if $A_0=0$ so that $A_1(Re_\tau)=o(1)$ dominates (\ref{Eq:EP2}), a wall-normal integration of (\ref{Eq:EP2}) leads to
 \begin{eqnarray}\label{Eq:log1c2}
\phi^+(y^\ast) = A_1(Re_\tau)\ln (y^\ast)+B_0.
\end{eqnarray}
Further, with $A_1(Re_\tau)=O(1/\ln Re_\tau)$, one has
\begin{equation}\label{eq:CS22A}
\phi^+(y^+_s)=A_1(Re_\tau)\ln y^+_s- A_1(Re_\tau)\ln(Re_\tau)+B_0=O(1),
\end{equation}
which means that $\phi^+$ is bounded at any specific $y^+_s$ location. Conversely, a bounded $\phi^+$ would exclude a constant logarithmic slope but require the slope to decrease with $Re_\tau$ according to $O(1/\ln Re_\tau)$.

One can easily check that (\ref{eq:CS:outer28}) falls into this category of (\ref{eq:CS22A}) as follows. Calculate the local logarithmic slope of (\ref{eq:CS:outer28}) yielding
\begin{eqnarray}\label{eq:CS:pp2:out3}
A_{local}(y^\ast)\equiv\frac{\partial \phi^+}{\partial \ln y^\ast}=(\beta_{\phi}/4) (y^\ast)^{1/4}=(\beta_{\phi}/4) (y^+)^{1/4} Re^{-1/4}_\tau,
\end{eqnarray}
which agrees with (\ref{eq:CS22A}) because $A_{local}\propto Re^{-1/4}_\tau=O(1/\ln Re_\tau)$ at any given position $y^+_s$. Moreover, as $y^\ast$ increases, $A_{local}$ in (\ref{eq:CS:pp2:out3}) becomes larger, indicating a steeper logarithmic slope towards the outer flow. This picture is indeed supported by data in the comparison below, in contrast to (\ref{eq:townsend}) with a constant slope valid in a narrower flow domain.

\subsection{Comparison with data}

\begin{figure}
\centering
\subfloat[]{\includegraphics[trim = 0.75cm 10cm 15.5cm 1cm, clip, width = 6.5 cm]{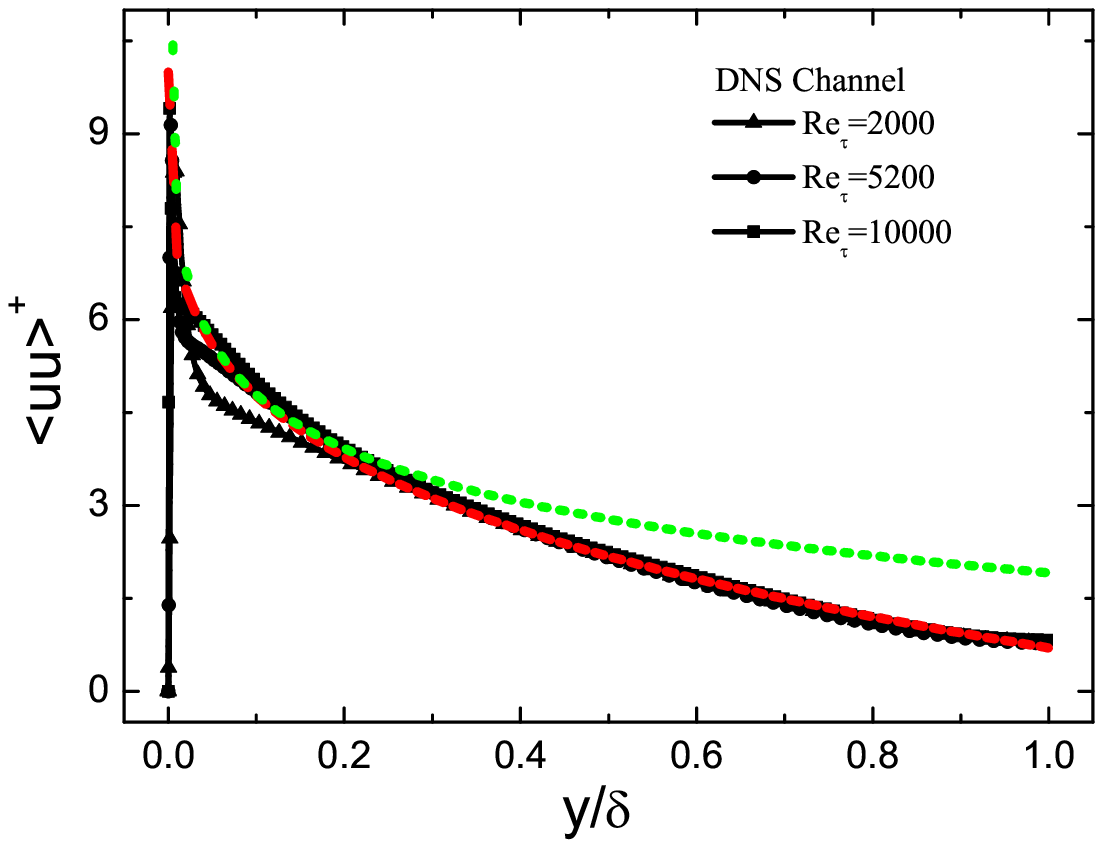}}
\subfloat[]{\includegraphics[trim = 0.75cm 10cm 15.5cm 1cm, clip, width = 6.5 cm]{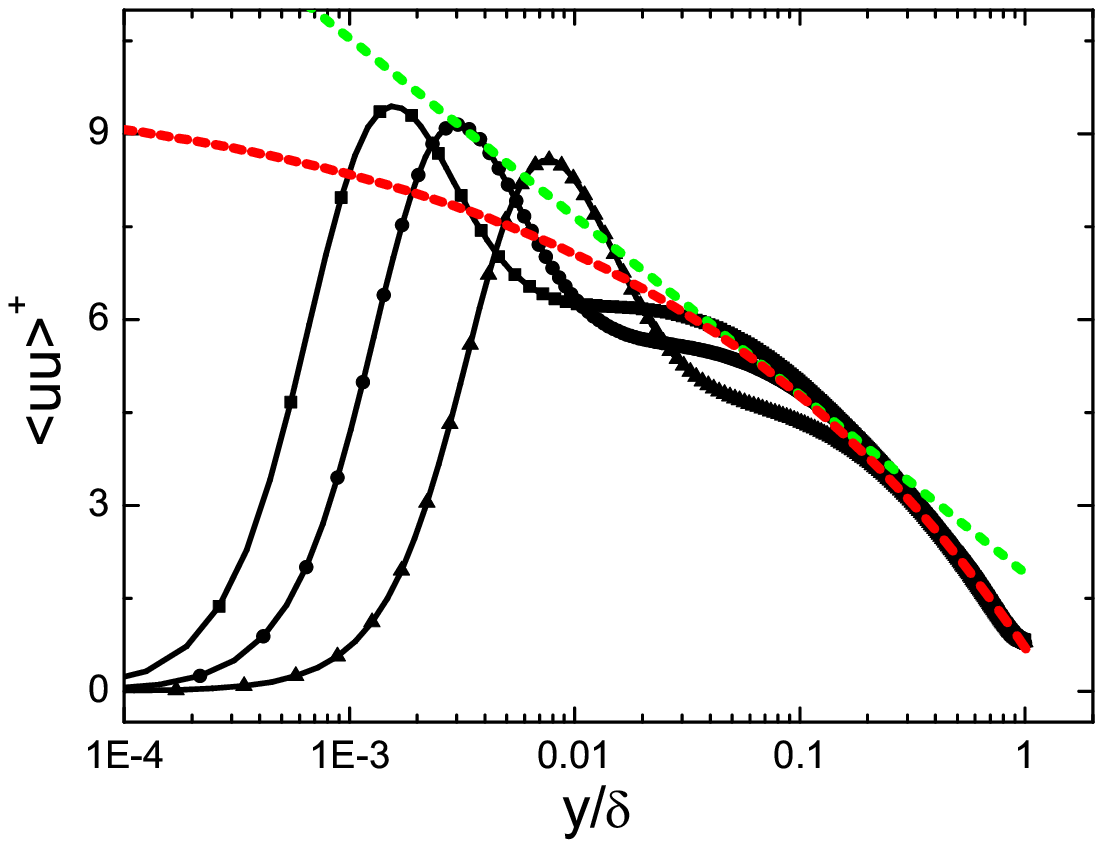}}\\
\subfloat[]{\includegraphics[trim = 0.75cm 10cm 15.5cm 1cm, clip, width = 6.5 cm]{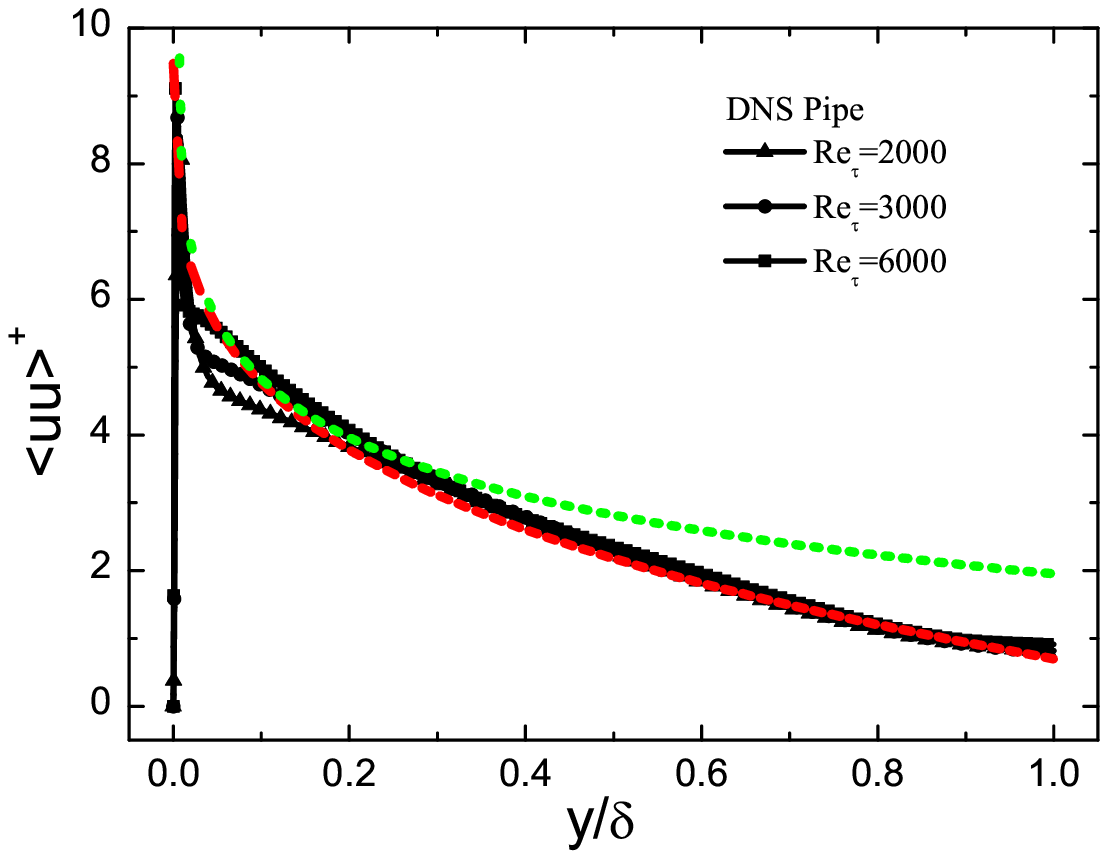}}
\subfloat[]{\includegraphics[trim = 0.75cm 10cm 15.5cm 1cm, clip, width = 6.5 cm]{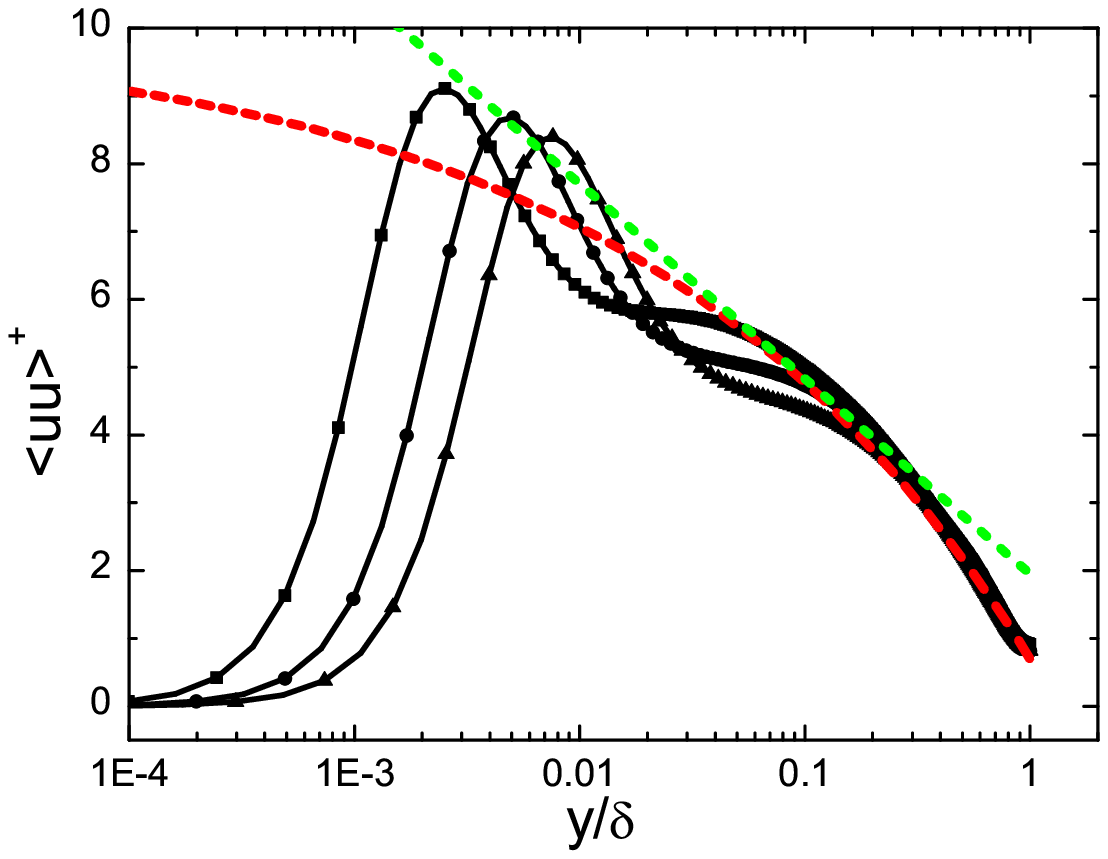}}\\
\subfloat[]{\includegraphics[trim = 0.75cm 10cm 15.5cm 1cm, clip, width = 6.5 cm]{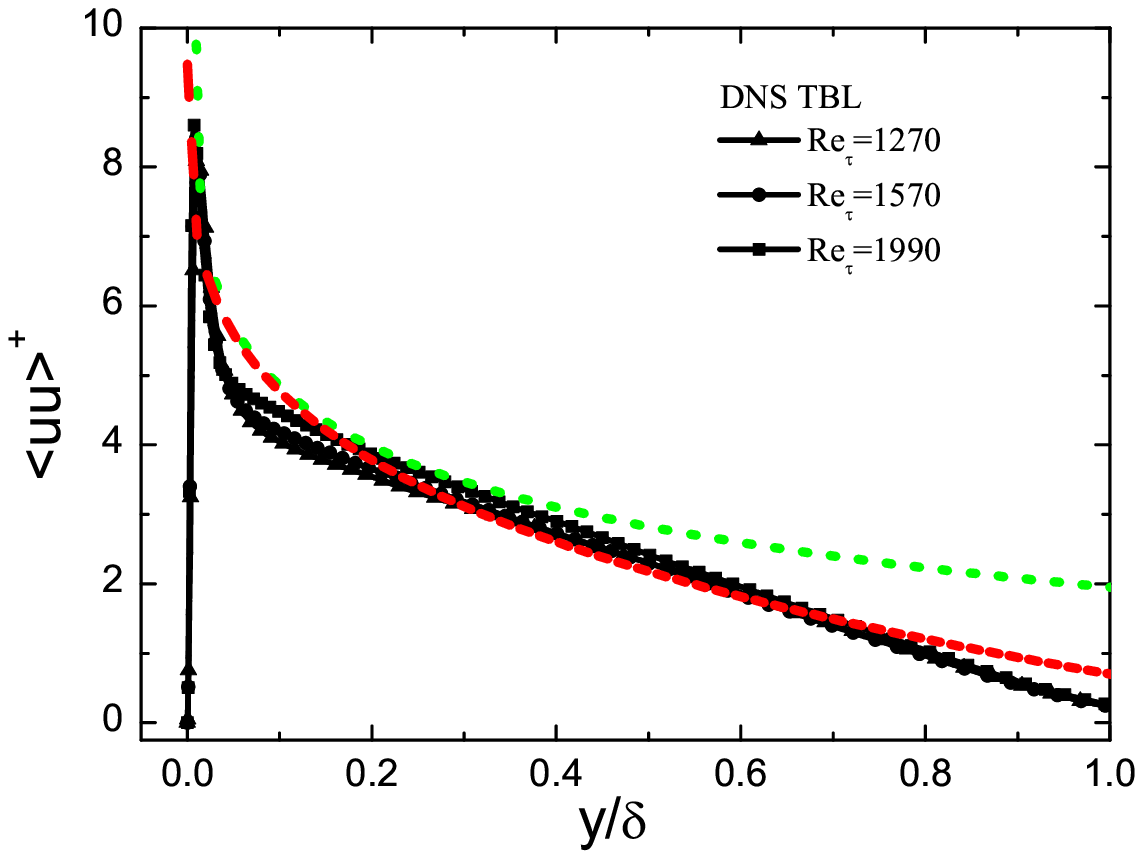}}
\subfloat[]{\includegraphics[trim = 0.75cm 10cm 15.5cm 1cm, clip, width = 6.5 cm]{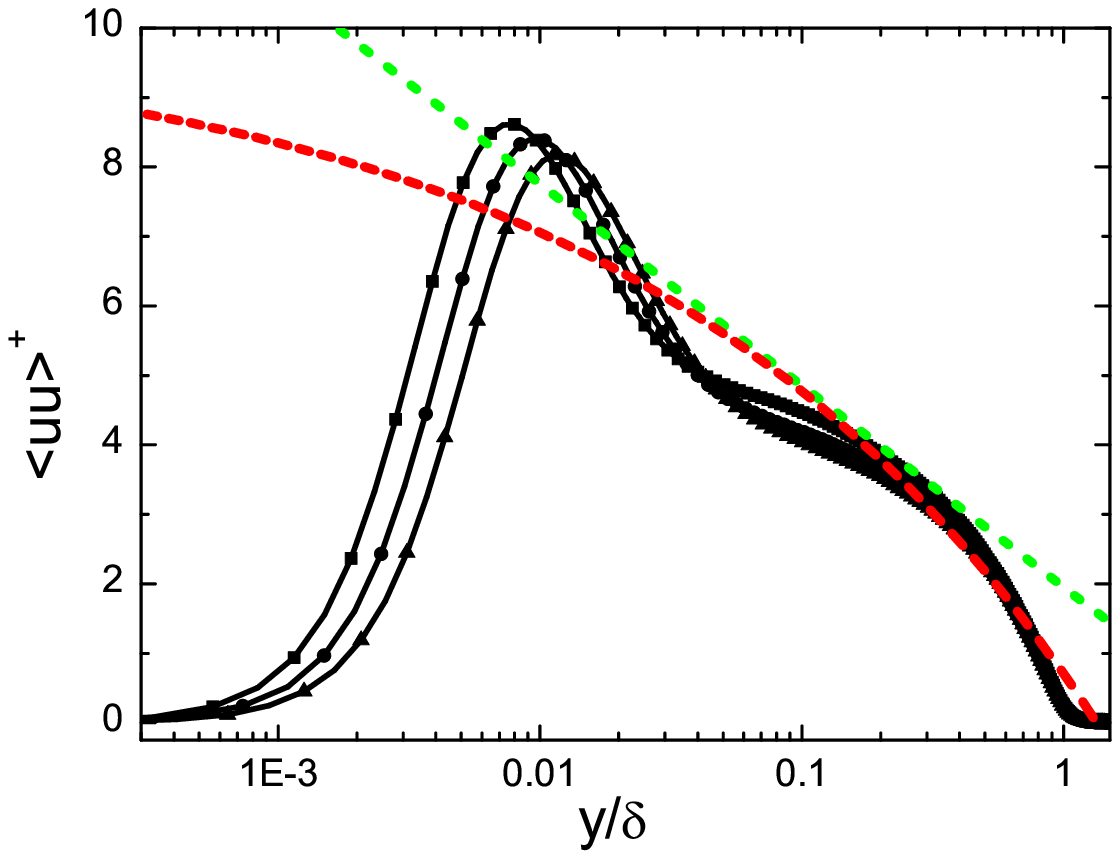}}\\
\caption{Variance of streamwise velocity fluctuation of DNS data at high $Re_\tau$. Channel data in top panels are $Re_\tau=2000, 5200$ from \cite{Moser2015}, $Re_\tau=10^4$ from \cite{hoyas2022}. Pipe data in middle panels from \cite{Pirozzoli2021}. Bottom panels for TBL, $Re_\tau=1270$ from \cite{SO2010}, $Re_\tau=1570,1990$ from \cite{Sillero2013}.  Left column is for abscissa in linear and the right in logarithmic outer units. Dotted (green) line in top panels indicate $1.61-1.25\ln(y^\ast)$ by \cite{Hultmark2012PRL}; $2.2-1.26\ln(y^\ast)$ in middle by \cite{Marusic2013} and $1.95-1.26\ln(y^\ast)$ in the bottom by \cite{Samie2018JFM}. Dashed (red) lines indicate the present relation (\ref{eq:CS:outer28}), i.e. $10-9.3(y^\ast)^{1/4}$, the same for all the flows here. }\label{fig:plog2uu2DNS}
\end{figure}

\begin{figure}
\centering
\subfloat[]{\includegraphics[trim = 0.75cm 10cm 15.5cm 1cm, clip, width = 6.5 cm]{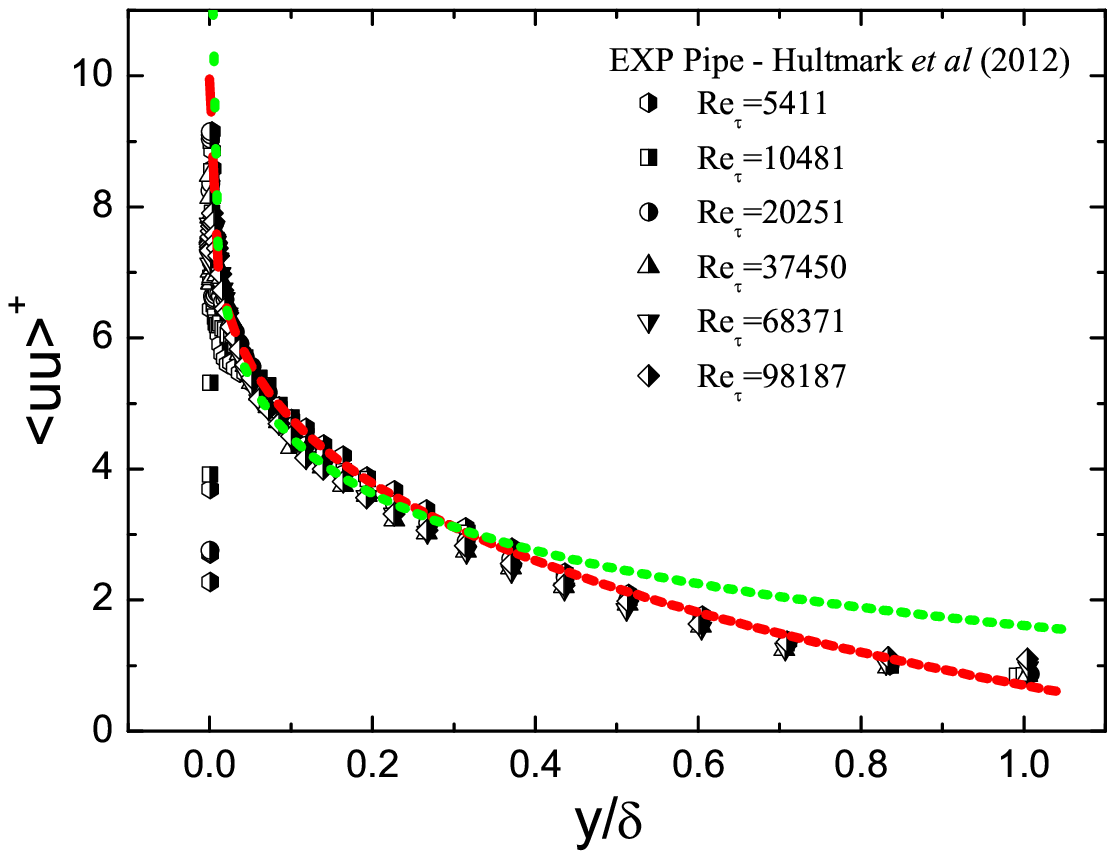}}
\subfloat[]{\includegraphics[trim = 0.75cm 10cm 15.5cm 1cm, clip, width = 6.5 cm]{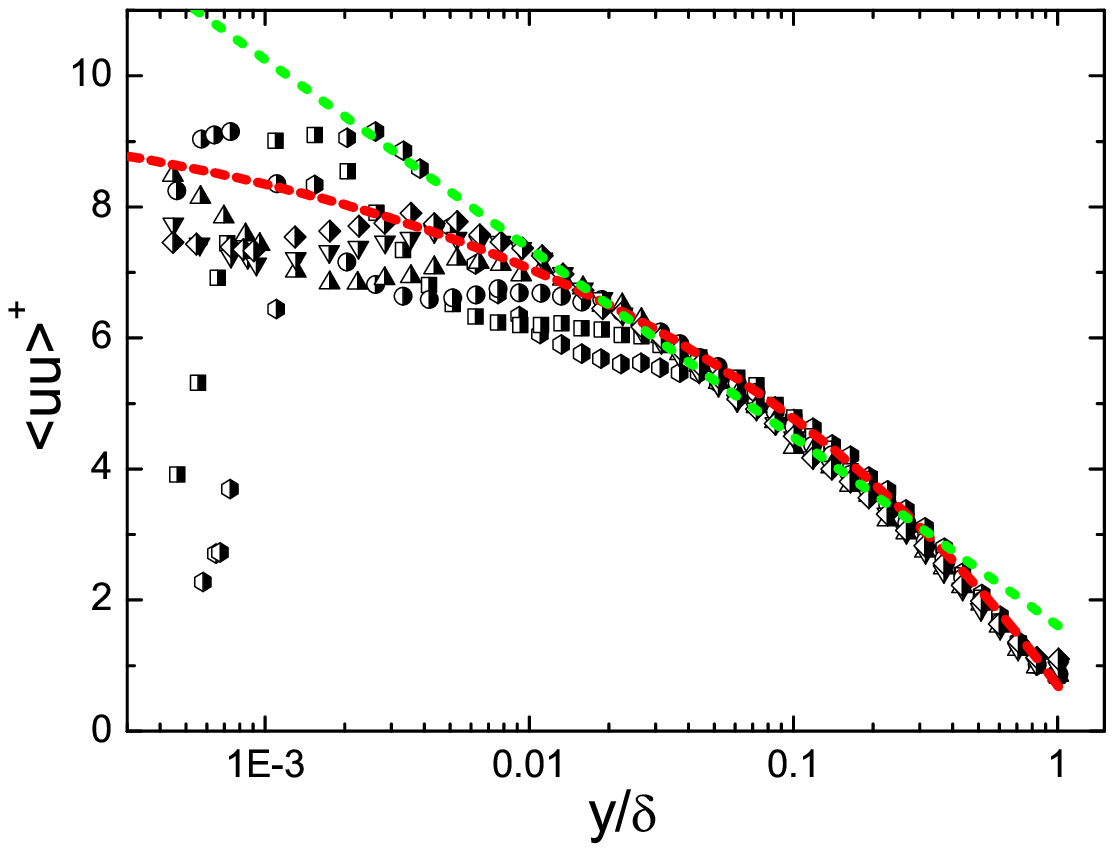}}\\
\subfloat[]{\includegraphics[trim = 0.75cm 10cm 15.5cm 1cm, clip, width = 6.5 cm]{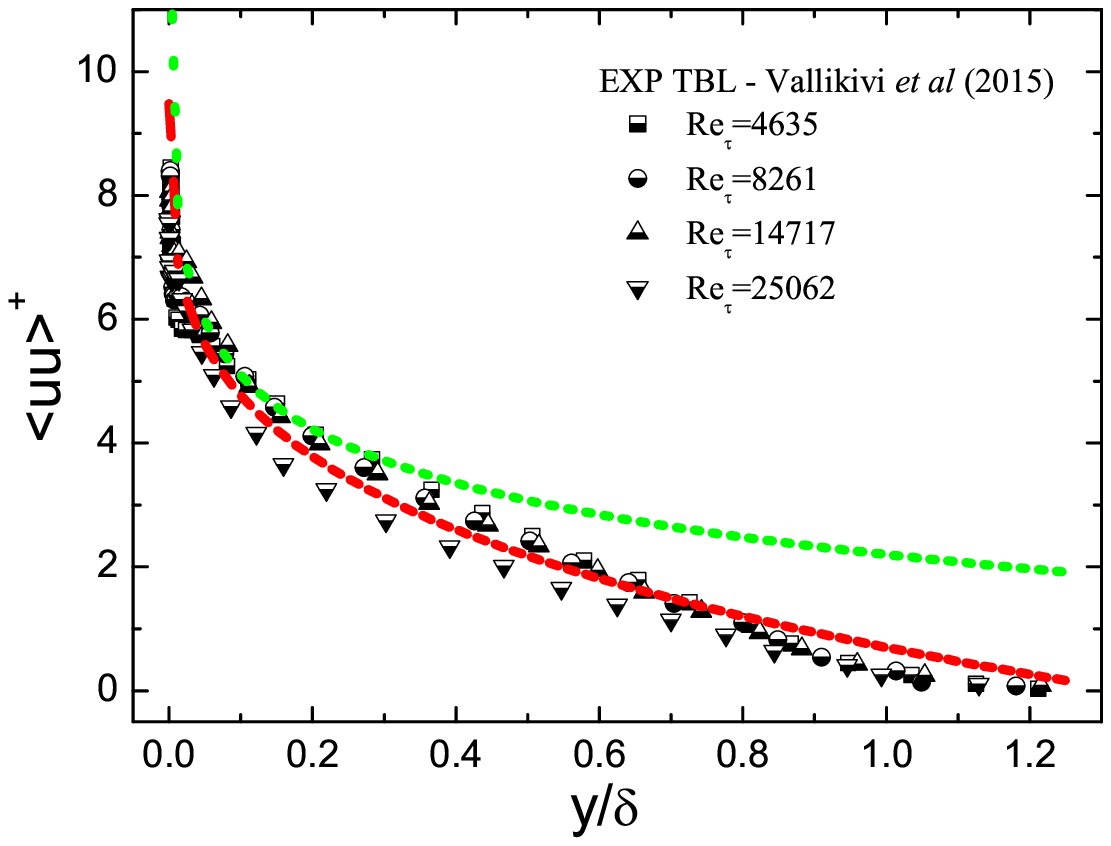}}
\subfloat[]{\includegraphics[trim = 0.75cm 10cm 15.5cm 1cm, clip, width = 6.5 cm]{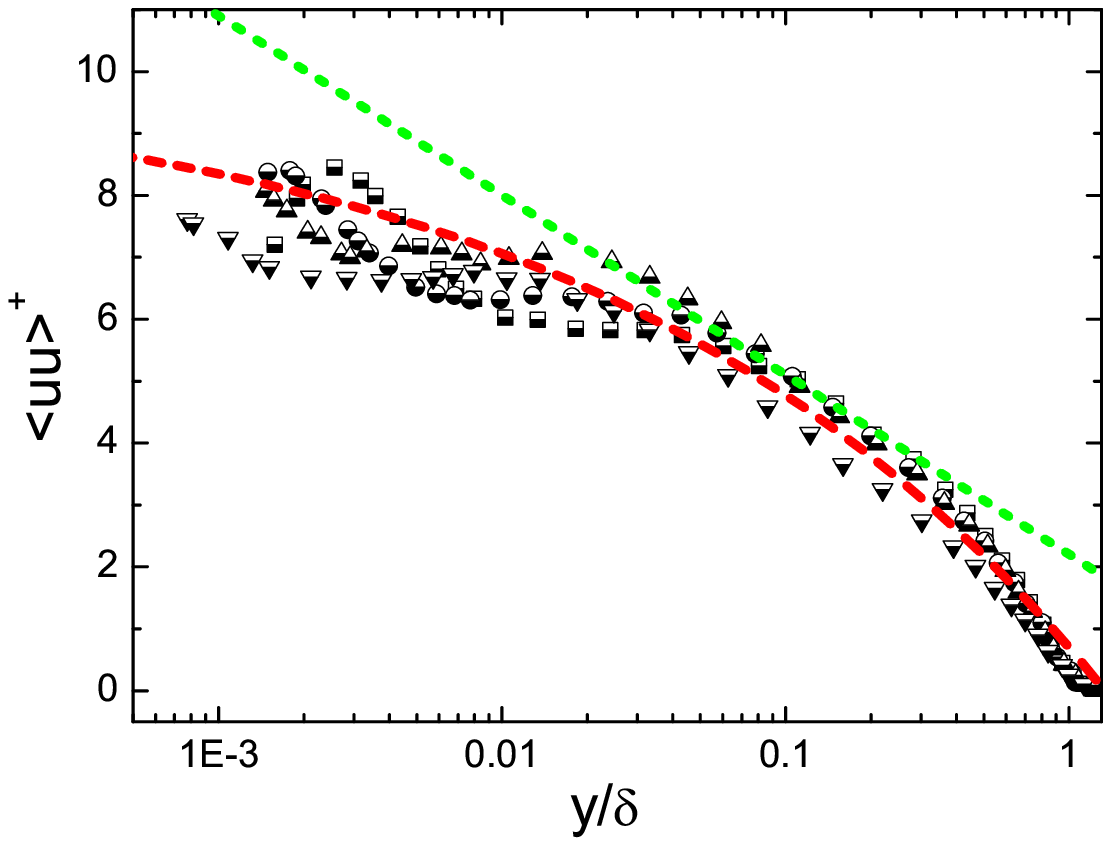}}\\
\subfloat[]{\includegraphics[trim = 0.75cm 10cm 15.5cm 1cm, clip, width = 6.5 cm]{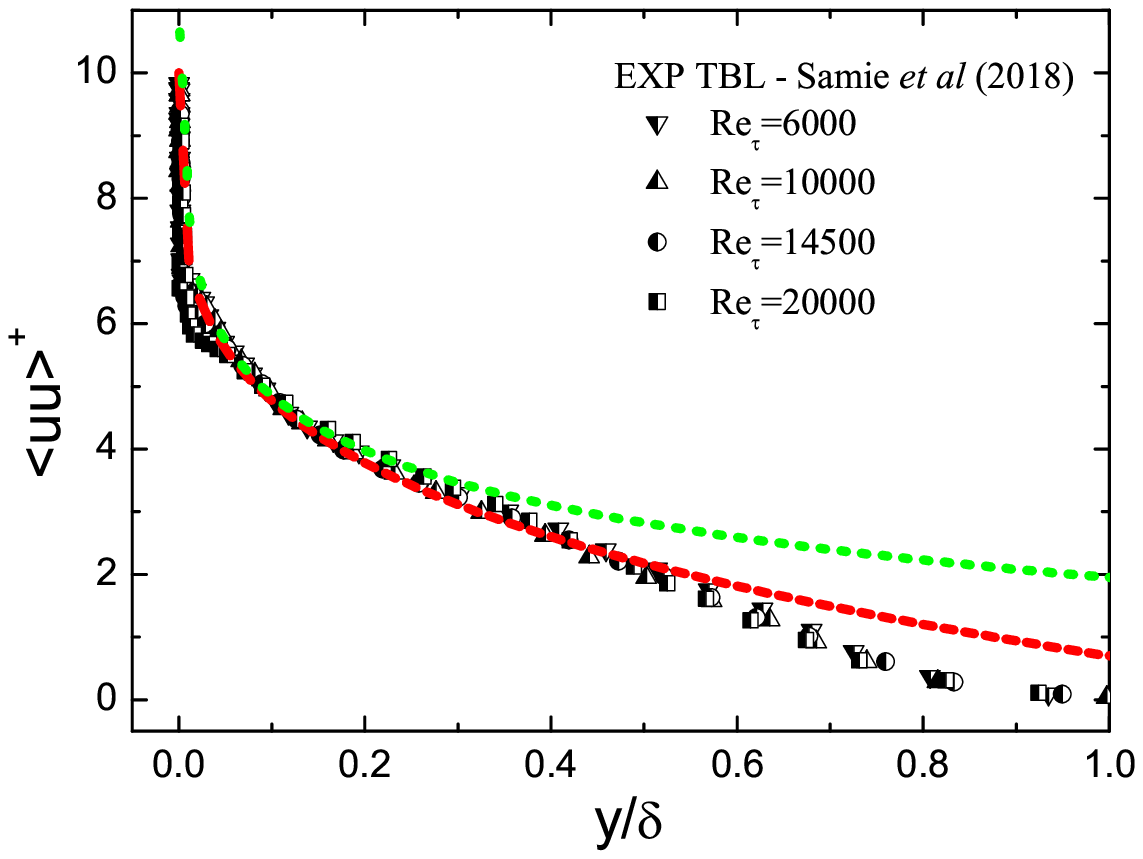}}
\subfloat[]{\includegraphics[trim = 0.75cm 10cm 15.5cm 1cm, clip, width = 6.5 cm]{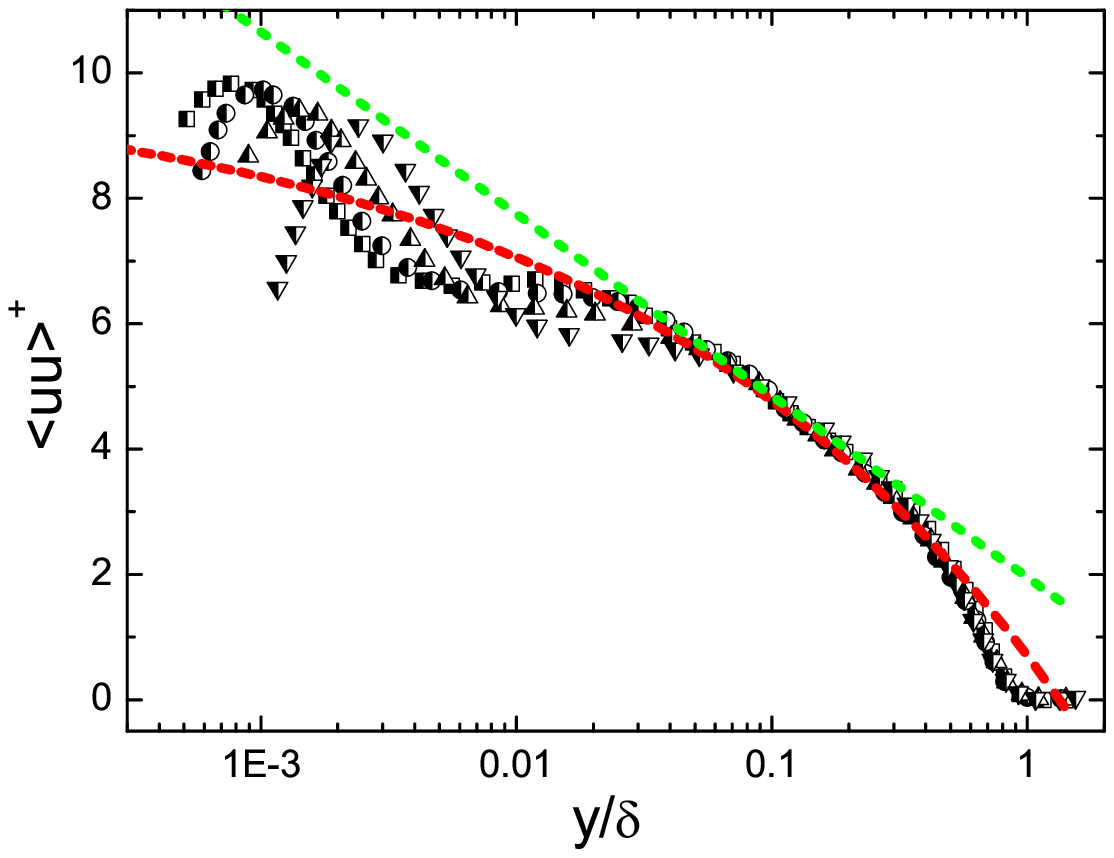}}
\caption{Variance of streamwise velocity fluctuation in high $Re_\tau$ experiments. Data in top panels from Princeton pipe \citep{Hultmark2012PRL}, middle from Princeton TBL \citep{vallikivi2015}, and bottom from Melbourne TBL \citep{Samie2018JFM}. Dotted (green) line in top panels indicate $1.61-1.25\ln(y^\ast)$ by \cite{Hultmark2012PRL}; $2.2-1.26\ln(y^\ast)$ in middle by \cite{Marusic2013} and $1.95-1.26\ln(y^\ast)$ in the bottom by \cite{Samie2018JFM}. Dashed (red) lines indicate (\ref{eq:CS:outer28}), i.e. $10-9.3(y^\ast)^{1/4}$ in all the figures. }\label{fig:plog2uu2}
\end{figure}

\begin{table}
\center
\begin{tabular}
{ l  c  c  c  c c c c}
  \hline
  Quantity $\phi$ \quad\quad & $\alpha^{(CH)}_{\phi}$ & $\alpha^{(Pipe)}_{\phi}$ & $\alpha^{(TBL)}_{\phi}$ & $\beta_\phi^{(CH)}$ & $\beta_\phi^{(Pipe)}$ & $\beta_\phi^{(TBL)}$ \\
  $\langle uu\rangle^+$   & 10 & 10 & 10  & 9.3 & 9.3 & 9.3   \\
  $\langle ww\rangle^+ $  & 3.9 & 3.9 & 3.6 & 3.45 & 3.45 & 2.6  \\
  $ p'^+ $ & 4.84 & 4.6 &4.65  & 4.1 & 3.7 & 3.15  \\
  \hline
\end{tabular}
    \caption{Parameters in (\ref{eq:CS:outer28}), for different fluctuations. Superscripts `CH', `Pipe' and `TBL' represent channel, pipe and boundary layer flows, respectively. Note that both $\alpha_\phi$ and $\beta_\phi$ vary only modestly among different flows, implying that essentially the same mechanisms applies for all flows. Moreover, $\beta_\phi$ is quite close to $\alpha_\phi$, as $\phi$ at the boundary layer edge $y^\ast=1$ is fairly small.}\label{tab:T1}
\end{table}

Figures 5-8 show the data comparisons with (\ref{eq:townsend}) and (\ref{eq:CS:outer28}) for $\langle uu\rangle^+$, $\langle ww\rangle^+$ and $ p'^+ $, respectively. Table 1 collects all the parameters for the three profiles, arising from these fits to the data, which will now be discussed in greater detail. 

Similar to figures 1-3, top panels of figures 5-8 are for channel, middle for pipe and bottom for boundary layer flows; the left column is for abscissa that are in linear units while right are for logarithmic units. A difference from figures 1-3 is that the data in figures 5-8 are denoted by symbols with lines (black), so that (\ref{eq:townsend}) and (\ref{eq:CS:outer28}) are better marked to guide the eye.  

Particularly for $\langle uu\rangle^+$, to avoid distractions by data scatter at small $Re_\tau$, we collect in figure 5 only high $Re_\tau$ profiles from DNS, namely, $Re_\tau$ from $2000$ to $10^4$ for channel; 2000 to 6000 for pipe; and 1270 to 1990 for TBL. Compared to figures 4 a,d,g, it is clear in figures 5 a,c,e that all high $Re_\tau$ profiles collapse on each other in the flow range for $0.1\lesssim y^\ast\lesssim1$, thus bearing out the inviscid similarity. This is confirmed again by experimental data in figure 6 corresponding to higher $Re_\tau$.

Note that the logarithmic behavior advanced in the literature \citep{Hultmark2012PRL,Marusic2013,Samie2018JFM} is indicated by green dotted line. Although it characterizes data in the region from $0.1\lesssim y^\ast\lesssim0.3$, the value of the intercept $B_\phi$ needs to be adjusted for the three flows from 1.61 to 2.2, while $A_\phi$ holds constant around 1.26. In contrast, the red dashed line represents (\ref{eq:CS:outer28}) with the same $\alpha_\phi=10$ and $\beta_\phi=9.3$, which reproduces data well for channel, pipe and TBL flows, covering not only the logarithmic range but also the so-called wake region, almost all the way to the centerline of channel and pipe flows.

\begin{figure}
\centering
\subfloat[]{\includegraphics[trim = 0.75cm 10cm 15.5cm 1cm, clip, width = 6.7 cm]{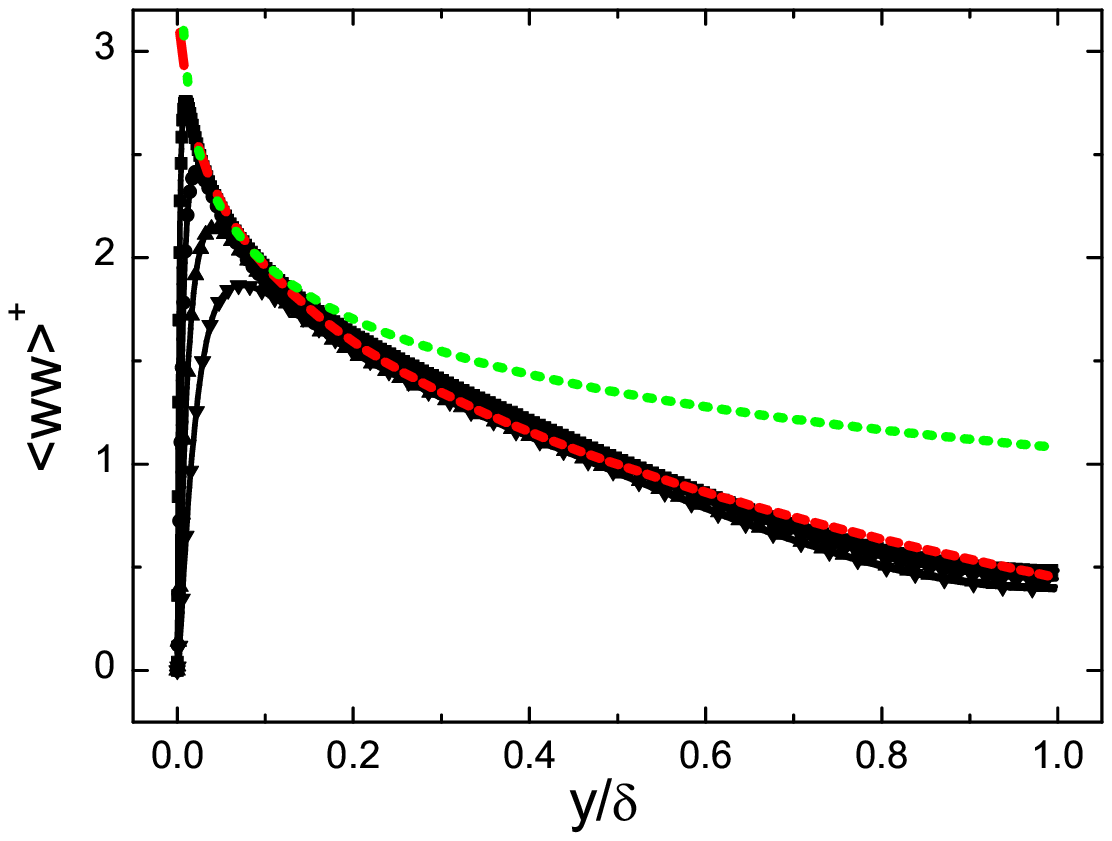}}
\subfloat[]{\includegraphics[trim = 0.75cm 10cm 15.5cm 1cm, clip, width = 6.7 cm]{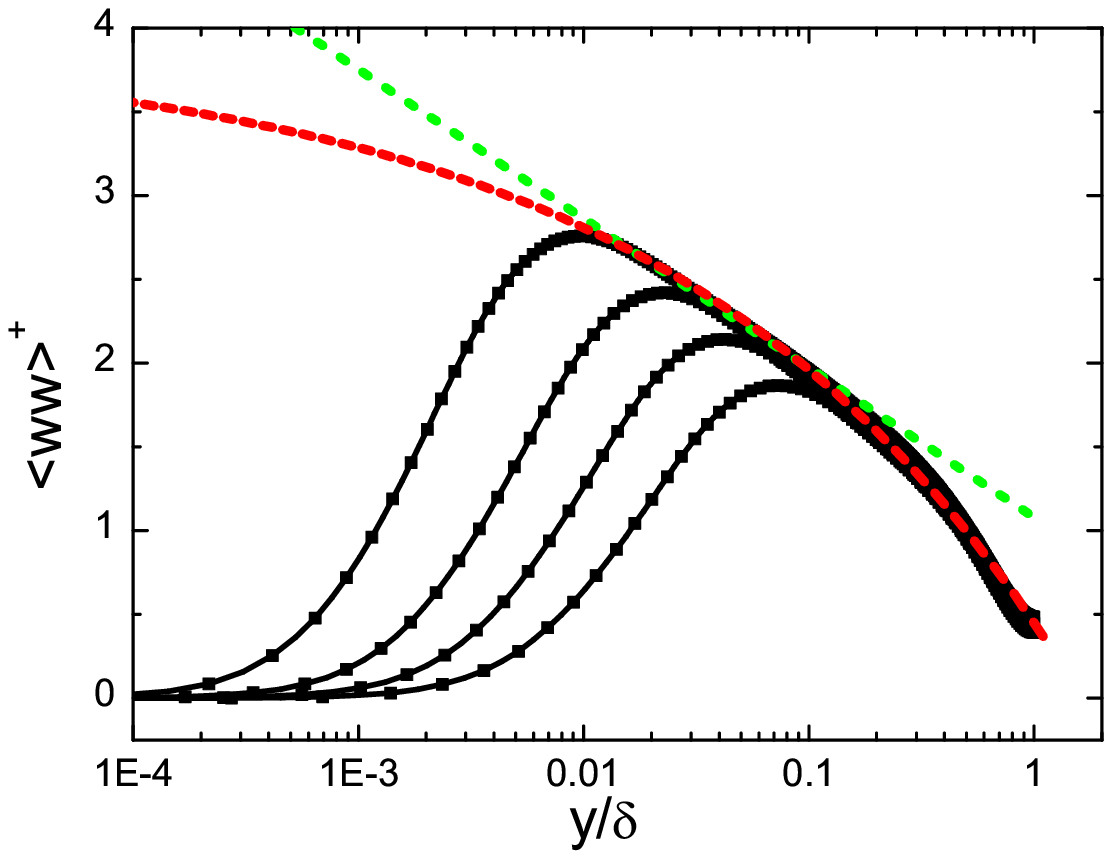}}\\
\subfloat[]{\includegraphics[trim = 0.75cm 10cm 15.5cm 1cm, clip, width = 6.7 cm]{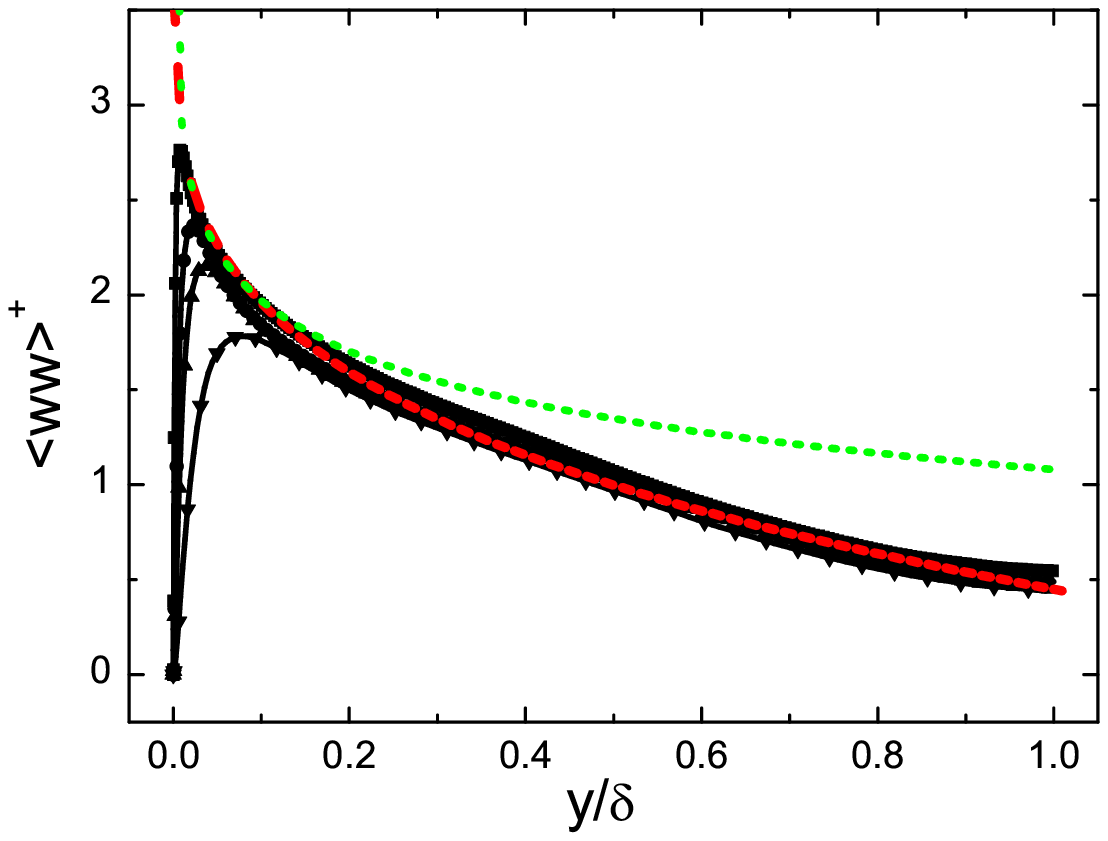}}
\subfloat[]{\includegraphics[trim = 0.75cm 10cm 15.5cm 1cm, clip, width = 6.7 cm]{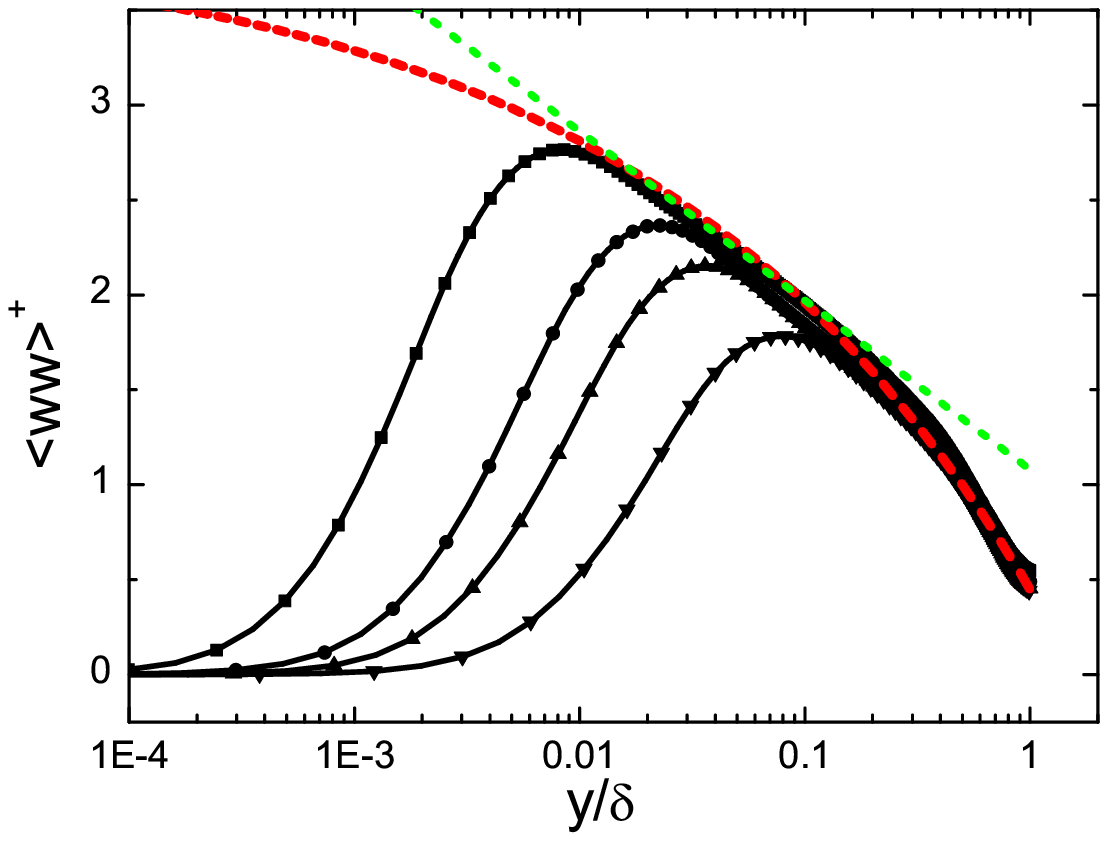}}\\
\subfloat[]{\includegraphics[trim = 0.75cm 10cm 15.5cm 1cm, clip, width = 6.7 cm]{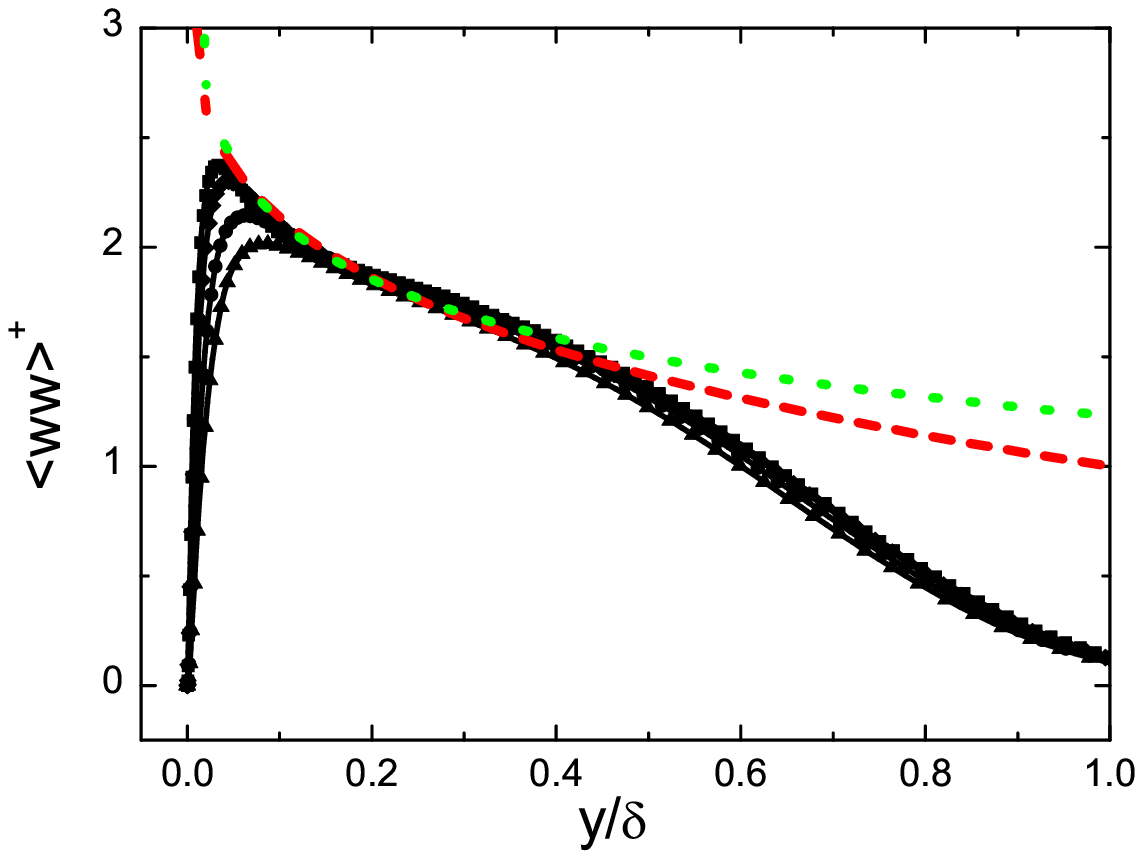}}
\subfloat[]{\includegraphics[trim = 0.75cm 10cm 15.5cm 1cm, clip, width = 6.7 cm]{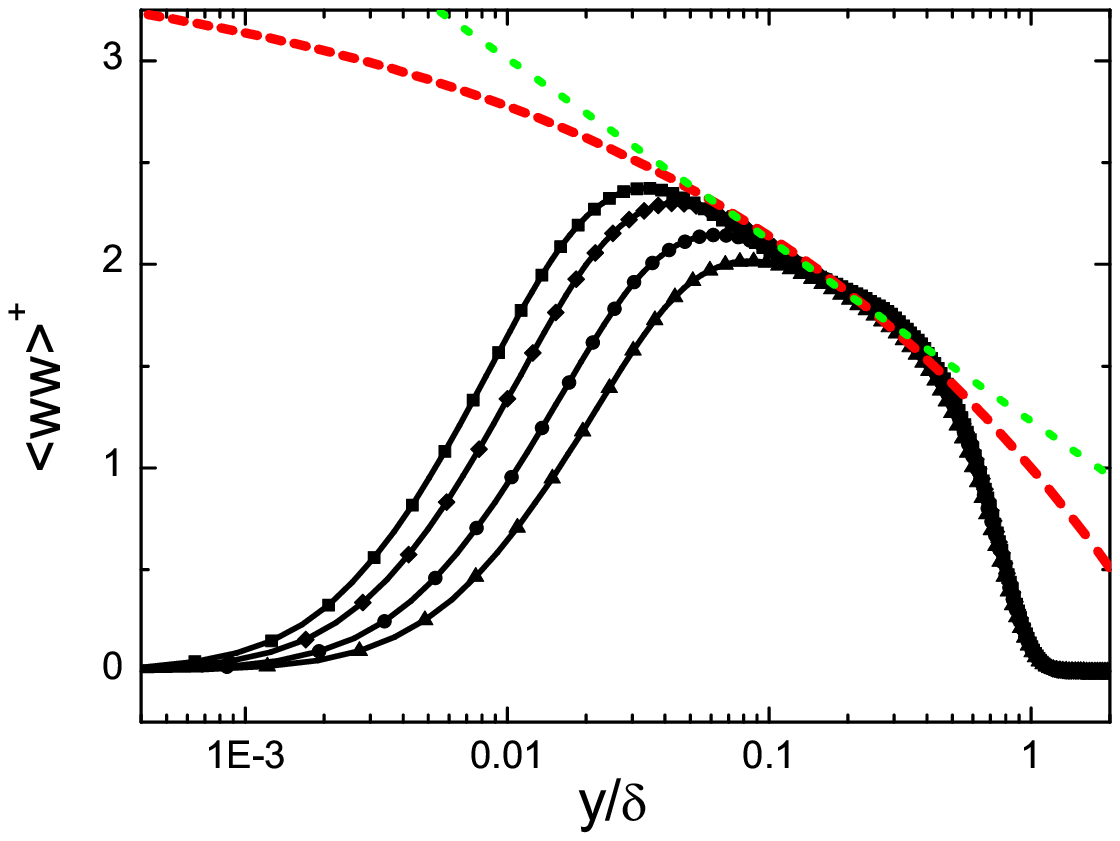}}\\
\caption{Variance of spanwise velocity fluctuation in channel (top), pipe (middle) and TBL (bottom) flows scaled in outer unit. Dashed (red) lines indicate the bounded decay (\ref{eq:CS:outer28}) with parameters in table 1. Dotted (green) lines indicate the logarithmic decay (\ref{eq:townsend}), i.e. $1.08-0.387\ln(y^\ast)$ for channel and pipe,  and $1.23-0.387\ln(y^\ast)$ for TBL. Symbols with lines are the same DNS data as in figure 1, i.e. channel from \cite{Moser2015}, pipe from \cite{Pirozzoli2021}, TBL from \cite{schlatter2009,schlatter2010simulations}.}\label{fig:plog2wvs}
\end{figure}

One may imagine that the data in figure 5 have not reached the asymptotic state and that (\ref{eq:townsend}) might agree with data better for higher $Re_\tau$. But experimental data from Princeton pipes \citep{Hultmark2012PRL} with $Re_\tau$ covering one more decade, e.g. $Re_\tau$ from about $5000$ to $10^5$, do not show any improvement of the fit to (\ref{eq:townsend}); see figures 6a-b. A similar observation is also true for the TBL, as shown in figures 6c-f.

Moreover, for $\langle ww\rangle^+$ and $p'^+$, (\ref{eq:CS:outer28}) extends almost all the way to the centerline of channel and pipe flows. Here, data in figures 7 and 8 are the same DNS groups as in figures 1 and contain those low $Re_\tau$ profiles. The agreement with (\ref{eq:CS:outer28}) is excellent at the smallest $Re_\tau\approx500$ for channel and pipe, in contrast to the log variation that agrees with data only for $Re_\tau\gtrsim2000$. Therefore, in both $y^\ast$ and $Re_\tau$ ranges, (\ref{eq:CS:outer28}) covers a wider range than (\ref{eq:townsend}).

\begin{figure}
\centering
\subfloat[]{\includegraphics[trim = 0.75cm 10cm 15.5cm 1cm, clip, width = 6.7 cm]{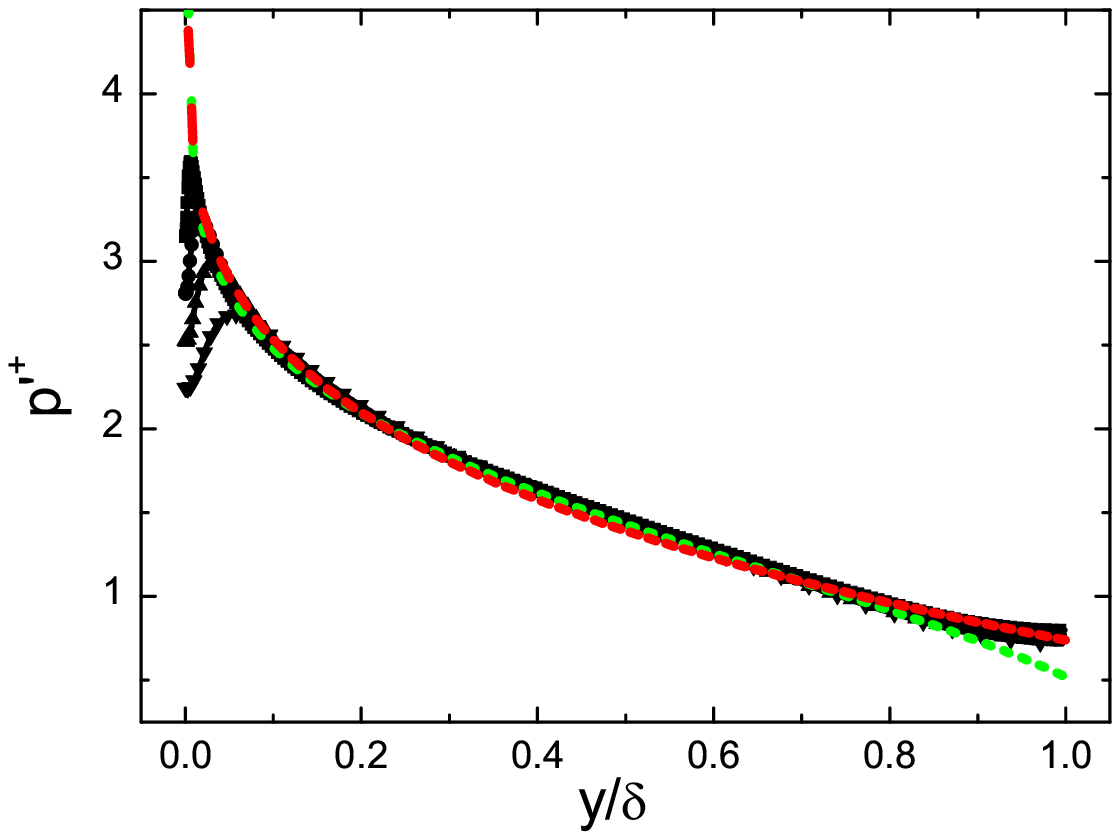}}
\subfloat[]{\includegraphics[trim = 0.75cm 10cm 15.5cm 1cm, clip, width = 6.7 cm]{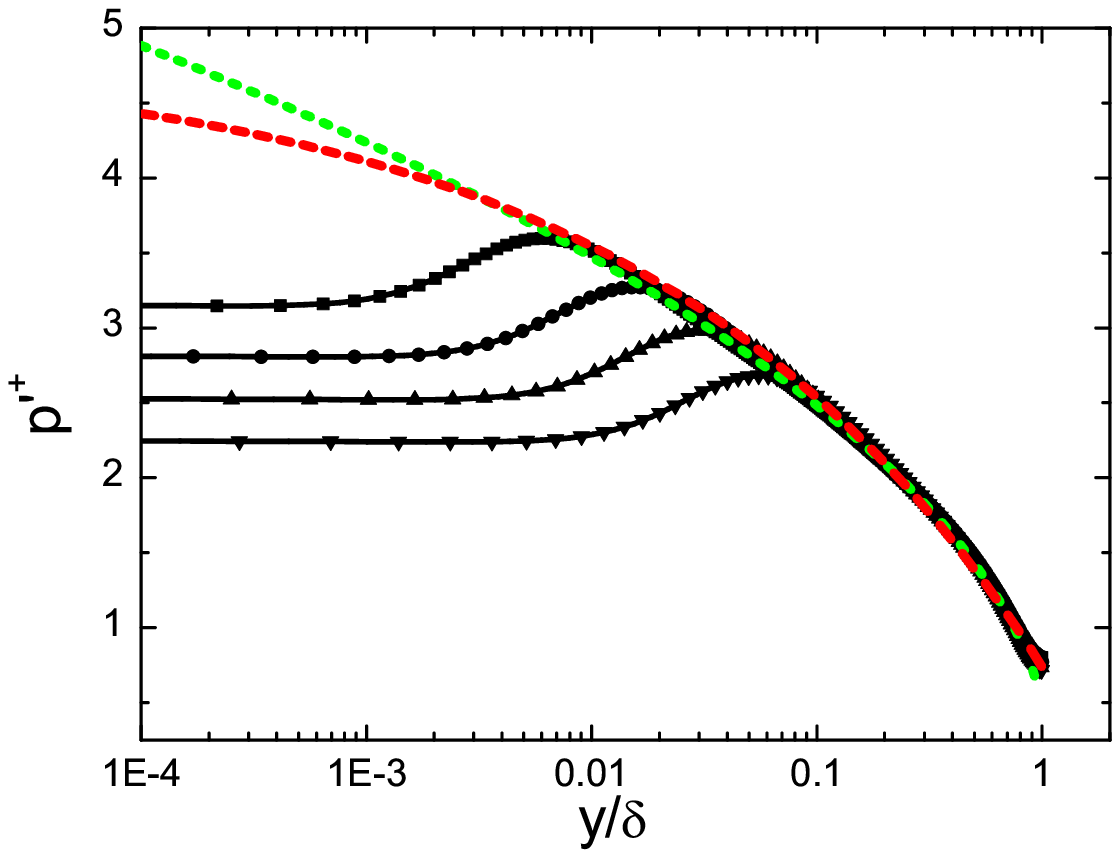}}\\
\subfloat[]{\includegraphics[trim = 0.75cm 10cm 15.5cm 1cm, clip, width = 6.7 cm]{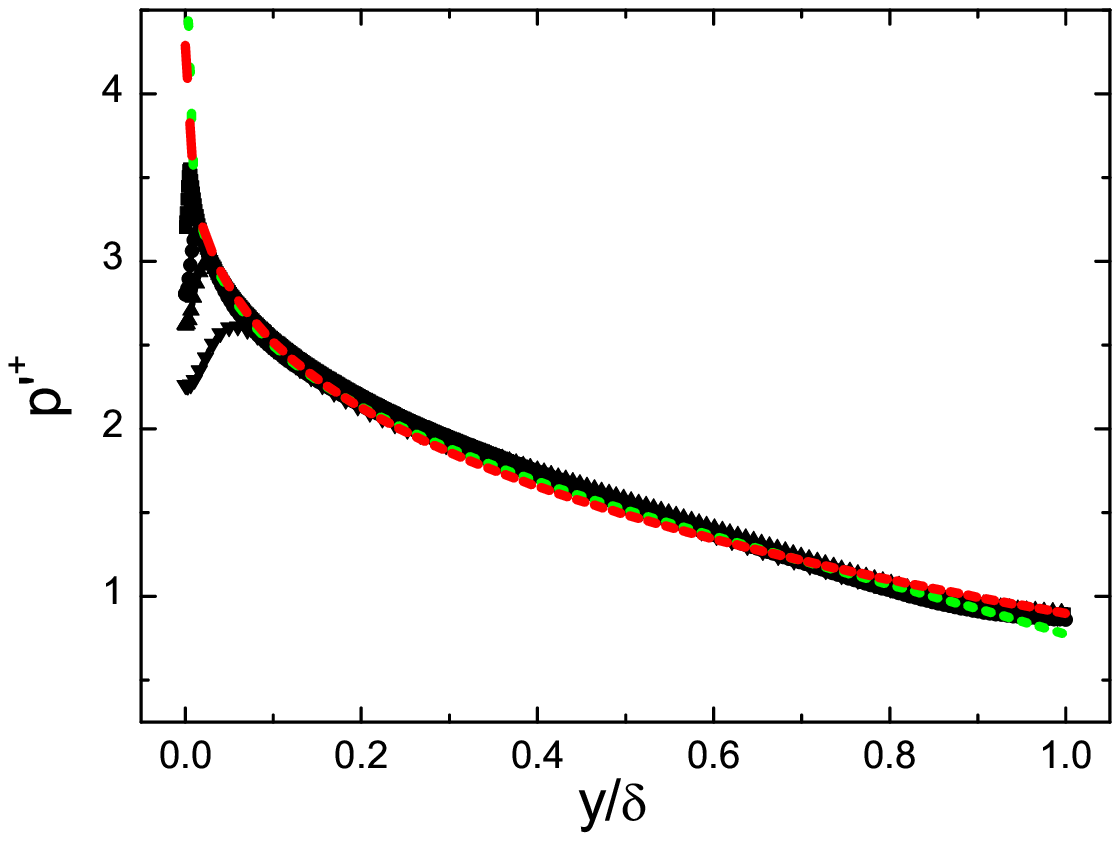}}
\subfloat[]{\includegraphics[trim = 0.75cm 10cm 15.5cm 1cm, clip, width = 6.7 cm]{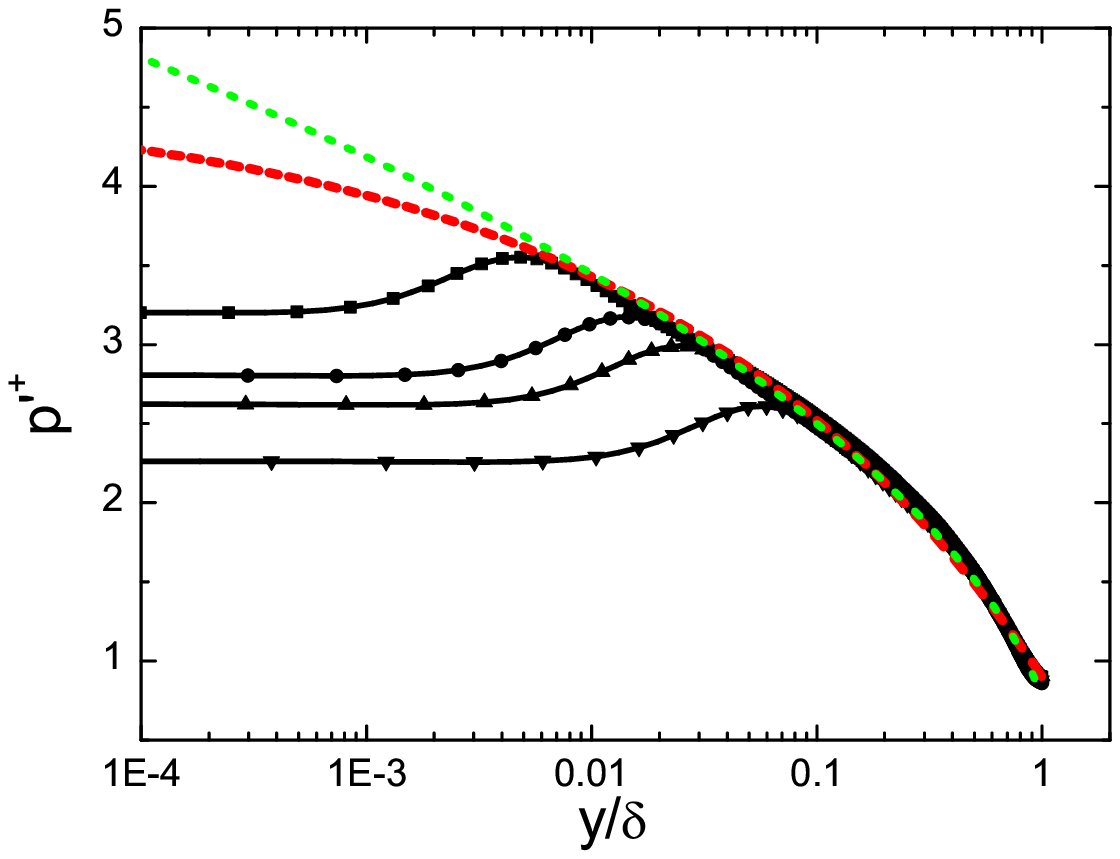}}\\
\subfloat[]{\includegraphics[trim = 0.75cm 10cm 15.5cm 1cm, clip, width = 6.7 cm]{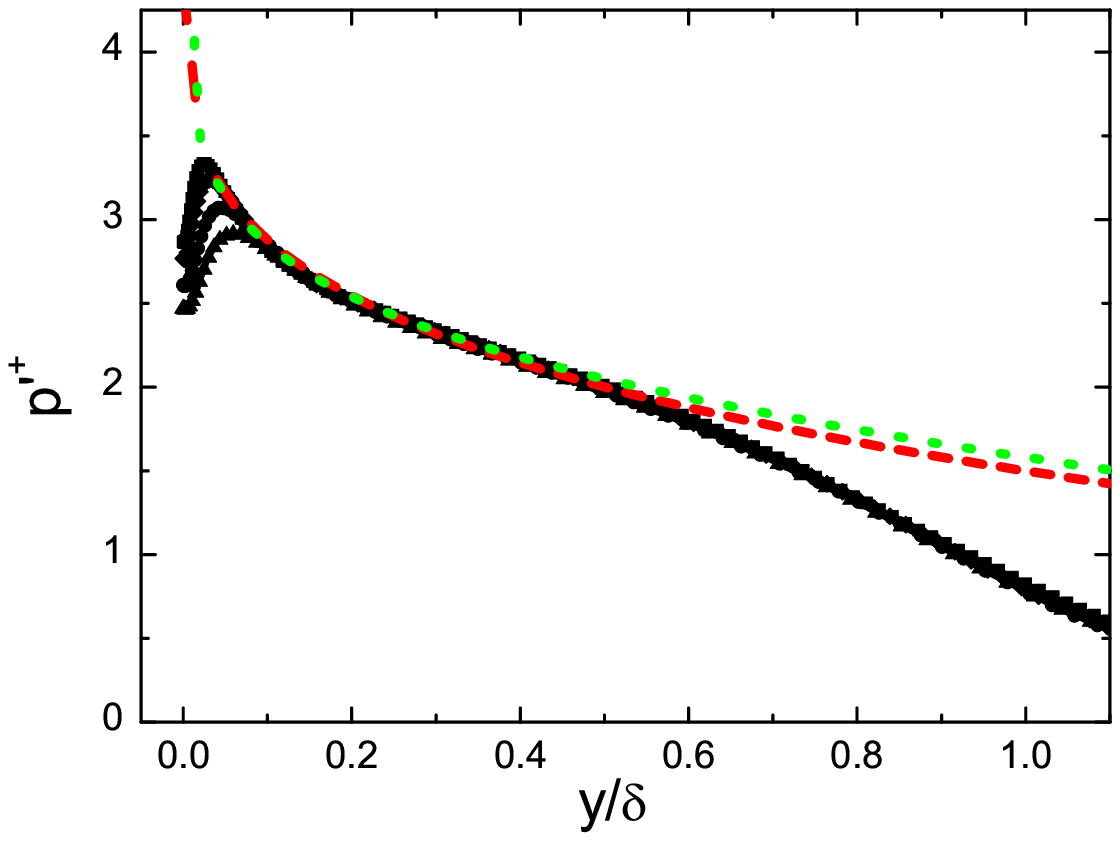}}
\subfloat[]{\includegraphics[trim = 0.75cm 10cm 15.5cm 1cm, clip, width = 6.7 cm]{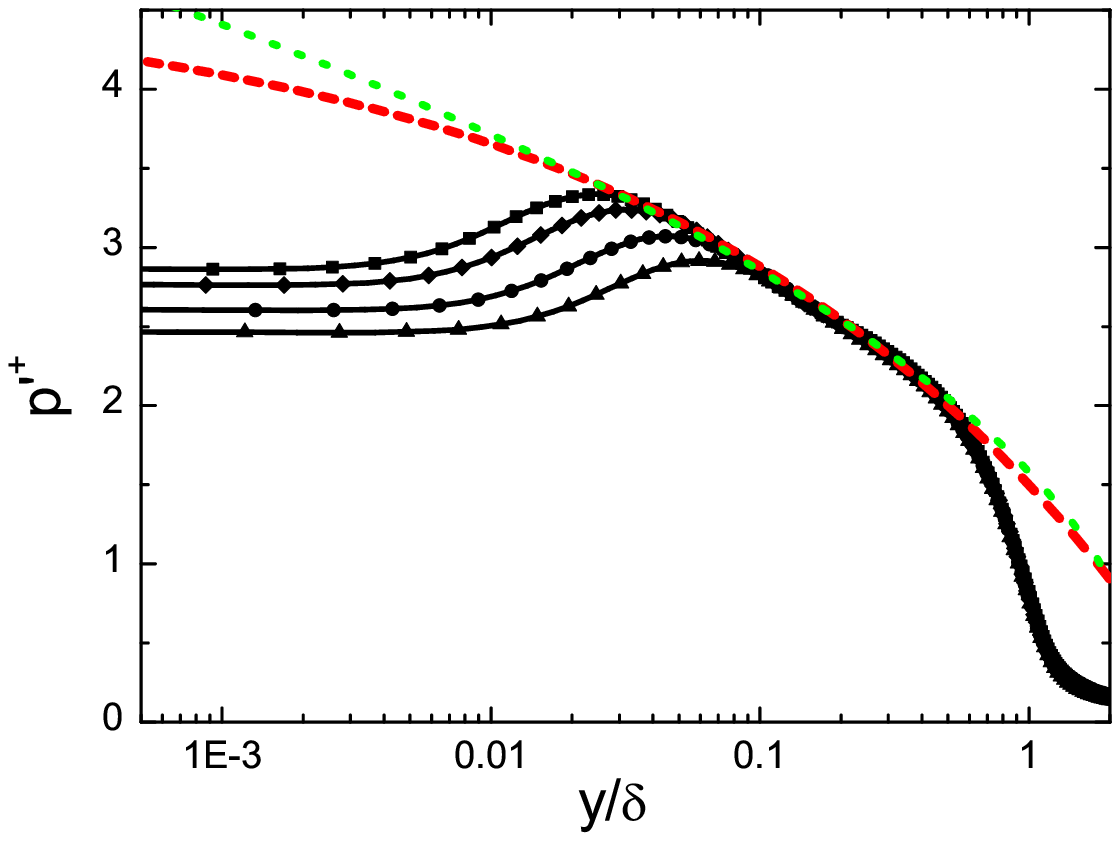}}\\
\caption{R.M.S. of pressure fluctuation in channel (top), pipe (middle) and TBL (bottom) flows scaled in the outer unit. Dashed (red) lines indicate the bounded decay (\ref{eq:CS:outer28}) with parameters in table 1. Dotted (green) lines indicate the logarithmic decay in (\ref{eq:log:AAB:pp1}), i.e. $\sqrt{0.27-2.56\ln(y^\ast)}$ for channel, $\sqrt{0.6-2.45\ln(y^\ast)}$ for pipe and $\sqrt{2.5-2.45\ln(y^\ast)}$ for TBL. Symbols with lines are the same DNS data as in figure \ref{fig:plog2wvs}.}\label{fig:plog2pvs}
\end{figure}

Three further points will |red{now} be discussed. First, the difference between (\ref{eq:CS:outer28}) and (\ref{eq:townsend}) is more vital for asymptotically high $Re_\tau$. For (\ref{eq:townsend}), an infinitely large of $\phi^+\propto\ln Re_\tau$ would arise as $y^\ast\rightarrow 0$ and $Re_\tau\rightarrow\infty$. 
In contrast, (\ref{eq:CS:outer28}) assigns a plateau of $\phi^+\approx \alpha_\phi$ in the same limit. Such an asymptotic plateau implies that turbulent eddies in the bulk would be in a quasi-equilibrium state in the sense that their contribution to $\phi^+$ is invariant when $y^\ast$ changes. Mimicking Townsend's terminology, the decreasing influence of attached eddies away from the wall would be compensated by the increasing influence of detached eddies, so that their total contribution to $\phi^+$ remains invariant. Note also that according to (\ref{eq:CS:outer28}), the outer peak of $\langle uu\rangle^+$, if it exists, should be bounded by $\langle uu\rangle^+\approx10$. Clarification of such differences of perspectives in the asymptotic state require future work.

Second, while (\ref{eq:CS:outer28}) adheres closely with TBL data of \cite{vallikivi2015} up to $y^\ast=1$ (figure 6c), it is slightly and uniformly higher than TBL data of \citet{Samie2018JFM} for $y^\ast>0.6$ (figure 6e). This is due to the fact that former set of data are obtained for flow over a flat plate mounted in the same pipe in which data of \cite{Hultmark2012PRL} are measured. Therefore, the data of \cite{vallikivi2015} in its outer wake region resemble the center behavior of pipe data by \cite{Hultmark2012PRL}, both in agreement with (\ref{eq:CS:outer28}) up to $y^\ast=1$. In contrast, TBL data of \cite{Samie2018JFM} are measured in the Melbourne wind tunnel with the normal free stream boundary condition, so that a vanishing $\langle uu\rangle^+\approx0$ towards boundary layer edge is observed (figures 6 e\&f), which is lower than (\ref{eq:CS:outer28}). Such a difference reflects the wake influence on $\langle uu\rangle^+$ in TBL. In fact, the wake influence is much sharper for $\langle ww\rangle^+$ and $p'^+$ in TBL (figures 7e \& 8e). This issue will be addressed in the section 4.2.

\begin{figure}
\centering
\subfloat[]{\includegraphics[trim = 1.0cm 10cm 15.5cm 1cm, clip, width = 6.5 cm]{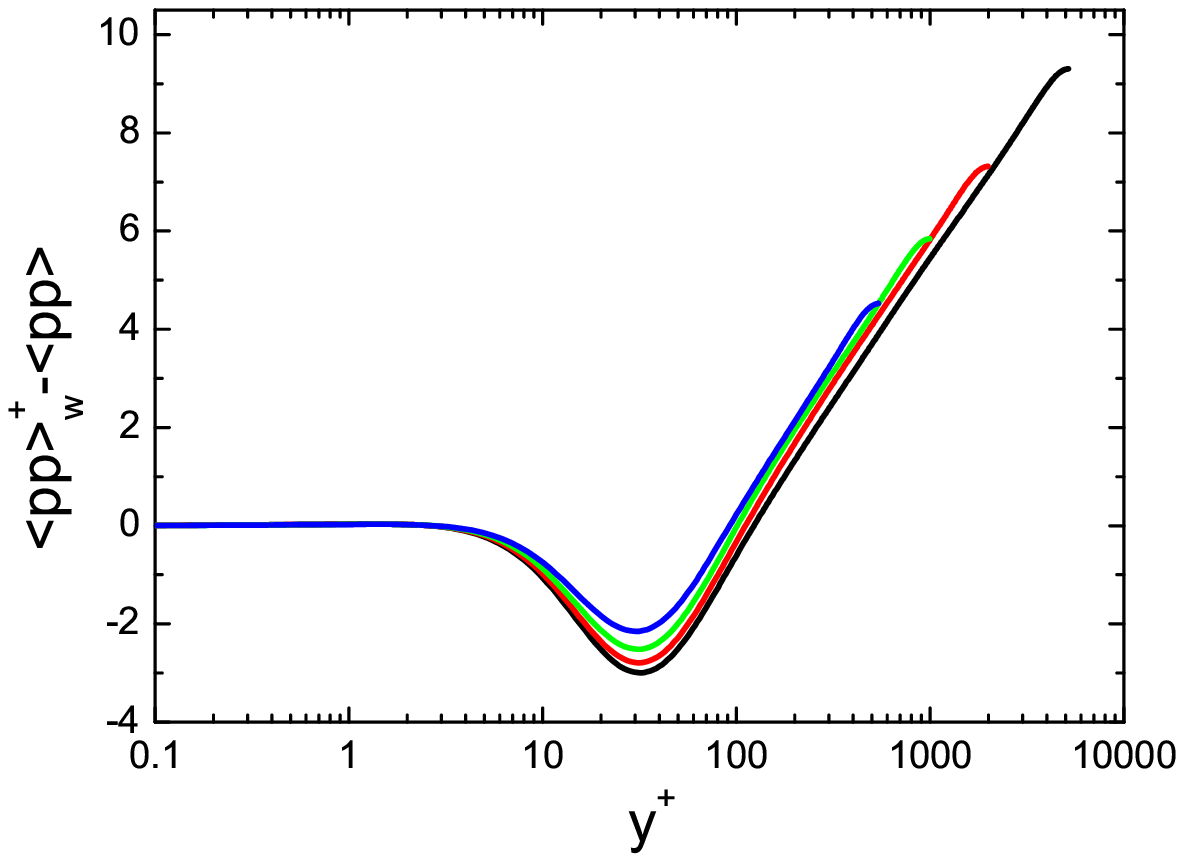}}
\subfloat[]{\includegraphics[trim = 1.0cm 10cm 15.5cm 1cm, clip, width = 6.5 cm]{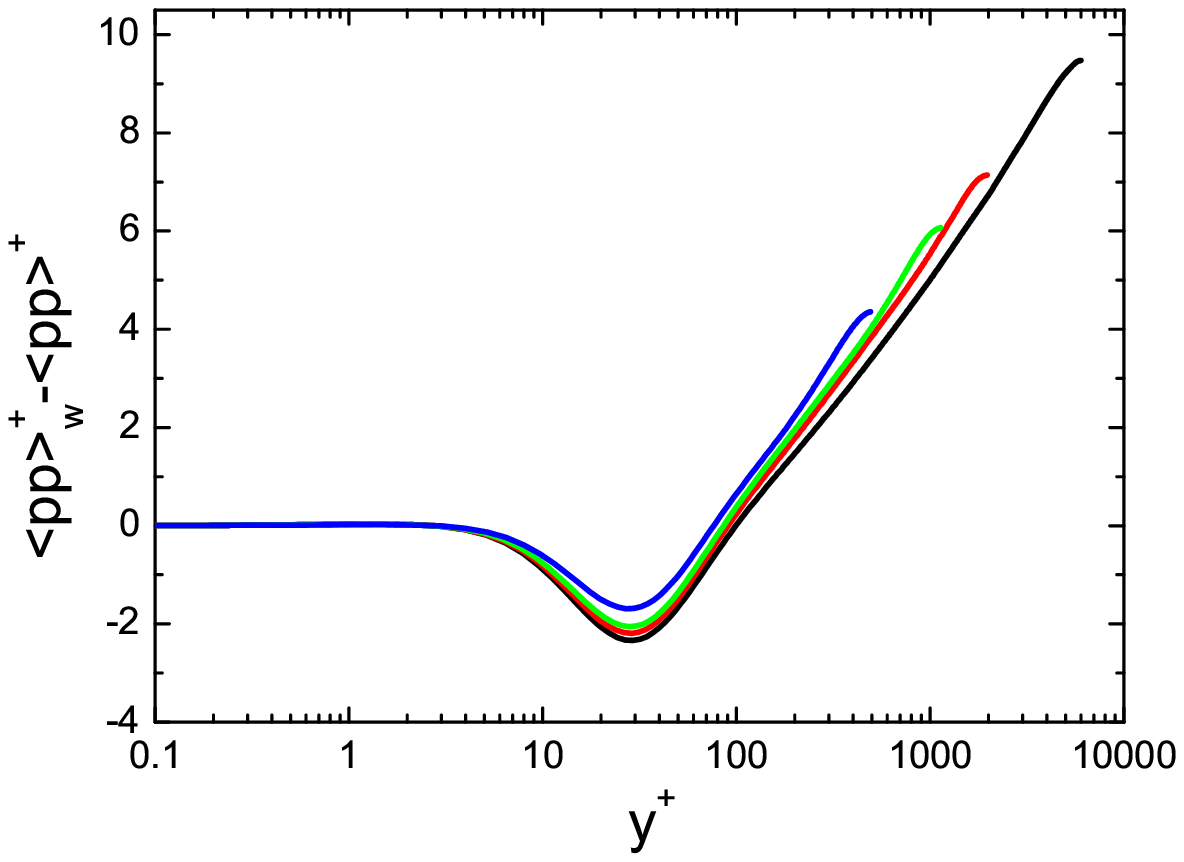}}\\
\subfloat[]{\includegraphics[trim = 1.2cm 10cm 15.5cm 1cm, clip, width = 6.5 cm]{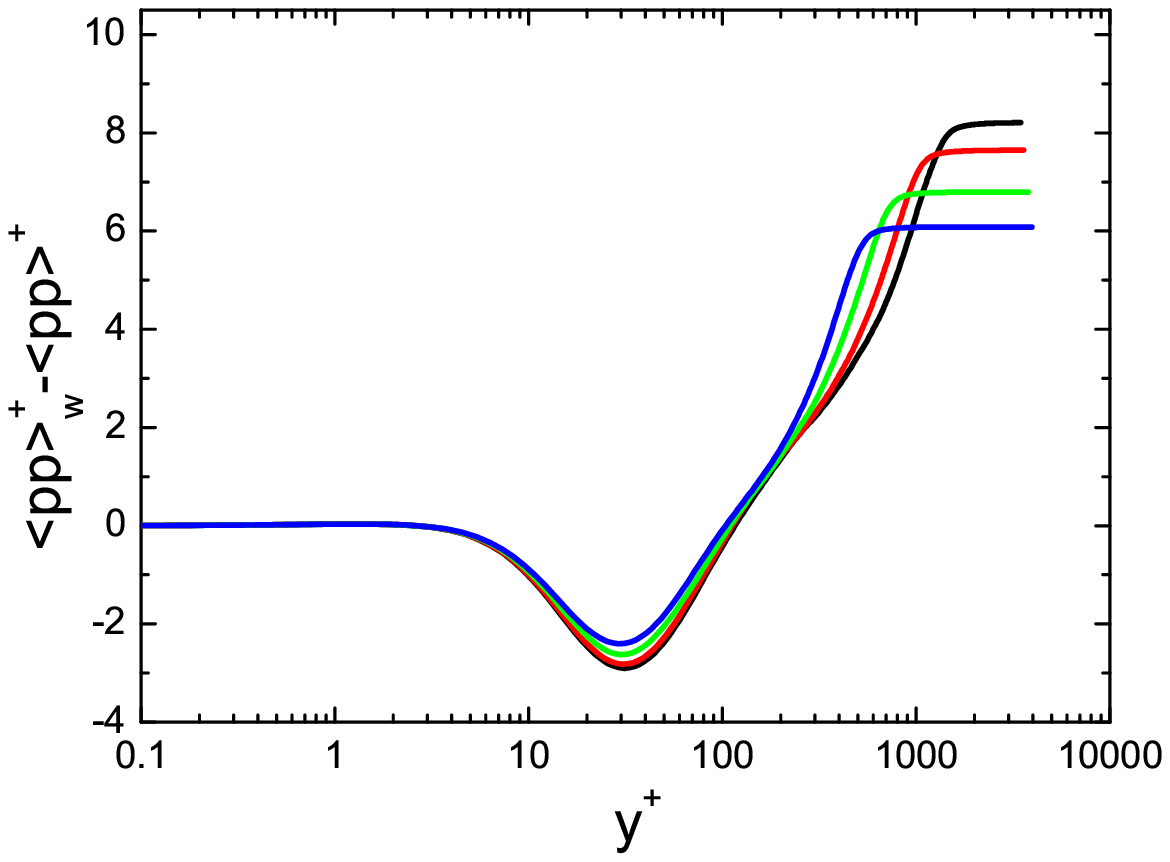}}
\subfloat[]{\includegraphics[trim = 1.0cm 10cm 15.5cm 1cm, clip, width = 6.5 cm]{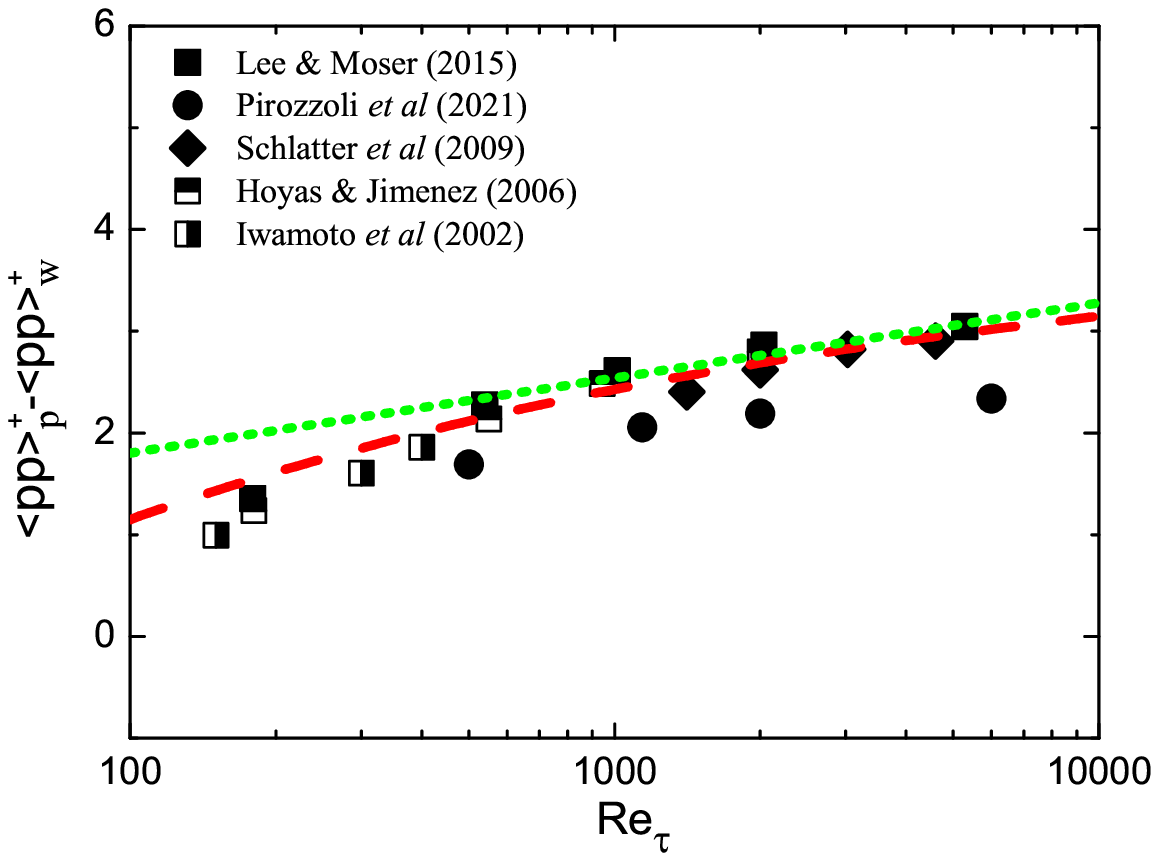}}
\caption{Wall-normal dependence of the variance of pressure fluctuations is shown in the plots of $\langle pp\rangle^+_w-\langle pp\rangle^+$ versus $y^+$, where $\langle pp\rangle^+_w$ is the wall-value of $\langle pp\rangle^+$. Lines are the same DNS data as in figure \ref{fig:p}, (a) for channels, (b) for pipes and (c) for TBLs. Note that lines depart markedly from each other with increasing $Re_\tau$ in the region $y^+>5$. (d) Difference between the peak and wall values of pressure variance, i.e. $\langle pp\rangle^+_p-\langle pp\rangle^+_w$, for channel, pipe and TBL flows for a series of $Re_\tau$ values. Dotted line (green) indicates the logarithmic growth by (\ref{eq:log:PLM1b}), i.e. $0.32\ln Re_\tau+0.33$. Dashed line (red) indicates the bounded variation of (\ref{eq:pw:p2}) according to CS. Symbols are DNS data, squares for channel \citep{Moser2015}, circles for pipe \citep{Pirozzoli2021} and diamonds for TBL \citep{schlatter2009}. Note that DNS channel data at $Re_\tau=150, 300, 400$ of \cite{Iwamoto2002}, and $Re_\tau=180, 550, 944, 2000$ of \cite{hoyas2006} are also included here.} \label{fig:pp}
\end{figure}

The third and final point is that for $p'^+$, the green dotted line in figure 8 represents
\begin{eqnarray}\label{eq:log:AAB:pp1}
p'^+(y^\ast)=\sqrt{ B_{\phi}-A_{\phi} \ln y^\ast },
\end{eqnarray}
which is the square root of (\ref{eq:townsend}) for pressure variance
\begin{eqnarray}\label{eq:log:AAB:pp2}
\langle pp\rangle^+(y^\ast)={ B_{\phi} - A_{\phi} \ln y^\ast}.
\end{eqnarray}
This equation is obtained by \cite{Panton2017} via an inner-outer matching (i.e. a viscous inner layer overlapping with an inviscid outer layer). It is almost indistinguishable from (\ref{eq:CS:outer28}) in figure 8. Nevertheless, as shown in figures \ref{fig:pp}a-c, $\langle pp\rangle^+_w-\langle pp\rangle^+$ versus $y^+$ produces no data collapse for $y^+>5$. Particularly for the trough located at about $y^+=30$, the data are markedly lower for increasing $Re_\tau$, thus creating a challenge for the inner-outer matching analysis. To reconcile this challenge, \cite{Panton2017} introduced two logarithmic slopes, i.e. $A^{CP}=2.56$ for the common part of presumed log profile in the overlap layer, and another $A^{w}=2.24$ for the $Re_\tau$ variation of the wall pressure. Following this fix, one can estimate
\begin{eqnarray}\label{eq:log:PLM1b}
\langle pp\rangle^+_p-\langle pp\rangle^+_w\propto (A^{CP}-A^{w})\ln Re_\tau\approx 0.32\ln Re_\tau,
\end{eqnarray}
which would break the wall scaling completely. 

As a comparison, figures 3b,d,f show that data of $p'^+_w-p'^+$ collapsed |red{well} up to the trough, better than $\langle pp\rangle^+_w-\langle pp\rangle^+$ in figures \ref{fig:pp}a-c. Moreover, via the bounding relation $p'^+_w\approx4.4-10.5/Re^{1/4}_\tau$ and $p'^+_p\approx4.84-10.5/Re^{1/4}_\tau$ given in CS, we have
\begin{equation}\label{eq:pw:p2}
\langle pp\rangle^+_p-\langle pp\rangle^+_w=p'^{+2}_p-p'^{+2}_w\approx4.07-9.24/Re_\tau^{1/4},
\end{equation}
which depicts data satisfactorily in a wider $Re_\tau$ range than (\ref{eq:log:PLM1b}) in figure \ref{fig:pp}d. 

\section{Discussion on flow geometry influence}

\subsection{Near wall universality}

\begin{figure}
\centering
\subfloat[]{\includegraphics[trim = 1.0cm 10cm 15.5cm 1cm, clip, width = 6.5 cm]{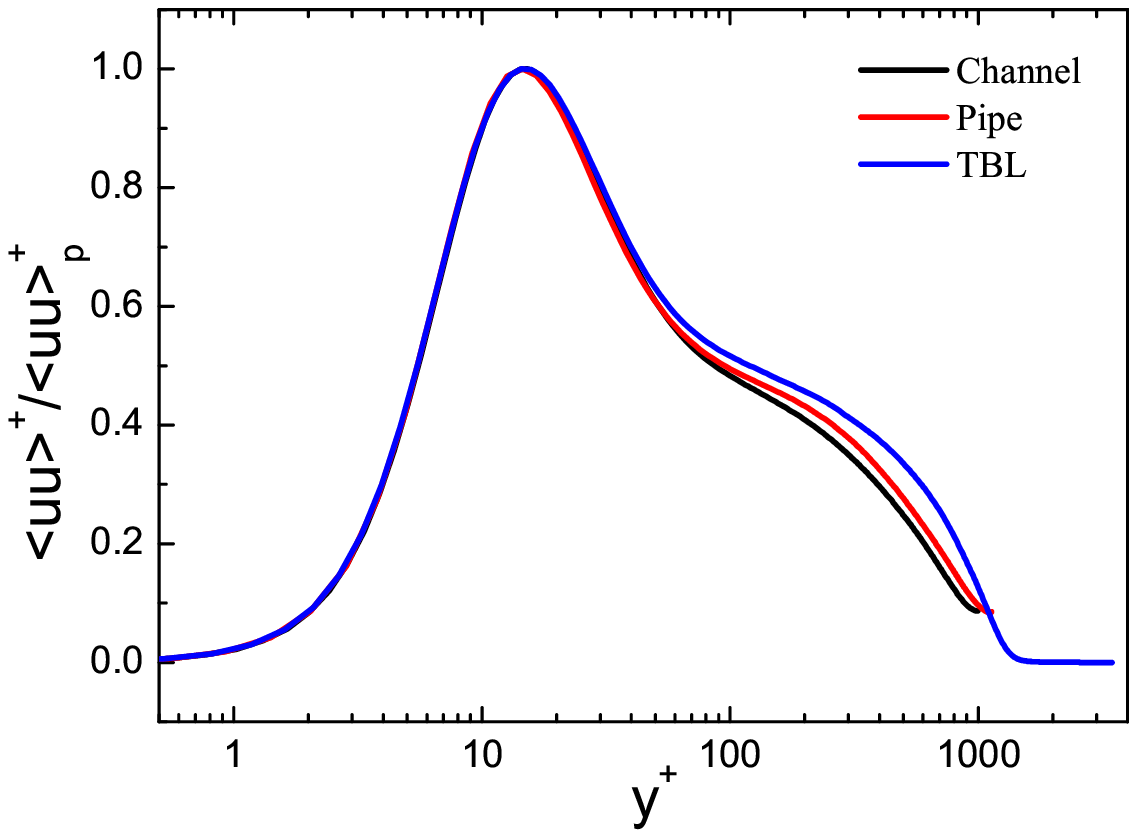}}\\
\subfloat[]{\includegraphics[trim = 1.0cm 10cm 15.5cm 1cm, clip, width = 6.5 cm]{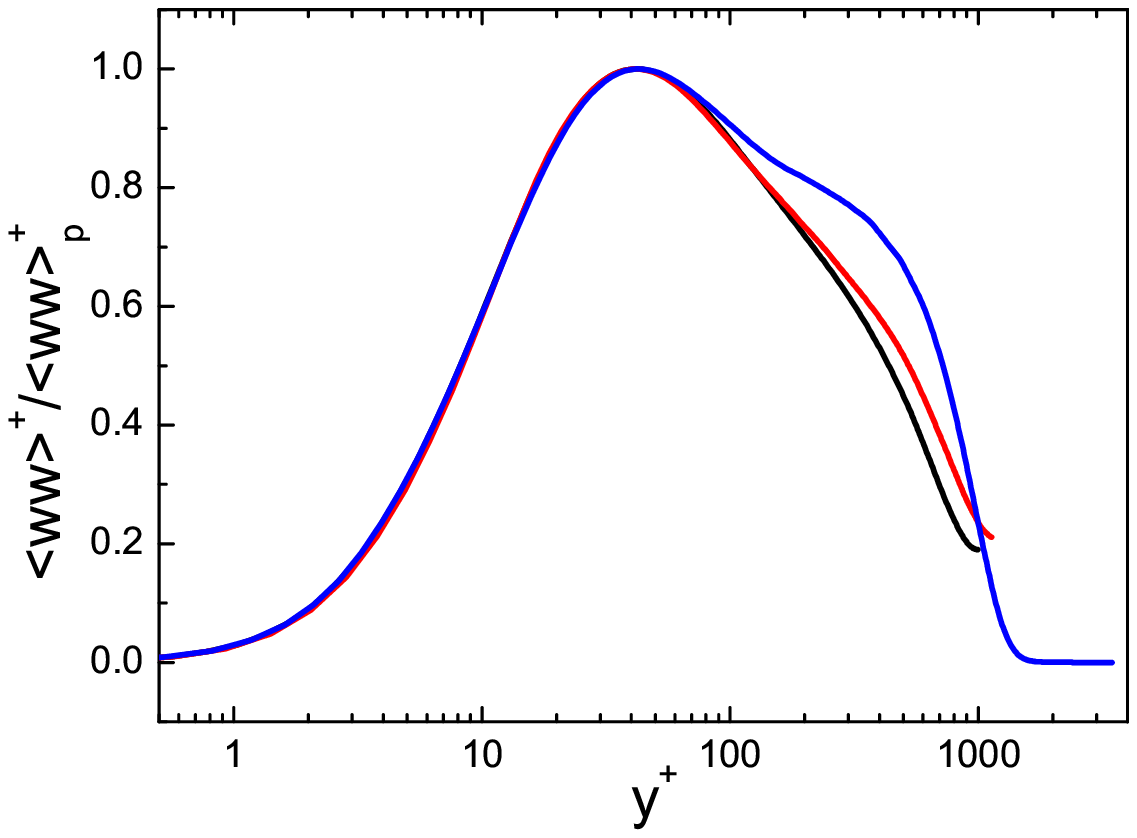}}
\subfloat[]{\includegraphics[trim = 1.0cm 10cm 15.5cm 1cm, clip, width = 6.5 cm]{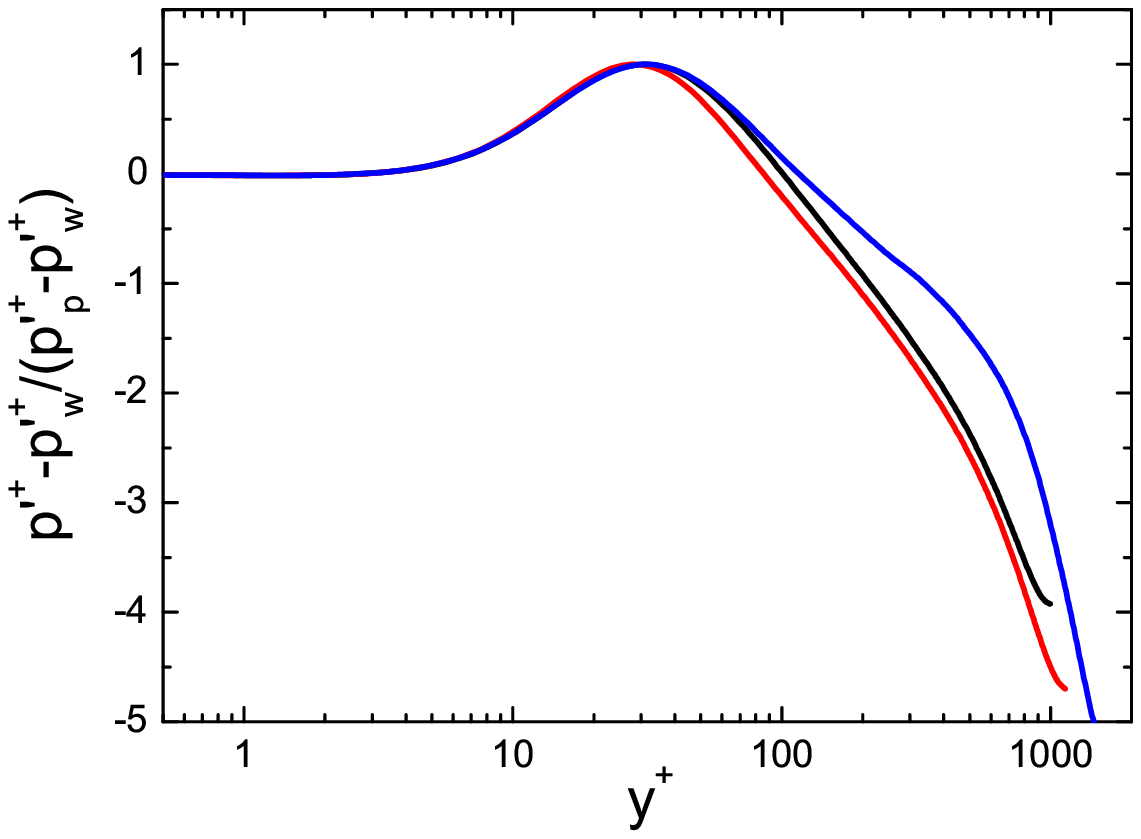}}
\caption{Wall-normal dependence for fluctuations after normalization by the corresponding peaks. (a) $\langle uu\rangle^+/\langle uu\rangle^+_p$; (b) $\langle ww\rangle^+/\langle ww\rangle^+_p$; (c) $p'^+_p-p'^+$. Lines are DNS data of channel at $Re_\tau=1000$ \citep{Moser2015}, of pipe at $Re_\tau=1140$ \citep{Pirozzoli2021}, and of TBL at $Re_\tau\approx1270$ \citep{schlatter2010simulations}, all of which collapse closely on each other.}\label{fig:cpt}
\end{figure}

We focus here on the geometry effects. First, a universal data collapse is summarized by unifying (\ref{Eq:CS21A}), (\ref{Eq:CS21B}) and (\ref{Eq:CS21C}) together, i.e.
\begin{equation}\label{Eq:universal}
\frac{\phi^+(y^+,Re_\tau)-\phi^+_w(Re_\tau)}{\phi^+_p(Re_\tau)-\phi^+_w(Re_\tau)}= f_\phi(y^+),
\end{equation}
where $f_\phi(y^+)$ depends on $y^+$ and $\phi$ but is independent of $Re_\tau$. For $\langle uu\rangle^+$, $\phi^+_w=0$ so that (\ref{Eq:universal}) reduces to (\ref{Eq:CS21A}); the same is true for $\langle ww\rangle^+$. For $p'^+$, as $\phi^+_p-\phi^+_w$ is a constant independent of $Re_\tau$ shown in figure 3, (\ref{Eq:universal}) reproduces (\ref{Eq:CS21C}) with $f_\phi=1-j(y^+)/j(0)$.

Moreover, as $y^+$ moves from the wall to the peak location, it is interesting to check whether $f_\phi(y^+)$ is universal for channel, pipe and TBL flows. This is indeed verified in figures \ref{fig:cpt}a, b \& c, for $\langle uu\rangle^+$,  $\langle ww\rangle^+$ and $p'^+$, respectively. Profiles from these three wall flows collapsed together from the wall to the peak location, which means that $Re_\tau$-dependence and geometry influence are canceled by the ratio of relative variations composed of $\phi^+_w$ and $\phi^+_p$. This is conceivable if the near wall region is viewed as a self-organized entity, so that superposition and modulation effects enforced by the outer flow structures are characterized to the first order by wall and peak values.

\subsection{Wake modification in TBL}

\begin{figure}
\centering
\subfloat[]{\includegraphics[trim = 0.75cm 10cm 15.5cm 1cm, clip, width = 6.7 cm]{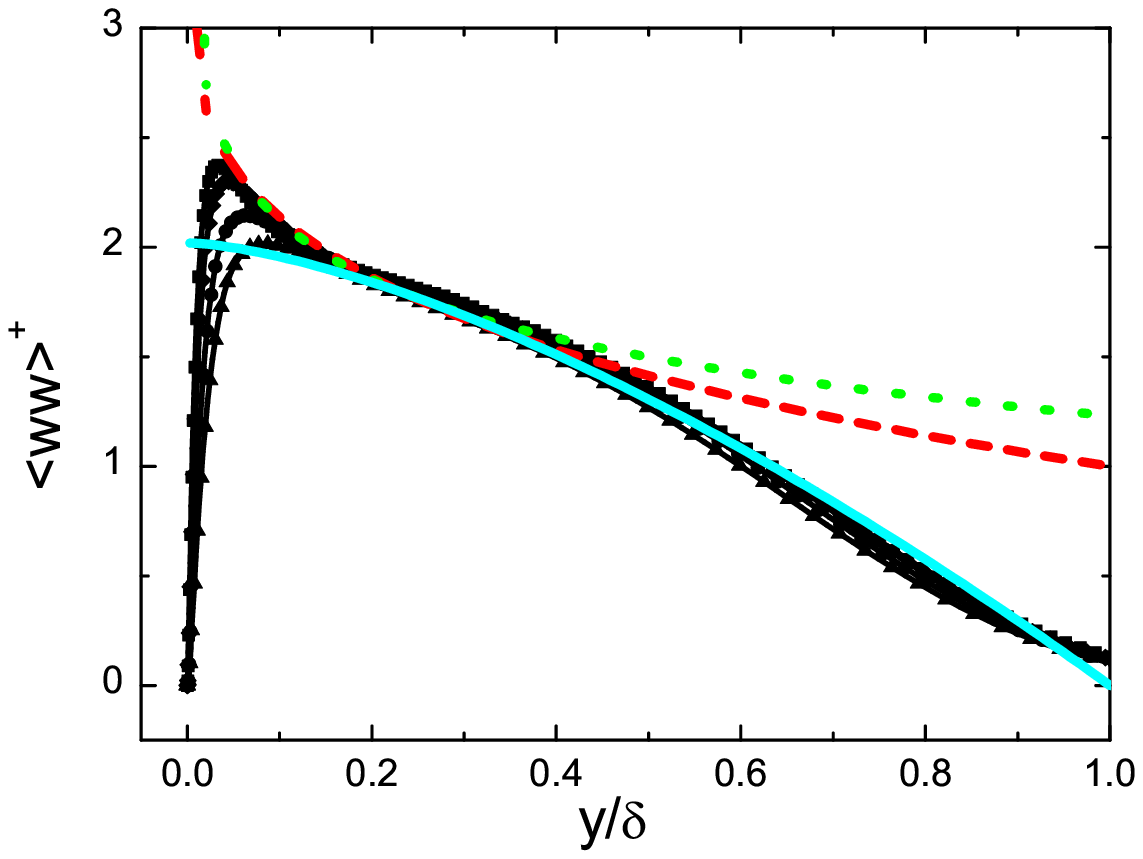}}
\subfloat[]{\includegraphics[trim = 0.75cm 10cm 15.5cm 1cm, clip, width = 6.7 cm]{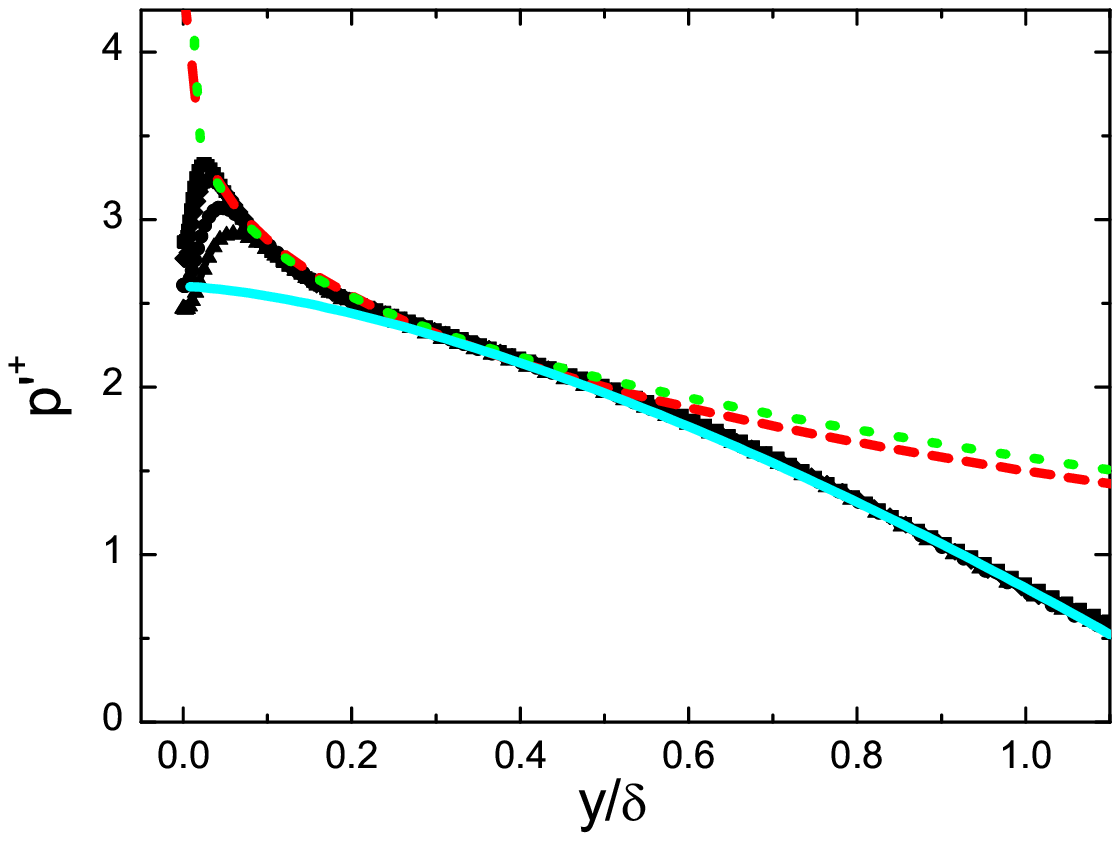}}
\caption{Same plot as figure \ref{fig:plog2wvs}(e) for $\langle ww\rangle^+$, and as \ref{fig:plog2pvs}(e) for $p'^+$, with newly added solid (cyan) lines (\ref{eq:tbl:edge2}) for wake modification in TBL. That is, $\langle ww\rangle^+=2.02[1-(y^\ast)^{3/2}]$ in the left panel and $p'^+=0.8+1.8[1-(y^\ast)^{3/2}]$ in the right.}\label{fig:geo}
\end{figure}

Note that in figures 7 and 8, $\langle ww \rangle^+$ and $p'^+$ depart from (\ref{eq:CS:outer28}) for $y^\ast \gtrsim 0.5$ in TBL, but the agreement persists all the way to the centerline of channel and pipe flows. Recall the findings by \cite{Chen2019JFM} that the total shear stress $\tau^+$ and the Reynolds shear stress $\langle -uv\rangle^+$ in the wake of the TBL differ notably from that in channel and pipe flows. 
The reason for this difference, according to \cite{Chen2019JFM}, is the nonzero mean momentum transport in the wall-normal direction of TBL, i.e. $V/V_e\propto (y^\ast)^{3/2}$ (the subscript $e$ indicates the value at the boundary layer edge, as noted earlier). The latter leads to $\langle -uv\rangle^+\approx \tau^+\approx 1-(y^\ast)^{3/2}$ in TBL, differing from $\langle -uv\rangle^+= 1-y^\ast$ in channel and pipe flows for which $V \equiv 0$.

Once we accept the difference in the $\langle -uv\rangle^+$ behavior, we may expect a similar wake modification on $\langle ww \rangle^+$ and $p'^+$ by the nonzero $V$ in TBL. That is, 
\begin{equation}\label{eq:tbl:edge}
\phi^+-\phi^+_e\propto \phi^+_{uv}-\phi^+_{uv,e},
\end{equation}
where $\phi^+$ represents $\langle ww \rangle^+$ or $p'^+$, and $\phi^+_{uv}$ represents $\langle -uv\rangle^+$. If so, substituting $\phi^+_{uv}\approx 1-(y^\ast)^{3/2}$ into (\ref{eq:tbl:edge}) yields
\begin{equation}\label{eq:tbl:edge2}
\phi^+(y^\ast)-\phi^+_e=c_{\phi} (\phi^+_{uv}-\phi^+_{uv,e})\approx c_{\phi}[1-(y^\ast)^{3/2}],
\end{equation}
where the proportionality coefficient $c_{\phi}$ is independent of $y^\ast$ but may depend on $\phi$.

Verification of (\ref{eq:tbl:edge2}) for TBL is provided in figure \ref{fig:geo}. The agreement with data is quite satisfactory for $y^\ast>0.2$, and the fitting parameters are  $\phi^+_e=0$ and $c_{\phi}=2.02$ for $\langle ww\rangle^+$, while $\phi^+_e=0.8$ and $c_{\phi}=1.8$ for $p'^+$. This model for the wake flow could also be applied to describe $\langle uu \rangle^+$ towards the free stream of TBL, but the deviation is fairly small as shown in figures 5e and 6e, and will not be considered further here.

\section{Perspective and conclusions}

New methods of analysis and generations of new experiments and simulations have revealed deeper layers of interesting questions on wall flow dynamics. Previously unthinkable questions, such as the universality of the K{\'a}rm{\'a}n constant in the mean flow description and the scaling of fluctuations in these flows, as well as the implications of the behavior of fluctuations for the mean velocity itself, can now be asked, and reasonable answers for them can be attempted. In contrast to the mean velocity, concerted effort to understand the scaling of fluctuations is relatively new.  This paper, when taken together with our earlier work \citep{CS2021JFM,CS2022JFM}, provides a self-consistent description of fluctuations in streamwise and spanwise velocity, as well as pressure fluctuations. One of the main qualitative conclusions of this work is that wall-normalized fluctuations are bounded even when $Re_\tau \to \infty$, thus restoring the validity of the standard law of the wall. The alternative scenario of attached eddy hypothesis and its consequences lead to a different conclusion. 

Aiming for an asymptotic description of fluctuations in canonical wall flows, we have obtained several new results, summarized as follows. First, excellent data collapse is achieved for the near-wall rms profiles of streamwise and spanwise velocity fluctuations ($\langle uu\rangle^+$ and $\langle ww\rangle^+$) as well as pressure fluctuations ($p'^+$). Their spatial variations and the Reynolds number dependence are decoupled via the normalization through peak values. With the defect law for the peaks given in CS, a universal near wall expansion (\ref{Eq:Mon1}) is obtained with the specific gauge function $g=Re^{-1/4}_\tau$, consistent with that developed by \cite{Monkewitz2021}.

Moreover, a defect decay (\ref{eq:CS:outer28}) is derived by matching (\ref{Eq:Mon1}) with the outer inviscid similarity (\ref{Eq:outer2}). Compared to the log-profile by Townsend's attached eddy hypothesis, it is shown that (\ref{eq:CS:outer28}) reproduces the data better, not only over a wider $Re_\tau$ domain but also in a larger flow region. As indicated by (\ref{eq:CS:outer28}), there would appear an asymptotic plateau as $y^\ast\rightarrow0$ and $Re_\tau\rightarrow\infty$, which implies a quasi-equilibrium state with contributions to fluctuations coming from all associated eddies that are invariant as wall-normal position changes. If so, the intriguing outer peak of streamwise fluctuation, if one exists, would be bounded by $\langle uu\rangle^+\approx10$.

Finally, a near wall universality (\ref{Eq:universal}) is obtained independent of both Reynolds number and flow geometries. In addition, a wake flow modification in TBL is introduced for $\langle ww\rangle^+$ and $p'^+$, which shows close agreement with data towards the boundary layer edge.

There is no gainsaying that more and better data are required to put all these results on a firmer foundation. It is exciting to await cleaner data at higher Reynolds numbers with improved resolution.\\


\emph{Acknowledgement.} We thank all the authors cited in the figures for making their data available. 
X. Chen acknowledges the support by the National Natural Science Foundation of China, No. 12072012, 11721202 and 91952302.

\bibliography{scibib}
\bibliographystyle{jfm}

\end{document}